\documentclass[aps,10pt,prd,notitlepage,showpacs,nofootinbib,preprintnumbers,superscriptaddress]{revtex4-1}
\usepackage{graphicx}
\usepackage{color,hyperref}
\usepackage[utf8]{inputenc} 
\usepackage{amsmath,amssymb,amsbsy}
\usepackage{float}
\usepackage{calc}
\usepackage{comment}
\usepackage{multirow}
\usepackage{slashed}
\usepackage[normalem]{ulem}
\usepackage{braket}
\usepackage{caption}
\usepackage{subcaption}
\usepackage{wrapfig}
\reversemarginpar
\usepackage{xcolor}

\usepackage{paralist}
\usepackage{comment}
\usepackage[T1]{fontenc}

\usepackage{enumitem}
\usepackage{cancel}

\newcommand{\makeheading}[1]%
        {\hspace*{-\marginparsep minus \marginparwidth}%
         \begin{minipage}[t]{\textwidth+\marginparwidth+\marginparsep}%
                {\large \bfseries #1}\\[-0.15\baselineskip]%
                 \rule{\columnwidth}{1pt}%
         \end{minipage}}

\usepackage{color,hyperref}
\definecolor{darkblue}{rgb}{0.0,0.0,0.3}
\definecolor{darkgreen}{rgb}{0.0,0.3,0.0}
\definecolor{blue(ncs)}{rgb}{0.0, 0.40, 0.80}
\hypersetup{colorlinks,breaklinks,
            linkcolor=darkblue,urlcolor=darkblue,
            anchorcolor=darkblue,citecolor=darkblue}

\usepackage{listings}

\definecolor{codegreen}{rgb}{0,0.6,0}
\definecolor{codegray}{rgb}{0.5,0.5,0.5}
\definecolor{codepurple}{rgb}{0.58,0,0.82}
\definecolor{backcolour}{rgb}{0.95,0.95,0.92}

\lstdefinestyle{mystyle}{
    backgroundcolor=\color{backcolour},   
    commentstyle=\color{codegreen},
    keywordstyle=\color{magenta},
    numberstyle=\tiny\color{codegray},
    stringstyle=\color{codepurple},
    basicstyle=\ttfamily\footnotesize,
    breakatwhitespace=false,         
    breaklines=true,                 
    captionpos=b,                    
    keepspaces=true,                 
    numbers=left,                    
    numbersep=5pt,                  
    showspaces=false,                    showstringspaces=false,
    showtabs=false,                  
    tabsize=2
}

\usepackage{xcolor}

\definecolor{linkcolor}{rgb}{0,0,0.5}

\newcommand {\ignore}[1]{}

\newcommand{\bea}{\begin{eqnarray}}
\newcommand{\eea}{\end{eqnarray}}

\def\gsim{\raise0.3ex\hbox{$\;>$\kern-0.75em\raise-1.1ex\hbox{$\sim\;$}}}
\def\lsim{\raise0.3ex\hbox{$\;<$\kern-0.75em\raise-1.1ex\hbox{$\sim\;$}}}

\usepackage{soul}

\definecolor{mightnightblue}{RGB}{25,25,112}

\definecolor{brown}{rgb}{0.59, 0.29, 0.0}

\def\21{$\mathrm{SU(2)_L \otimes U(1)_Y}$}

\begin{document}
\bibliographystyle{unsrt}   


\title{Testing electroweak scale seesaw models at $e^{-} \gamma$ and $\gamma \gamma$ colliders}

\author{Arindam Das}
\email{adas@particle.sci.hokudai.ac.jp}
\affiliation{Institute for the Advancement of Higher Education, Hokkaido University, Sapporo 060-0817, Japan}
\affiliation{Department of Physics, Hokkaido University, Sapporo 060-0810, Japan}

\author{Sanjoy Mandal}
\email{smandal@kias.re.kr}
\affiliation{Korea Institute for Advanced Study, Seoul 02455, Korea}

\author{Sujay Shil}
\email{sujayshil1@gmail.com}
\affiliation{Instituto de F\'isica, Universidade de S\~ao Paulo, R. do Mat\~ao 1371, 05508-090 S\~ao Paulo, Brazil}
\date{\today}
\begin{abstract}
We investigate the possibilities of probing the electroweak scale seesaw scenarios such as type-I, type-II and type-III seesaw at $e^-\gamma$ and $\gamma\gamma$ colliders. For the case of type-I seesaw, the heavy neutrinos can be produced at $e^{-}\gamma$ colliders in association with a $W$ boson. We study a variety of final states in this case including single and multilepton modes in association with jets to estimate bounds on the light-heavy neutrino mixing angle. In case of type-II seesaw, doubly charged multiplets of the SU$(2)_L$ triplet scalar can be produced in pair at $\gamma \gamma$ collider. We study the multi-leptonic decay modes coming from this pair production of doubly charged Higgs and show how one can probe neutrino mass hierarchy. We also study same sign $W$ boson production from the doubly charged Higgs to study multilepton modes in association with missing energy. From the type-III seesaw, we study same sign dilepton+jets and trilepton+jets modes at $e^-\gamma$ collider which are coming from the neutral and charged component of the triplet fermion in association with a $W$ boson and $Z$ boson, respectively. Due to the existing limits on the triplet fermions from the LHC we choose heavier mass so that the gauge boson originated from the decay of a neutral multiplet can be sufficiently boosted producing a fat-jet signature in association with same sign dilepton and trilepton. Finally we estimate bounds on the light neutrino-heavy triplet fermion mixing angle and compare with the existing bounds. 
\end{abstract}
\maketitle
\preprint{EPHOU-22-017}
\setcounter{page}{1}
\setcounter{footnote}{0}
\section{Introduction}
The tiny neutrino mass and flavor mixing from different experiments \cite{ParticleDataGroup:2020ssz} address a long standing puzzle of the massless neutrino sector in the Standard Model~(SM). However, within the SM the answer of the origin of the tiny neutrino mass and flavor mixing is hitherto unknown despite a variety of proposals to explain these experimental anomalies. 
The most amicable idea was introduced by a dimension five operator in the context of SM which is commonly known as the Weinberg operator \cite{Weinberg:1979sa} which literally inaugurated a simple but powerful direction of beyond the SM (BSM) physics followed by the introduction of SM singlet heavy Majorana type Right Handed Neutrinos (RHNs) \cite{Schechter:1980gr,Minkowski:1977sc,Gell-Mann:1979vob,Yanagida:1979as,Sawada:1979dis,Mohapatra:1979ia} where tiny neutrino mass is generated by the suppression of a lepton number violating high mass scale. This scenario is called the canonical seesaw or type-I seesaw scenario. These RHNs can have masses from eV upto a very heavy scale \cite{Abazajian:2012ys}.
The sterile neutrinos having mass in the eV to keV scale may lead to effects from the short distance neutrino oscillation experiments \cite{BOREXINO:2014pcl,MiniBooNE:2020pnu,Denton:2021czb,Bolton:2019pcu,Machado:2019oxb,MicroBooNE:2021zai} to cosmology \cite{Arguelles:2019tum,Dasgupta:2021ies}. 
The RHNs having mass in the MeV to GeV scale may lead to the effects which can be studied in neutrino beam-dump \cite{Jodlowski:2020vhr,Bakhti:2020szu,Jho:2020jfz}, near detector \cite{Ballett:2019bgd,Carbajal:2022zlp} and meson decay \cite{Cvetic:2012hd,Cvetic:2014nla,Dib:2014iga,Cvetic:2015ura,Mandal:2016hpr,Mandal:2017tab,Dib:2017iva,Cvetic:2017vwl,Chun:2019nwi,Tapia:2021gne,Tastet:2020tzh,Abada:2016plb} experiments. Heavy neutrinos in the GeV to TeV scale has been studied extensively in the context of the high energy colliders from a variety of initial states via prompt and displaced vertex searches, see Refs.~\cite{Tsai:1971vv,Gourdin:1979qr,Gourdin:1979qq,Petcov:1984nf,Maalampi:1991fx,Buchmuller:1992wm,Asaka:2005pn,Han:2006ip,Gorbunov:2007ak,delAguila:2008cj,Mitra:2011qr,Das:2012ze,BhupalDev:2012zg,Dev:2013wba,Das:2014jxa,Das:2015toa,Antusch:2015gjw,Das:2016hof,Das:2017gke,Bhardwaj:2018lma,Antusch:2018bgr,Arganda:2015ija,Das:2017zjc,Das:2017rsu,Chakraborty:2018khw,Das:2018usr,Liu:2020vur,Das:2017nvm,Liu:2019ayx,Mondal:2016kof,Helo:2018qej,Drewes:2013gca,Drewes:2019fou,Drewes:2019vjy,Hirsch:2020klk,Gao:2021one,Abada:2019bac,Abada:2018sfh,Abada:2018qok,DeRomeri:2016gum,Abada:2017jjx,Abada:2015oba,Tastet:2021vwp,Mekala:2022cmm,delAguila:2008hw,delAguila:2007qnc,delAguila:2005pin,delAguila:2005ssc,Atwood:2005bf,Atwood:2004mj,delAguila:2008ir,Bar-Shalom:2008bhg,Atwood:2007zza,Bar-Shalom:2006osy,Cottin:2021tfo,Belanger:2021slj,Cottin:2021lzz,Abada:2022wvh,R:2020odv,Cottin:2021lzz,Dib:2019tuj,Cottin:2018nms,Cottin:2018kmq,Barducci:2020icf,Deppisch:2015qwa,Das:2018hph,Bandyopadhyay:2020mnp,He:2009ua,Mekala:2023diu,Kwok:2023dck,Li:2023tbx}. 

\hspace{0.5cm}Type-II seesaw scenario \cite{Schechter:1981cv,Magg:1980ut,Cheng:1980qt,Lazarides:1980nt,Mohapatra:1980yp,Mandal:2022ysp,Mandal:2022zmy} is another attractive  possibility to study neutrino mass generation mechanism which can explain the origin of tiny neutrino mass and flavor mixing. This is a UV-complete scenario of the dimension five Weinberg operator, where a triplet scalar field with the hypercharge $Y=+2$ is introduced to the SM where we find an extended scalar sector \cite{Arhrib:2011uy,BhupalDev:2013xol,Das:2016bir,Bonilla:2015eha,Chun:2019hce,Primulando:2019evb}. In this case the complex scalar transforms as a triplet under the SU$(2)_L$ group which interacts with the SM gauge bosons, the lepton doublets and the SM Higgs doublet. The neutral component of the triplet scalar acquires a vacuum expectation value (VEV) $v_\Delta$ which allows to generate eV scale light neutrino masses
through the Yukawa interaction between lepton doublets and triplet scalar fields. The type-II seesaw model can naturally allow large neutrino Yukawa couplings simultaneously with a light seesaw scale even below TeV scale. This can be realized with relatively small value of induced triplet VEV $(v_\Delta)$ which naturally
generates $\text{eV}$ scale neutrino masses with even $\mathcal{O}(1)$ neutrino Yukawa coupling. Also, in this scenario, the same Yukawa interaction has a significant impact on charged Higgs phenomenology. The most intriguing element of this model is the presence of a doubly charged Higgs, which can have multiple decay modes whose branching ratios are determined by neutrino oscillation data \cite{deSalas:2020pgw,10.5281/zenodo.4726908}. As a result, the detection of this exotic particle could have a `smoking gun' signature of this model. Several experimental searches to study a variety of signatures from the doubly-charged Higgs at high energy colliders have been conducted in \cite{ATLAS:2017xqs,CMS:2017pet,ATLAS:2018ceg,CMS:2014mra,DELPHI:2002bkf}. Theoretical aspects of the type-II seesaw and triplet scalar models at different colliders have been discussed in \cite{Akeroyd:2005gt,Padhan:2019jlc,Crivellin:2018ahj,Perez:2008ha,Melfo:2011nx,Chakrabarti:1998qy,Aoki:2011pz,Akeroyd:2011zza,Chun:2013vma,delAguila:2013mia, Banerjee:2013hxa,kang:2014jia,Han:2015hba,Han:2015sca,Babu:2016rcr,FileviezPerez:2008jbu,Ashanujjaman:2021txz,Rizzo:1980gz,Keung:1984hn,Djouadi:1997rp}. Depending on the magnitude of the triplet VEV $v_\Delta$, the doubly-charged Higgs mainly decay to same sign dileptons~($v_\Delta\leq 10^{-4}$ GeV) or gauge bosons~($v_\Delta > 10^{-4}$ GeV). Studying these leptonic or gauge boson decay modes, constraints on doubly charged Higgs properties can been probed by the LHC search. For small triplet VEV $v_\Delta\leq 10^{-4}$~GeV, the constraint is $m_{H^{\pm\pm}}>870$ GeV \cite{CMS:2017pet,ATLAS:2017xqs}, whereas for $v_\Delta>10^{-4}$ GeV, the constraint is rather loose, $m_{H^{\pm\pm}}>220$ GeV \cite{ATLAS:2018ceg}. On the other hand, for the case of singly charged and neutral scalars, the official LHC searches have been conducted in the context of type-II seesaw model through the gluon-fusion channels in \cite{CMS:2015lsf,ATLAS:2018gfm,ATLAS:2018ntn,ATLAS:2017eiz,CMS:2018rmh,ATLAS:2018sbw} which are further suppressed by $\mathcal{O}(v_\Delta^2)$, rendering them irrelevant for constraining the type-II seesaw model.

\hspace{0.5cm}Apart from the type-I and type-II seesaw scenarios, type-III seesaw can be considered as another interesting framework which is realized extending the SM by an SU$(2)_{\rm L}$ triplet fermion with zero hypercharge. Such a set-up can generate small neutrino mass \cite{Foot:1988aq} and flavor mixing through the seesaw mechanism. The triplet fermion contains singly charged and charge neutral multiplets. The light neutrino mass is generated by the vacuum expectation value of the charge-neutral multiplet after the seesaw mechanism. These charge neutral multiplets can mix with the SM neutrinos through which they interact with the SM bosons. The charged multiplets of the triplet fermions can also interact with the SM gauge bosons through such mixings at the time of their association with the SM leptons. Therefore such neutral and charged multiplets of the triplet fermions can be produced at the high energy colliders through their interactions with the SM gauge bosons and such production modes are generally suppressed by the light-heavy mixing square. On the other hand the charged multiplets of the fermion triplets can be pair produced directly (i.e., not being suppressed by light-heavy mixing) from the neutral SM boson mediated processes. Type-III seesaw scenario has been studied at the high energy colliders to study a variety of phenomenological aspects like prompt and displaced signatures from multi-lepton and multi-jet channels \cite{Franceschini:2008pz,Biggio:2011ja,Bandyopadhyay:2011aa,Aguilar-Saavedra:2013twa,Bandyopadhyay:2009xa,Klemz:2005nj,Bandyopadhyay:2010wp,Goswami:2017jqs,Jana:2020qzn,Das:2020gnt,Das:2020gnt,Das:2020uer,Ashanujjaman:2021zrh,Ashanujjaman:2021jhi,Li:2009mw}. Recently LHC has studied these triplets from the multi-lepton mode at $\sqrt{s}=13$ TeV with 138 (139) fb$^{-1}$ luminosity at CMS (ATLAS) experiment setting a limit on their masses as 1 TeV from flavor democratic scenario \cite{ATLAS:2022yhd,CMS:2022nty}. 

\hspace{0.5cm}These seesaw scenarios can be tested at different high energy colliders from a variety of final states. Among these high energy machines lepton colliders are interesting ones, however, they can also be modified in an interesting way. Commonly we use electron and positron beams to construct a lepton collider. 
In some proposals muons and anti-muons are also considered to be a future collider where muon mass helps to achieve a high center of mass energy \cite{Geer:2009zz,Adey:2015iha,Han:2020uid,AlAli:2021let}. On the other hand in the context of electron positron colliders, positron beam can be replaced by a back scattered photon to think about an electron photon collider which can have extremely rich physics potential \cite{Arteaga-Romero:1971pai,Budnev:1975poe,Ginzburg:1981ik,Ginzburg:1982bs,Ginzburg:1982yr,Sessler:1998yg,Gluza:1997ts,Bray:2005wv,ECFADESYPhotonColliderWorkingGroup:2001ikq,Velasco:2001fsi,Choudhury:2002aw,Brodsky:2005wk,Bechtel:2006mr,Telnov:1989sd,Telnov:2006cj,Jikia:1992mt,Berge:2000cb,Honkavaara:2005fn,Sekaric:2005ma,Boos:2000gr,Krawczyk:2003yz,DeRoeck:2003cjp,Belusevic:2004pz,Muhlleitner:2005pr,Bechtel:2006mr,Kanemura:2009xs,Mawatari:2011cu,Telnov:2018klr,Demirci:2020zgt,Telnov:2009vq, Choudhury:1994vi,Choudhury:2002hs} at different projected luminosities \cite{Telnov:1999tb,Telnov:2000zx,Pak:2003jq,Makarenko:2003xg} including conceptually interesting beam dump facilities \cite{Shekhtman:2004qg,Shekhtman:2014jfa}. In addition to that both electron and positron can be replaced by back scattered photons to propose photon photon collider. Taking $e^-\gamma$ and $\gamma \gamma$ colliders under consideration we propose some interesting and unique signatures from the seesaw models which can be significant to test neutrino mass generation mechanism in future.

\hspace{0.5cm}In case of type-I seesaw the SM singlet heavy neutrino can be produced at $e^-\gamma$ colliders in association with $W$ boson. The heavy neutrino then can decay into charged lepton and $W$ boson. The leptonic decay of the associated $W$ can produce same sign dilepton (SSDL) and hadronic decay can produce multi-jet signatures provided that the $W$ boson from the heavy neutrino decays into jets. In $e^-\gamma$ colliders we directly produce the first generation heavy neutrino which couples with electron dominantly. The SSDL signature is unlikely to observe from a single heavy neutrino production at the electron positron colliders. As a result this mode could be an interesting complementarity check of the neutrino mass generation mechanism. We also consider the opposite sign dilepton (OSDL) signal to study limits on the mixing angle and compare with those obtained from the SSDL scenarios. 

\hspace{0.5cm}In case of type-II seesaw the triplet scalar has single and double charged multiplets. Doubly charged multiplet can be produced at $e^-\gamma$ collider in association with charged leptons and at $\gamma \gamma$ collider in pair.  The doubly charged multiplet can decay into pair of $W$ bosons or pair of charged leptons giving lepton number and flavor violating signatures which could be probed at these colliders. The leptonic decay of the doubly charged Higgs involve the Yukawa coupling which depends on the neutrino mass hierarchy. Investigation of such decay modes from the charged multiplets of the triplet scalar may enlighten the order of the light neutrino mass at the $e^- \gamma$ and $\gamma \gamma$ colliders. 

\hspace{0.5cm}In case of type-III seesaw the triplet fermion  consisting of neutral and charged multiplets can be produced at the $e^- \gamma$ colliders. Due to the high mass bound on the triplet fermion from the LHC, we investigate the boosted decay of the gauge bosons from the neutral and charged multiplets of the triplet fermion. In case of neutral multiplet we study the SSDL signature with boosted $W$ boson from the neutral multiplet of the triplet fermion. On the other hand we investigate the production of the charged multiplet in association with $Z$ boson. Considering the leptonic decay mode of the associated $Z$ boson and boosted $Z$ boson decay from the charged multiplet of the triplet fermion, we investigate a trilepton signature with a fat-jet. Due to the presence of the visible final states, we  may reconstruct the triplet fermion involved in the neutrino mass generation mechanism. 

\hspace{0.5cm}The paper is arranged in the following way. In Sec.~\ref{model} we discuss the testable seesaw models at the tree level. We calculate the production cross sections of the BSM particles in Sec.~\ref{egm} in the context of electron photon and photon photon colliders. In Sec.~\ref{Analysis} we discuss collider analyses of different production modes of the BSM particles. Finally we conclude the article in Sec.~\ref{conc}.
\section{Testable seesaw models at the tree level}
\label{model}
Mechanism of neutrino mass generation has been proposed in many ways which can explain the origin of tiny neutrino mass and flavor mixing leading to a variety of reach phenomenology. The common scenarios are known as type-I seesaw where SM is extended by SM singlet heavy Majorana neutrinos, type-II seesaw where SM is extended by SU$(2)_{\rm L}$ triplet scalar and type-III seesaw where SM is extended by SU$(2)_{\rm L}$ triplet fermion respectively. In the following section we give a comprehensive review of type-I, II and III seesaw mechanisms:
\subsection{Singlet fermion induced tree level seesaw scenario}
In case of type-I seesaw scenario, SM singlet RHNs~$(N_R^{\beta})$ are introduced. The RHNs directly couple with the SM lepton doublets $(\ell_{L}^{\alpha})$ and the SM Higgs doublet $(H)$. The relevant part of the interaction Lagrangian is written below as 
\bea
\mathcal{L}_{\rm int} \supset -y_D^{\alpha\beta} \overline{\ell_L^{\alpha}}H N_R^{\beta} 
                   -\frac{1}{2} m_N^{\alpha \beta} \overline{N_R^{\alpha C}} N_R^{\beta}  + H. c. .
\label{typeI}
\eea
where $\alpha$ and $\beta$ are flavor indices. After the spontaneous electroweak symmetry breaking through the VEV of the Higgs field, $H =\begin{pmatrix} \frac{v}{\sqrt{2}} &  0 \end{pmatrix}^T$, we obtain the Dirac mass matrix as $M_{D}= \frac{y_D v}{\sqrt{2}}$. We write down the neutrino mass matrix below using the Dirac and Majorana mass matrices as 
\bea
M_{\nu}=\begin{pmatrix}
0&&M_{D}\\
M_{D}^{T}&&M_N
\end{pmatrix}.
\label{typeInu}
\eea
Hence diagonalizing Eq.~\ref{typeInu}, we obtain the seesaw formula for the light Majorana neutrinos as 
\bea
m_{\nu} \simeq - M_{D} M_N^{-1} M_{D}^{T}.
\label{seesawI}
\eea
For example using $M_N\simeq 100$ GeV, we may obtain $y_{D} \sim 10^{-6}$ with $m_{\nu}\simeq 0.1$ eV. However, in the general parameterization for the seesaw formula \cite{Casas:2001sr}, Dirac Yukawa coupling $y_{D}$ can be $\mathcal{O}(1)$, and in this paper we consider such a scenario.


Assuming $M_D M_N^{-1} \ll 1$, the light neutrino flavor eigenstates~($\nu$) can be expressed as a combination of the light ($\nu_m$) and heavy ($N_m$) neutrino mass eigenstates in the following way 
\bea 
  \nu \simeq {\cal N} \nu_m  +  V N_m,  
\label{fla}  
\eea 
where 
\bea
  V = M_D M_N^{-1}, \; 
 {\cal N} =  \left(1 - \frac{1}{2} \epsilon \right) U_{\rm PMNS} ,
\eea
 with $\epsilon = V^* V^T$,  and $U_{\rm PMNS}$ is the usual neutrino mixing matrix 
 by which the mass matrix $m_\nu$ is diagonalized as  
\bea
   U_{\rm PMNS}^T m_\nu U_{\rm PMNS} = {\rm diag}(m_1, m_2, m_3). 
\eea
In the presence of $\epsilon$, the mixing matrix ${\cal N}$ is not unitary \cite{Antusch:2006vwa,Abada:2007ux,Antusch:2014woa, Antusch:2016brq}. 
Replacing the flavor eigenstates from Eq.~\ref{fla} in the charged current (CC) interaction of the SM we obtain the modified form the $\mathcal{L}_{\rm CC}$ in the following as 
\bea 
\mathcal{L}_{\rm CC}= 
 -\frac{g}{\sqrt{2}} W_{\mu}
  \overline{e} \gamma^{\mu} P_L 
   \left( {\cal N} \nu_m+  V N_m \right) + \text{h.c.}, 
\label{CC}
\eea
where $e$ denotes the three generations of the charged 
 leptons in the vector form, and 
$P_{\rm L(R)} =\frac{1}{2} (1\mp \gamma_5)$ are the projection operators. 
Similarly, replacing the flavor eigenstate of the light neutrinos in terms of the mass eigenstates from Eq.~\ref{fla} in the SM neutral current (NC) interaction we obtain
\bea 
\mathcal{L}_{\rm NC}= 
 -\frac{g}{2 c_w}  Z_{\mu} 
\left[ 
  \overline{\nu_m} \gamma^{\mu} P_L ({\cal N}^\dagger {\cal N}) \nu_m 
 +  \overline{N_m} \gamma^{\mu} P_L ( V^\dagger  V) N_m 
+ \left\{ 
  \overline{\nu_m} \gamma^{\mu} P_L ({\cal N}^\dagger   V) N_m 
  + h.c. \right\} 
\right] , 
\label{NC}
\eea
 where $c_w=\cos \theta_w$ is the weak mixing angle. 
Because of non-unitarity of the matrix ${\cal N}$, 
 ${\cal N}^\dagger {\cal N} \neq 1$,  
 the flavor-changing neutral current occurs. 

The dominant decay modes of the heavy neutrino when it's heavier than the SM gauge and scalar bosons are $N \to \ell W$, $\nu_{\ell} Z$, $\nu_{\ell} h$, respectively and the corresponding partial decay widths are respectively given by
\bea
\Gamma(N_{i} \rightarrow \ell_\alpha W) 
 &=& \frac{g^2 |V_{i \alpha}|^{2}}{64 \pi} 
 \frac{ (M_N^2 - M_W^2)^2 (M_N^2+2 M_W^2)}{M_N^3 M_W^2} ,
\nonumber \\
\Gamma(N_{i} \rightarrow \nu_\alpha Z) 
 &=& \frac{g^2 |V_{i \alpha}|^{2}}{128 \pi c_w^2} 
 \frac{ (M_N^2 - M_Z^2)^2 (M_N^2+2 M_Z^2)}{M_N^3 M_Z^2} ,
\nonumber \\
\Gamma(N_{i} \rightarrow \nu_\alpha h) 
 &=& \frac{ |V_{i \alpha}|^2 (M_N^2-M_h^2)^2}{32 \pi M_N} 
 \left( \frac{1}{v }\right)^2 .
\label{widths}
\eea 
The partial decay width of the heavy neutrino into charged gauge bosons being twice as large as neutral one owing to the two degrees of freedom of $(W^{\pm})$.
When heavy neutrinos are in the mass range $10 \text{ GeV} \leq M_{N} < M_W$, they undergo mainly through three body decay channels. In this case, the partial decay widths of $N_i$ can be approximately given by
\bea
 \Gamma(N_i \to \ell_\alpha^- \ell_\beta^+  \nu_{\ell_\beta}) & = \Gamma(N_i \to \ell_\alpha^+ \ell_\beta^- \bar \nu_{\ell_\beta}) \simeq  |V_{\alpha i}|^2 \frac{G_F^2}{192 \pi^3} M_{N_i}^5 \quad (\alpha \neq \beta), \\
 \Gamma(N_i \to \ell_\beta^- \ell_\beta^+  \nu_{\ell_\alpha}) & = \Gamma(N_i \to \ell_\beta^+ \ell_\beta^- \bar \nu_{\ell_\alpha}) \nonumber \\ 
& \simeq  |V_{ \alpha i}|^2 \frac{G_F^2}{192 \pi^3} M_{N_i}^5 \left( \frac14 \cos^2 2 \theta_W + \sin^4 \theta_W \right) \quad (\alpha \neq \beta), \\ 
\label{int}
 \Gamma(N_i \to \ell_\alpha^- \ell_\alpha^+  \nu_{\ell_\alpha}) & = \Gamma(N_i \to \ell_\alpha^+ \ell_\alpha^- \bar \nu_{\ell_\alpha}) \nonumber \\
& \simeq  |V_{ \alpha i}|^2 \frac{G_F^2}{192 \pi^3} M_{N_i}^5 \left( \frac14 \cos^2 2 \theta_W + \cos 2 \theta_W + \sin^4 \theta_W \right), \\
\Gamma(N_i \to \nu_\beta \bar \nu_\beta  \nu_{\ell_\alpha}) & = \Gamma(N_i \to \nu_\beta \bar \nu_\beta \bar \nu_{\ell_\alpha}) \simeq   |V_{\alpha i}|^2 \frac{1}{4} \frac{G_F^2}{192 \pi^3} M_{N_i}^5  , \\
 \Gamma(N_i \to \ell_\alpha^-  q_a \bar q_b)  &= \Gamma(N_i \to \ell_\alpha^+  \bar q_a  q_b) \simeq  N_c |V_{ \alpha i}|^2 |V_{\rm CKM}^{ab} |^2 \frac{G_F^2}{192 \pi^3} M_{N_i}^5, \\
\Gamma(N_i \to q_a \bar q_a  \nu_{\ell_\alpha}) & = \Gamma(N_i \to q_a \bar q_a \bar \nu_{\ell_\alpha}) \simeq N_c  |V_{ \alpha i}|^2 \frac{G_F^2}{192 \pi^3} M_{N_i}^5 2 \left( |g_V^q|^2 +  |g_A^q|^2 \right) , 
\eea
where
\bea
g_V^u=\frac{1}{2} -\frac{4}{3} \sin^2 \theta_W, \, \, g_A^u= -\frac{1}{2}, \nonumber \\
g_V^d = -\frac{1}{2}+\frac{2}{3} \sin^2 \theta_W, \, \, g_A^u= \frac{1}{2} .
\eea
respectively which come from the interaction between $Z$ boson and the quarks. $N_c=3$ is the color factor for the quarks. 
\begin{figure}[h]
\centering
\includegraphics[width=0.75\textwidth]{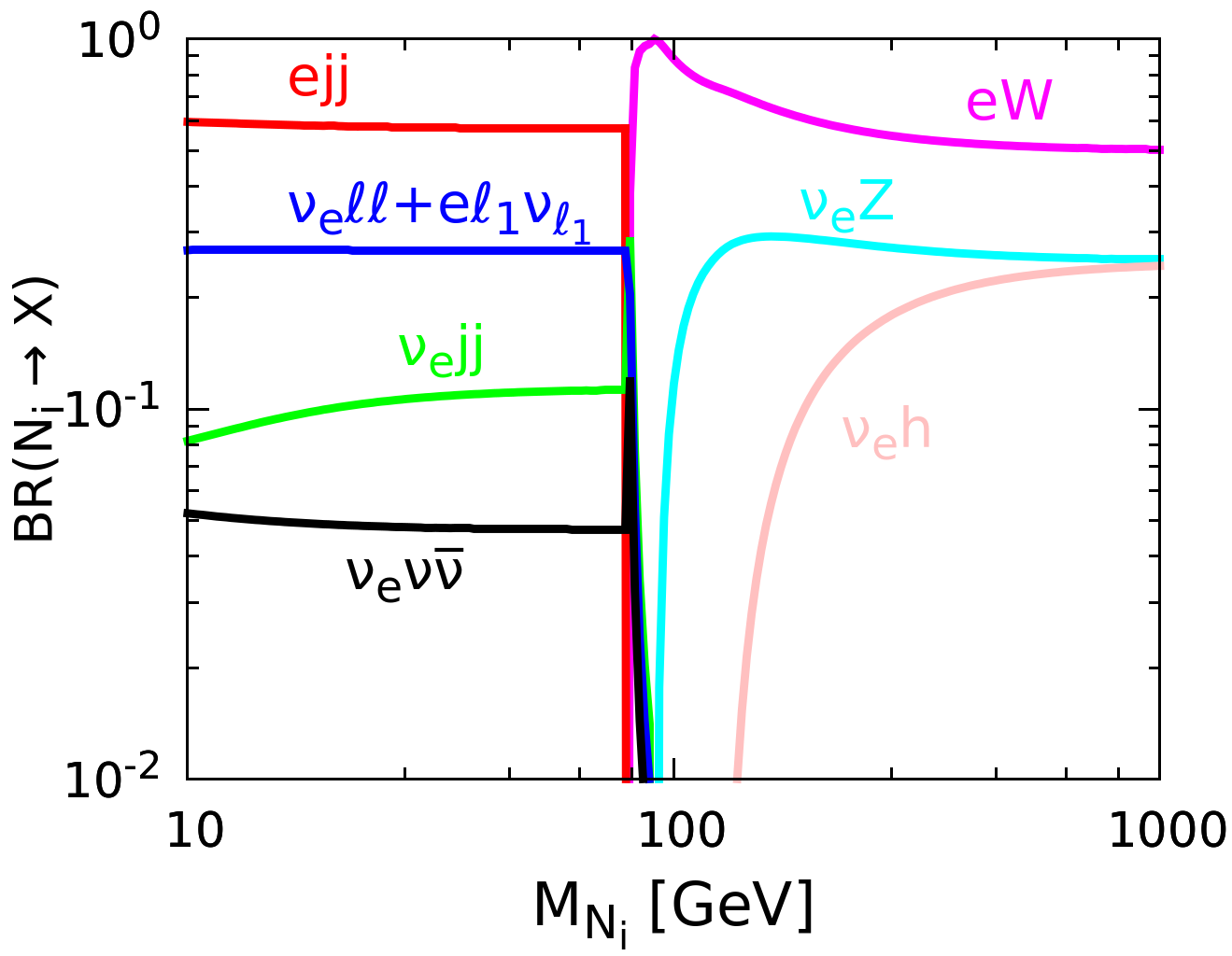}
\caption{Branching ratios of $N_i$ to different final states under the assumption $V_{eN_i}=1, V_{\mu N_i}=0$ and $V_{\tau N_i}=0$. We show the branching ratios to three body leptonic channels $e\ell_1\nu_{\ell_1}+\nu_e\ell\ell$, $\nu_e\nu\bar{\nu}$ and semi-leptonic channels such as $ejj$ and $\nu_e jj$, where $\ell=e,\mu,\tau$, $\ell_1=\mu,\tau$. For relatively large $M_{N_i}$ the two body decay channels such as $eW$, $\nu_e Z$ and $\nu_e h$ dominates. }
\label{fig:NiBR}
\end{figure}
In Fig.~\ref{fig:NiBR}, we show the branching ratios of heavy neutrinos $N_i$ to various final states with the assumption $V_{eN_i}=1, V_{\mu N_i}=0$ and $V_{\tau N_i}=0$. For mass range $M_{N_i}<M_W$, three body leptonic or semi-leptonic final states dominates, whereas for $M_{N_i}>M_W$, two body decay channels starts to dominate. Note that for large values of $M_N$ with $|V_{\ell N_i}|\neq 0$, the branching ratios can be obtained as
\begin{align*}
\text{BR}(N_i\to\ell W):\text{BR}(N_i\to\nu_{\ell}Z):\text{BR}(N_i\to\nu_{\ell}h)=2:1:1.
\end{align*} 
Note that for sufficiently small mixing and relatively small $M_{N_i}$, the heavy neutrinos are long-lived particles that can travel macroscopic distances before they decay, giving rise to displaced vertex signatures. In our study we consider mass and mixing angle to be free parameter and the mixing angle is always large enough to have prompt decay.
\subsection{Triplet scalar induced tree level seesaw scenario}
Here we focus on type-II seesaw scenario \cite{Schechter:1981cv,Magg:1980ut,Cheng:1980qt,Lazarides:1980nt,Mohapatra:1980yp}. It introduces one heavy $SU(2)_{L}$ triplet scalar $\Delta=(\Delta^{++},\Delta^{+},\Delta^{0})^{T}$, with hypercharge $Y_{\Delta}= 2$. It is convenient to describe $\Delta$ in its matrix form as
\begin{equation}
\Delta=\left(\begin{array}{cc}
\Delta^{+}/\sqrt{2} & \Delta^{++}\\
\Delta^{0} & -\Delta^{+}/\sqrt{2}
\end{array}\right).
\end{equation}
The new triplet scalar field $\Delta$, being a triplet under $SU(2)_L$ interacts with the SM gauge bosons. In
addition to the gauge interactions, $\Delta$ has Yukawa interaction with the SM lepton doublet. The relevant Lagrangian reads as
\begin{equation}
{\cal L}_{{\rm type\ II}}=\left[iY_{\Delta\alpha\beta}L_{\alpha}^{T}C^{-1}
    \tau_2\Delta L_{\beta}+\text{h.c.}\right]+\left(D_{\mu}\Phi\right)^{\dagger}\left(D^{\mu}\Phi\right) +\left(D_{\mu}\Delta\right)^{\dagger}\left(D^{\mu}\Delta\right)-V(\Phi,\Delta),
    \label{type2}
\end{equation}
where $Y_{\Delta}^{\alpha\beta}$ is a symmetric complex matrix, $L_\alpha$ are lepton doublets, $C$ is the charge conjugation operator, and $D_\mu$ is the covariant derivative of the related scalar field.
The scalar potential $V(\Phi,\Delta)$ is given as, 
\begin{equation}
\begin{aligned}
 V(\Phi,\Delta) = & -m_{\Phi}^{2}\Phi^{\dagger}\Phi + \frac{\lambda}{4}(\Phi^{\dagger}\Phi)^{2} +  \tilde{M}_{\Delta}^{2}{\rm Tr}\left[\Delta^{\dagger}\Delta\right]+\lambda_{2}\left[{\rm Tr}\Delta^{\dagger}\Delta\right]^{2}+\lambda_{3}{\rm Tr}\left[\Delta^{\dagger}\Delta\right]^{2}\\
  &+ \left[\mu \Phi^{T}i\sigma_{2}\Delta^{\dagger}\Phi+\text{h.c.}\right]+\lambda_{1}(\Phi^{\dagger}\Phi){\rm Tr}\left[\Delta^{\dagger}\Delta\right]+\lambda_{4}\Phi^{\dagger}\Delta\Delta^{\dagger}\Phi.\label{eq:potential}
\end{aligned}
\end{equation}
A tiny induced VEV for the triplet characterizes the Higgs triplet UV-completion of the Weinberg operator. Minimization of the total potential $V(\Phi,\Delta) $ leads to the relations
\begin{align}
 \tilde{M}_\Delta^2&=\displaystyle M_\Delta^2-\frac{1}{2}\left[2v_\Delta^2(\lambda_2+\lambda_3)+v_\Phi^2(\lambda_1+\lambda_4)\right],\,\,\text{ with } M_\Delta^2\equiv \frac{v_\Phi^2\mu}{\sqrt{2}v_\Delta}.  \label{Tadpole1} \\
m_\Phi^2&=\displaystyle \frac{1}{2}\left[\frac{v_\Phi^2\lambda}{2}+v_\Delta^2(\lambda_1+\lambda_4)-2\sqrt{2}\mu v_\Delta\right].
\label{Tadpole2}
\end{align}
In the limit $M_{\Delta} \gg v_{\Phi} $, which is consistent with all the existing constraints, we can solve  Eq. (\ref{Tadpole1}) for $v_{\Delta}$.
Keeping terms of $\mathcal{O}(v_{\Phi}/M_{\Delta})$ we get the small induced triplet VEV
\begin{equation}
v_{\Delta} \approx \frac{\mu v_{\Phi}^{2}}{\sqrt{2}\tilde{M}_{\Delta}^{2}}.
\label{V-triplet-approx}
\end{equation}
The light neutrino mass can be produced from the type-II scenario using the Yukawa interaction between the scalar triplet and SM lepton doublets given in Eq.~\ref{type2} as 
\bea
m_\nu = \sqrt{2}Y_\Delta v_\Delta = Y_\Delta \frac{\mu v_{\Phi}^2}{\tilde{M}_{\Delta}^{2}}.
\eea
From the observed light neutrino mass constraints, the triplet Yukawa and the triplet VEV can be written using the U$_{\rm PMNS}$ matrix depending on the neutrino oscillation data in the following way 
\bea
Y_\Delta= \frac{1}{\sqrt{2} v_\Delta} U_{\rm{PMNS}} m_\nu^{\rm diag} U_{\rm{PMNS}}^{T},
\label{eq:YDelta}
\eea
where $m_\nu^{\rm diag}$ is the neutrino mass eigenvalues which further depends on the neutrino oscillation data and the normal and inverted orderings of the light neutrino masses. Eq.~(\ref{V-triplet-approx}) explicitly shows that the smallness of the triplet VEV can be induced either by a small $\mu$, or by a large value for $\tilde{M}_\Delta$. We can write the doublet and triplet neutral fields as follows after spontaneous symmetry breaking:
\begin{equation}
\Delta = \frac{1}{\sqrt{2}}\begin{pmatrix}
 \Delta^{+}& \sqrt{2}\Delta^{++}\\ 
 v_{\Delta} + h_{\Delta} + i\eta_{\Delta}& - ~\Delta^{+}
\end{pmatrix},~~~~~~~
\Phi = \frac{1}{\sqrt{2}}\begin{pmatrix}
\sqrt{2}{\Phi}^{+}\\ 
v + h_{\Phi} + i\eta_{\Phi}
\end{pmatrix}.
\end{equation}
Scalar sector contains 10 degrees of freedom, among which, after EW breaking, seven remain as physical fields with definite masses: $H^{\pm\pm}$, $H^{\pm}$, and the neutral Higgs bosons $h$, $H^0$ and $A^0$.   The doubly-charged Higgs $H^{\pm\pm}$ is simply the $\Delta^{\pm\pm}$ present in $\Delta$. The physical masses of doubly charged Higgs bosons $H^{\pm \pm}$ can be written as:
\begin{align}
m_{H^{++}}^2=M_\Delta^2-v_\Delta^2\lambda_3-\frac{\lambda_4}{2}v_\Phi^2. \label{eq:mhpp}
\end{align}
The other mass eigenstates can be obtained by rotating the gauge eigenstates as shown below:
\begin{eqnarray*}
&&\left( \begin{array}{c} \Phi^\pm \\ \Delta^\pm \end{array} \right) = R(\beta_\pm) \left( \begin{array}{c} H^\pm \\ G^\pm \end{array} \right),~ \left( \begin{array}{c} h_\Phi \\ h_\Delta \end{array} \right) = R(\alpha) \left( \begin{array}{c} h \\ H^0 \end{array} \right),
\\
&&\left( \begin{array}{c} \eta_\Phi \\ \eta_\Delta \end{array} \right) = R(\beta_0) \left( \begin{array}{c} A^0 \\ G^0 \end{array} \right),~ R(\theta) =\left( \begin{array}{cc} \cos\theta & -\sin\theta \\ \sin\theta & \cos\theta \\ \end{array} \right)~,
\end{eqnarray*}
where $\beta^{\pm},\beta_0$ and $\alpha$ are the rotation angles with $\tan\beta^{\pm}=\frac{\sqrt{2}v_\Delta}{v_\Phi}$, $\tan\beta_0=\frac{2v_\Delta}{v_\Phi}$ and $\tan(2\alpha)=\frac{2B}{A-C}$ where,
\begin{eqnarray}
  A &=& \frac{\lambda}{2}{v_\Phi^2}, \; \;
  B =v_\Phi ( -\sqrt{2}\mu+(\lambda_1+\lambda_4)v_\Delta) , \; \; 
  C = M_\Delta^2+2(\lambda_2+\lambda_3)v_\Delta^2 .
\label{eq:ABC}
\end{eqnarray}
The two scalar fields $\Phi^{\pm}$ and $\Delta^{\pm}$ from $\Phi$ and $\Delta$ mix, giving $H^{\pm}$ and the unphysical charged Goldstone $G^{\pm}$. The physical masses of singly-charged Higgs bosons $H^{\pm}$  can be written as:
    \begin{eqnarray}
m_{H^+}^2= \left(M_\Delta^2-\frac{\lambda_4}{4}v_\Phi^2\right)\left(1+\frac{2v_\Delta^2}{v_\Phi^2}\right).\label{eq:mhp}
\end{eqnarray} 
Similarly, $\eta_{\Delta}$ and $\eta_{\Phi}$ will mix and give rise to the CP-odd scalar $A^0$ and the neutral Goldstone boson $G^0$ which becomes the longitudinal mode of $Z$ boson. The mass of this CP-odd Higgs field $A^0$ has the following mass:
\begin{eqnarray}
m_{A^{0}}^2 &= &M_\Delta^2\left(1+\frac{4v_\Delta^2}{v_\Phi^2}\right). \label{mA}
\end{eqnarray}  
Finally, CP-even fields $h_\Delta$ and $h_\Phi$ will mix and give rise to the SM Higgs boson $h$ and a heavy Higgs boson $H^0$. The masses of $h$ and $H^0$ have the following physical masses:
\begin{eqnarray}
m_{h}^2&=&\frac{1}{2}[A+C-\sqrt{(A-C)^2+4B^2}], \label{eq:mh0}\\
m_{H^0}^2&=&\frac{1}{2}[A+C+\sqrt{(A-C)^2+4B^2}]. \label{eq:mH0}
\end{eqnarray}
The triplet VEV $v_\Delta$ is subject to constraint coming from $\rho$ parameter. The $\rho$ parameter leads to upper bound of $\mathcal{O}(1\text{ GeV})$ on the triplet VEV $v_\Delta$~\cite{ParticleDataGroup:2020ssz}. For $v_\Delta\ll v_\Phi$, the masses of the physical Higgs bosons can be approximated as follows:
\begin{align}
m_{H^{\pm\pm}}^2\simeq M_\Delta^2-\frac{\lambda_4}{2}v_\Phi^2,\,\,\,\, m_{H^\pm}^2\simeq M_\Delta^2-\frac{\lambda_4}{4}v_\Phi^2,\,\,\,\, m_h^2\simeq 2\lambda v_\Phi^2\,\,\,\,\text{and}\,\,\,\,m_{H^0}^2\approx m_{A^0}^2\simeq M_{\Delta}^2 ,
\end{align}
so their mass-squared differences are given by:
\begin{align}
m_{H^\pm}^2-m_{H^{\pm\pm}}^2\approx m_{H^0/A^0}^2-m_{H^\pm}^2\approx \frac{\lambda_4}{4}v_\Phi^2.
\end{align}
We further define the two mass-splittings as follows:
\begin{align}
\delta m_1=m_{H^0}-m_{H^\pm},\,\,\,\,\,\,\,\, \delta m_2= m_{H^\pm}-m_{H^{\pm\pm}}.
\label{eq:dm}
\end{align}
With the assumptions $v_\Delta\ll v_\Phi$ and $M_{\Delta}^2\gg |\lambda_4| v_\Phi^2$, the two mass-splittings $\delta m_{1,2}$ can be approximated as 
\begin{align}
\Delta m\equiv \delta m_{1,2}\approx \frac{\lambda_4}{8}\frac{v_\Phi^2}{M_\Delta}.
\end{align}
Hence, the masses of all the physical Higgs states can be written in terms of just two parameters: $m_{H^{\pm\pm}}$ and $\Delta m$. Note that $\lambda_4$ should be small as a relatively large value of this quartic coupling at the electroweak scale can become non-perturbative at high energies even below the Planck scale. Also the mass splitting $\Delta m=m_{H^\pm}-m_{H^{\pm\pm}}$ affects the electroweak precision data observables, such as $S, T$ and $U$ parameters. These put a tight constraints on the mass splitting $|\Delta m|\lesssim 40$ GeV~\cite{Aoki:2012jj,Chun:2012jw,Primulando:2019evb,Das:2016bir}. Note that, three different mass spectra are expected depending on the value (sign) of $\lambda_4$, as:
$(1).\,\, \lambda_4=0:\,\,\Delta m\approx 0\,( m_{H^{\pm\pm}}\simeq m_{H^\pm}\simeq m_{H^0/A^0}) $,
$(2).\,\, \lambda_4<0:\,\,\Delta m<0\,( m_{H^{\pm\pm}} > m_{H^\pm} > m_{H^0/A^0})$ and
$(3).\,\, \lambda_4>0:\,\,\Delta m>0\,( m_{H^{\pm\pm}} < m_{H^\pm} < m_{H^0/A^0})$. 
These will be important for our later discussions. The Higgs triplet has, of course, interaction with gauge bosons, as well as Yukawa couplings to the SM fermions. This opens up a number of possible decay modes. We write down all possible partial decay widths of the charged multiplets $H^{\pm\pm}$ and $H^\pm$ of the triplet scalar. The relevant partial decay widths of $H^{\pm\pm}$ in the case of degenerate scenario ($\delta m\approx 0$) are written below
\begin{align}
&\Gamma (H^{\pm \pm} \to l^{\pm}_i l^{\pm} _j)=\frac{m_H^{\pm \pm} } {(1+\delta_{ij}) 8 \pi}   \left |\frac{m_{ij}^{\nu}}{v_{\Delta}} \right |^2, \, \, m^{\nu}=Y_{\Delta} v_{\Delta}/\sqrt{2},\\
&\Gamma (H^{\pm \pm} \to W^{\pm} W^{\pm})=\frac{g^2 v^2_{\Delta}}{8 \pi m_{H^{\pm \pm}}} \sqrt{1- \frac{4}{r^2_W}} \left[ \left (2+(r_W/2-1)^2 \right ) \right ],
\end{align}
where $r_W=\frac{m_{H^{\pm \pm}}}{M_W}$. Here $m^{\nu}$ denotes the neutrino mass matrix, $i,j$ are the generation indices. For the case of negative mass splitting $(\delta m < 0)$, one must also consider the additional decay channel:
\begin{align}
\Gamma(H^{\pm\pm} \to H^\pm W^{\pm *})=
\frac{9g^4m_{H^{\pm \pm}}\cos^2\beta_\pm}{128\pi^3} G\left(\frac{m_{H^\pm}^2}{m_{H^{\pm \pm}}^2},\frac{m_W^2}{m_{H^{\pm \pm}}^2}\right),
\end{align} 
where $\tan\beta_{\pm}=\frac{\sqrt{2}v_\Delta}{v_\Phi}$ and the functions $\lambda(x,y)$, $G(x,y)$ are given as
\begin{align}
&\lambda(x,y)=(1-x-y)^2-4xy~,
\\
&G(x,y)=\frac{1}{12y}\Bigg[2\left(-1+x\right)^3-9\left(-1+x^2\right)y+6\left(-1+x\right)y^2 -6\left(1+x-y\right)y\sqrt{-\lambda(x,y)}\Bigg\{\tan^{-1}\left(\frac{1-x+y}{\sqrt{-\lambda(x,y)}}\right) \notag
\\
&+\tan^{-1}\left(\frac{1-x-y}{\sqrt{-\lambda(x,y)}}\right)\Bigg\}-3\left(1+\left(x-y\right)^2-2y\right)y\log x\Bigg]~,
\end{align}
The relevant partial decay widths of $H^{\pm}$ in the case of degenerate scenario ($\delta m\approx 0$) are written below
\begin{align}
&\Gamma(H^\pm \to q\bar{q}')=\frac{3m_{H^\pm}^3\sin^2\beta_\pm}{8\pi v_\Phi^2} 
\left[\left(\frac{m_q^2}{m_{H^\pm}^2}+\frac{m_{q'}^2}{m_{H^\pm}^2}\right)\left(1-\frac{m_q^2}{m_{H^\pm}^2}-\frac{m_{q'}^2}{m_{H^\pm}^2}\right)-4\frac{m_q^2}{m_{H^\pm}^2}\frac{m_{q'}^2}{m_{H^\pm}^2}\right] \lambda^{\frac{1}{2}}\left(\frac{m_q^2}{m_{H^\pm}^2},\frac{m_{q'}^2}{m_{H^\pm}^2}\right),
\\
&\Gamma(H^\pm \to \ell_i^\pm\nu_j)=\frac{m_{H^\pm}}{8\pi v_\Phi^2} \left( \delta_{ij}m_i^2\sin^2\beta_\pm +|Y^{ij}_\Delta|^2v_\Phi^2\cos^2\beta_\pm \right) \left(1-\frac{m_i^2}{m_{H^\pm}^2}\right)^2,
\\
&\Gamma(H^\pm \to W^\pm Z) =\frac{g^4v_\Delta^2\cos^2\beta_\pm}{32\pi \cos^2\theta_w m_{H^\pm}}\left[\lambda\left(\frac{m_W^2}{m_{H^\pm}^2},\frac{m_Z^2}{m_{H^\pm}^2}\right)\right]^{1/2}\left[2+\frac{m_{H^\pm}^4}{4m_W^2m_Z^2}\left(1-\frac{m_W^2}{m_{H^\pm}^2}-\frac{m_Z^2}{m_{H^\pm}^2}\right)^2\right].
\end{align}
For the case of non-zero mass splitting $|\delta m|\neq 0$, one also has the following decay channels for $H^{\pm}$: 
\begin{align}
&\Gamma(H^\pm \to H^0/A W^{\pm *}) =\frac{9g^4m_{H^\pm}}{512\pi^3}\xi_{H^\pm W^\mp H^0/A}^2G\left(\frac{m_{H^0/A}^2}{m_{H^\pm}^2},\frac{m_W^2}{m_{H^\pm}^2}\right);\text{  for  }\delta m<0,
\\
&\Gamma(H^\pm \to H^{\pm \pm} W^{\mp *})=
\frac{9g^4m_{H^\pm}\cos^2\beta_\pm}{128\pi^3} G\left(\frac{m_{H^{\pm \pm}}^2}{m_{H^\pm}^2},\frac{m_W^2}{m_{H^\pm}^2}\right); \text{  for  }\delta m > 0,
\end{align} 
where
$\xi_{H^\pm W^\mp \hat{\varphi}}=\cos\alpha\sin\beta_\pm-\sqrt{2}\sin\alpha\cos\beta_\pm$, $\sin\alpha\sin\beta_\pm+\sqrt{2}\cos\alpha\cos\beta_\pm$, $\sin\beta_0\sin\beta_\pm+\sqrt{2}\cos\beta_0\cos\beta_\pm$, $\text{ for } \hat{\varphi}= h^0,~H^0,~A$, $\tan(2\alpha)=\frac{2B}{A-C}\text{ and }\tan\beta_0=\frac{2v_\Delta}{v_\Phi}.$
\begin{figure}[h]
\centering
\includegraphics[width=0.45\textwidth]{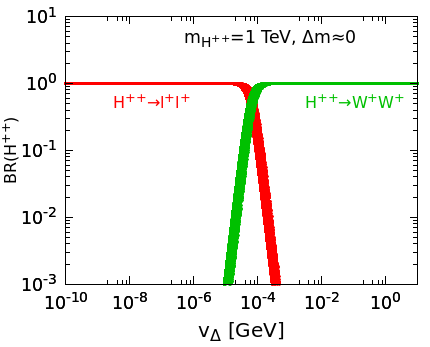}
\includegraphics[width=0.45\textwidth]{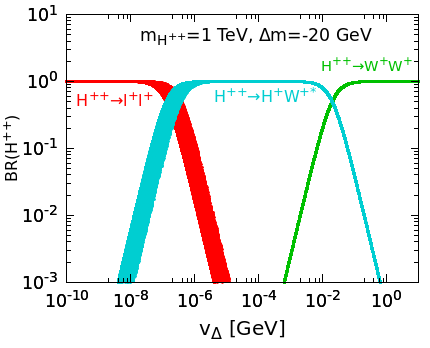}
\caption{Branching ratios of $H^{\pm\pm}$ for mass $m_{H^{\pm\pm}}=1$~TeV with mass splitting $\Delta m\approx 0$~(left-panel) and $\Delta m=-20$~GeV~(right-panel), respectively. The red, green and cyan points stand for the decay mode $H^{\pm\pm}\to \ell^{\pm}\ell^{\pm}$, $H^{\pm\pm}\to W^{\pm} W^{\pm}$ and $H^{\pm\pm}\to H^{\pm}W^{\pm *}$, respectively. Note that the decay mode $H^{\pm\pm}\to H^{\pm}W^{\pm *}$ is only open when mass-splitting is negative and further dominates in the intermediate triplet VEV region when mass-splitting is relatively large. The Dirac phase $\delta_{\text{CP}}$ is varied in the range $[-\pi:\pi]$ and other oscillation parameters are varied within their $3\sigma$ ranges whereas the lightest neutrino mass is varied in the range $10^{-4}\,\text{eV}\leq m_{\text{lightest}}<0.1\,\text{eV}$.}
\label{fig:HppBR}
\end{figure}
In Fig.~\ref{fig:HppBR}, we display the branching ratios of $H^{\pm\pm}$ for $m_{H^{\pm\pm}}=1$~TeV with mass splitting $\Delta m\approx 0$~(left panel) and $\Delta m=-20$~GeV~(right panel), respectively. For the case of small mass splitting, depending on the magnitude of the triplet VEV $v_\Delta$, $H^{\pm\pm}$ mainly decay to same-sign dileptons~($v_\Delta\leq 10^{-4}$~GeV) or gauge bosons~($v_\Delta > 10^{-4}$~GeV). On the other hand from the right panel of Fig.~\ref{fig:HppBR} we see that with relatively large negative mass splitting, the cascade decay $H^{\pm\pm}\to H^{\pm} W^{\pm *}$ quickly dominate over the leptonic and diboson decay modes in the intermediate triplet VEV region.
\begin{figure}[h]
\includegraphics[width=0.45\textwidth]{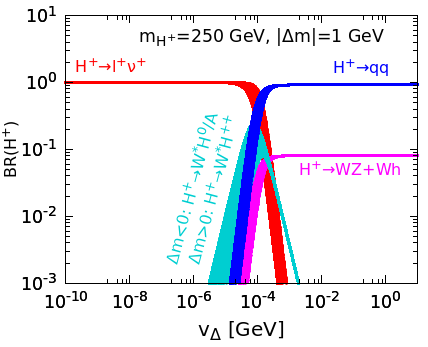}
\includegraphics[width=0.45\textwidth]{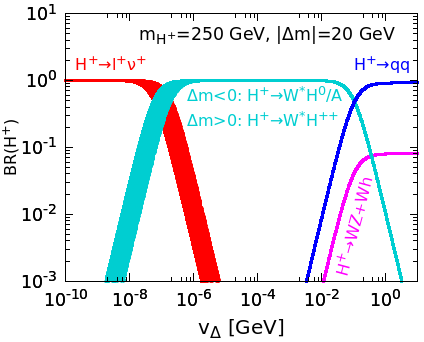}
\includegraphics[width=0.45\textwidth]{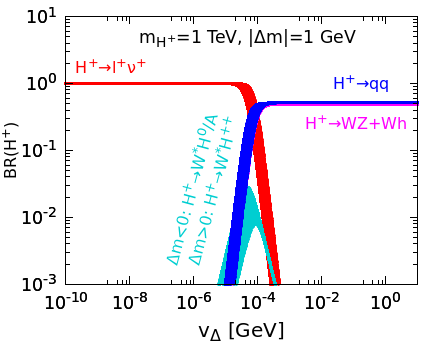}
\includegraphics[width=0.45\textwidth]{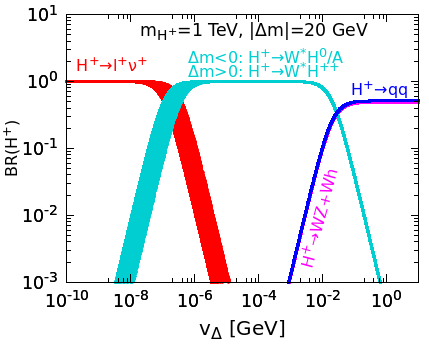}
\caption{Branching ratios of $H^{\pm}$ for mass $m_{H^{\pm}}=250$~GeV~(top-panel) and $m_{H^{\pm}}=1$~TeV~(bottom-panel) with mass splitting $|\Delta m|= 1\,\text{GeV}$~(left-panel) and $|\Delta m|=20$~GeV~(right-panel). The singly-charged Higgs boson $H^{\pm}$ has four decay modes: (i) $\ell^{\pm}\nu$~(red), (ii) $qq$~(blue), (iii) $W^{\pm}Z+W^{\pm}h$~(magenta) and (iv) $H^0/A W^{*}\,(\Delta m<0)$~(cyan) or $H^{\pm\pm}W^{\mp *}\,(\Delta m>0)$~(cyan). For large mass of $H^{\pm}$, hadronic and dibosonic decay modes almost coincides. The oscillation parameters are varied again in same manner as in Fig.~\ref{fig:HppBR}.}
\label{fig:HpBR}
\end{figure}
The branching ratio of the singly-charged Higgs $H^{\pm}$ into various channels is depicted in Figs.~\ref{fig:HpBR}. $H^\pm$ has four decay modes: (i) leptonic decay, i.e. $\ell^\pm\nu$, (ii) hadronic decay, i.e. $q\bar{q'}$, (iii) diboson decay, i.e. $W^\pm Z$, $W^\pm h$ and (iv) cascade decay, i.e. $H^0/A W^{\pm*}$~($\delta m<0$) or $H^{\pm\pm}W^{\mp*}$~($\delta m>0$). When comparing the left and right panels in Fig.~\ref{fig:HpBR}, we can see that cascade decays dominate in the intermediate triplet VEV area for relatively significant mass splitting. The branching ratio pattern for singly-charged scalar decays is extremely sensitive to mass splitting and the triplet VEV, as it is for doubly-charged scalar decays. In both Figs.~\ref{fig:HppBR} and \ref{fig:HpBR}, the band for various decay modes are due to the variation of oscillation parameters. The lightest neutrino mass is varied in the range $10^{-4}\text{ eV}\leq m_{\rm lightest}\leq 0.1\text{ eV}$. The Dirac phase $\delta_{\text{CP}}$ is varied in the range $[-\pi:\pi]$ and other oscillation parameters are varied within their $3\sigma$ ranges~\cite{deSalas:2020pgw,10.5281/zenodo.4726908}.

The tightest constraint on triplet scalar comes basically from doubly charged Higgs scalar searches at LHC. One clearly sees from Fig.~\ref{fig:HppBR} that the collider search strategy for the doubly charged Higgs boson crucially depends on the value of the triplet VEV $v_\Delta$ and mass splitting $\Delta m$. The direct limit on $m_{H^{\pm\pm}}$ has been derived from collider searches of multi-lepton final states assuming $\Delta m\approx 0$. Stringent constraints on $m_{H^{\pm\pm}}$ have been placed by the 13 TeV LHC searches by analysing Drell-Yan production $pp\to H^{++}H^{--}$ and subsequent decay in the $H^{\pm\pm}\to\ell^{\pm}\ell^{\pm}$ channel. In addition to this LHC has also studied the associated production of $H^{\pm\pm}H^{\mp}$ through s-channel $W^{\pm}$ exchange, followed by $H^{\pm\pm}$ decay to charged lepton pair and $H^{\pm}\to\ell^{\pm}\nu$. This combined channel of Drell-Yan production and associated production gives the constarint $m_{H^{\pm\pm}}>820$~GeV~\cite{CMS:2017pet} at $95\%$ C.L. ATLAS searches only include the Drell-Yan production and the bound is $m_{H^{\pm\pm}}>870$~GeV at $95\%$ C.L~\cite{ATLAS:2017xqs}. LEP has also searched for $H^{\pm\pm}H^{\mp\mp}$ pair production through s-channel $\gamma/Z$ exchange, with subsequent decay of $H^{\pm\pm}$ into charged lepton pairs and they put constraint as $m_{H^{\pm\pm}}>97.3$~GeV~\cite{DELPHI:2002bkf} at $95\%$ C.L. These limits hold only for small triplet VEV, $v_\Delta< 10^{-4}$ GeV. For large triplet VEV $v_\Delta > 10^{-4}$ GeV, $H^{\pm\pm}$ decay dominantly to diboson mode $W^{\pm}W^{\pm}$ and hence a search via pair-production $H^{\pm\pm}H^{\mp\mp}$ , with subsequent decay into gauge boson and further to leptonic final states is required. The ATLAS collaboration has studied this channel and constrained the doubly-charged Higgs mass around 220 GeV~\cite{ATLAS:2018ceg}. Searches for singly-charged and neutral scalars~\cite{CMS:2015lsf,ATLAS:2018gfm,ATLAS:2018ntn,ATLAS:2017eiz,CMS:2018rmh,ATLAS:2018sbw}, on the other hand, do not apply to the type-II seesaw because the involved couplings are $v_\Delta^2$ suppressed.
\subsection{Triplet fermion induced tree level seesaw scenario}
Type-III seesaw is an unique possibility to describe the origin of tiny neutrino mass and flavor mixing where SM is extended by zero hypercharge 
SU$(2)_{\rm L}$ triplet fermion $(\Psi)$. It allows us to generate tiny but nonzero mass term form the light neutrinos through the seesaw mechanism. We write down the Lagrangian in the following way
\begin{align}
\mathcal{L}=\mathcal{L}_{\text{SM}}+ \text{Tr}(\overline{\Psi}i \gamma^\mu D_\mu \Psi)-\frac{1}{2}M \text{Tr}(\overline{\Psi}\Psi^c+\overline{\Psi^c}\Psi)-\sqrt{2}(\overline{\ell_L}Y_D^\dagger \Psi H + H^\dagger \overline{\Psi} Y_D \ell_L).
\label{L}
\end{align}
Here $D_\mu$ and $M$ denote the covariant derivative and Majorana mass term respectively. $\mathcal{L}_{\text{SM}}$ is the relevant part of the SM Lagrangian involved in the type-III seesaw scenario. In this analysis we consider three generations of the triplet fremion which are degenerate in mass. Hence we consider $M$ is proportional to $\bf{1}_{3\times3}$. $Y_D$ is the Dirac Yukawa coupling among the triplet fermion $(\Psi)$, SM lepton doublet $(\ell_L)$ and SM Higgs doublet $(H)$ respectively. We suppress the generation indices for simplicity. We represent the SM candidates, the triplet fermion and its charged conjugate $(\Psi^c = C\overline{\Psi}^T)$ involved in the type-III seesaw mechanism as 
\bea
\ell_L
 =
 \begin{pmatrix}
  \nu_{L}\\
  e_{L} \\
 \end{pmatrix},\,\,\, 
H = 
 \begin{pmatrix}
 h^0\\
  h^-\\
 \end{pmatrix},\,\,\, 
\Psi=
 \begin{pmatrix}
  \Sigma^0/\sqrt{2}  &  \Sigma^+ \\
 \Sigma^-           &   -\Sigma^0/\sqrt{2}  \\
 \end{pmatrix}\,\,\text{and}\,\,
  \Psi^c=
 \begin{pmatrix}
  \Sigma^{0c}/\sqrt{2}  &  \Sigma^{-c} \\
 \Sigma^{+c}           &   -\Sigma^{0c}/\sqrt{2}  \\
 \end{pmatrix} .
 \label{L2}
\eea
When electroweak symmetry is broken, $h^0$ acquires a VEV as  $h^0=\frac{v+h}{\sqrt{2}}$ with $v=246$ GeV.
To study the mixing between $\Sigma^\pm$ and the SM charged leptons we write the four degrees of freedom of each $\Sigma^\pm$ in terms of a Dirac spinor such as 
$\Sigma=\Sigma_R^-+\Sigma_R^{+c}$ where as $\Sigma^0$ are two component fermions with two degrees of freedom. After electroweak symmetry breaking the corresponding Lagrangian can be written as 
\begin{align}
 -\mathcal{L}_{\text{mass}}=
  \begin{pmatrix}
  \overline{e}_L & \overline{\Sigma}_L \\
 \end{pmatrix}
 \begin{pmatrix}
  m_\ell & Y_D^\dagger v\\
  0 & M \\
 \end{pmatrix}
 \begin{pmatrix}
  e_R \\
  \Sigma_R \\
 \end{pmatrix} 
+
 \frac{1}{2}\begin{pmatrix}
  \overline{\nu_L^c} & \overline{\Sigma_R^0} \\
 \end{pmatrix}
 \begin{pmatrix}
  0 & Y_D^T \frac{v}{\sqrt{2}} \\
  Y_D\frac{v}{\sqrt{2}} & M \\
 \end{pmatrix}
 \begin{pmatrix}
  \nu_L \\
  \Sigma_R^{0c} \\
 \end{pmatrix}
 +\text{h. c.},
 \label{n1}
\end{align}
where $m_\ell$ is the Dirac mass term of the SM charged lepton. The $3\times3$ Dirac mass of the triplets can be written as 
\bea
M_D=\frac{Y_D^T v}{\sqrt{2}}.
\label{mDI}
\eea

\begin{figure}[]
\centering
\includegraphics[width=0.46\textwidth]{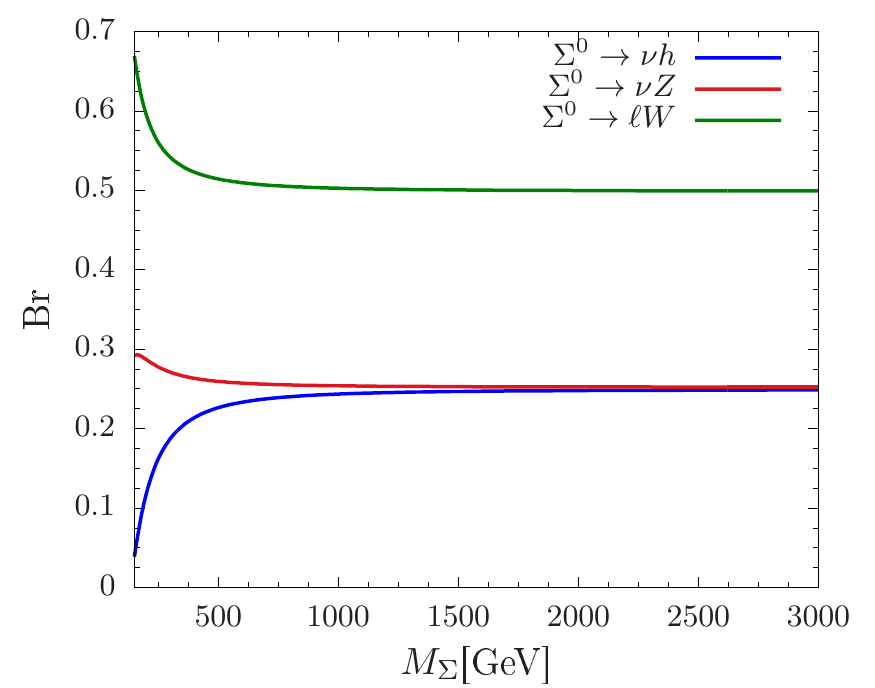}
\includegraphics[width=0.46\textwidth]{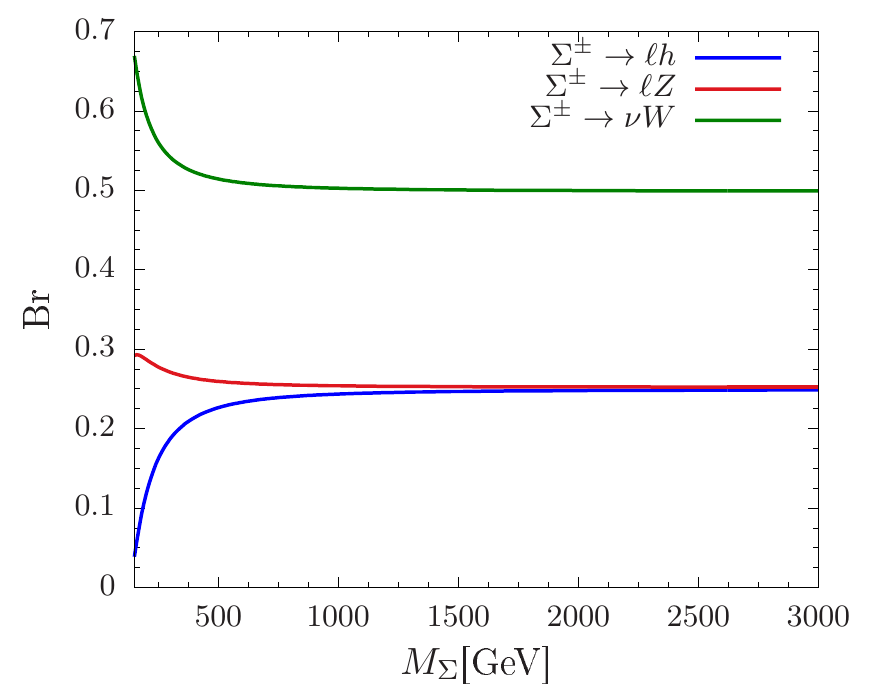}\\
\caption{Branching ratio (Br) of $\Sigma^0$~(left) and $\Sigma^{\pm}$~(right) into the SM particles as a function of $M_\Sigma$ for $V_e=0.019$, $V_\mu=0$ and $V_\tau=0$.}
\label{branching ratio1}
\end{figure}

\begin{figure}[]
\centering
\includegraphics[width=0.46\textwidth]{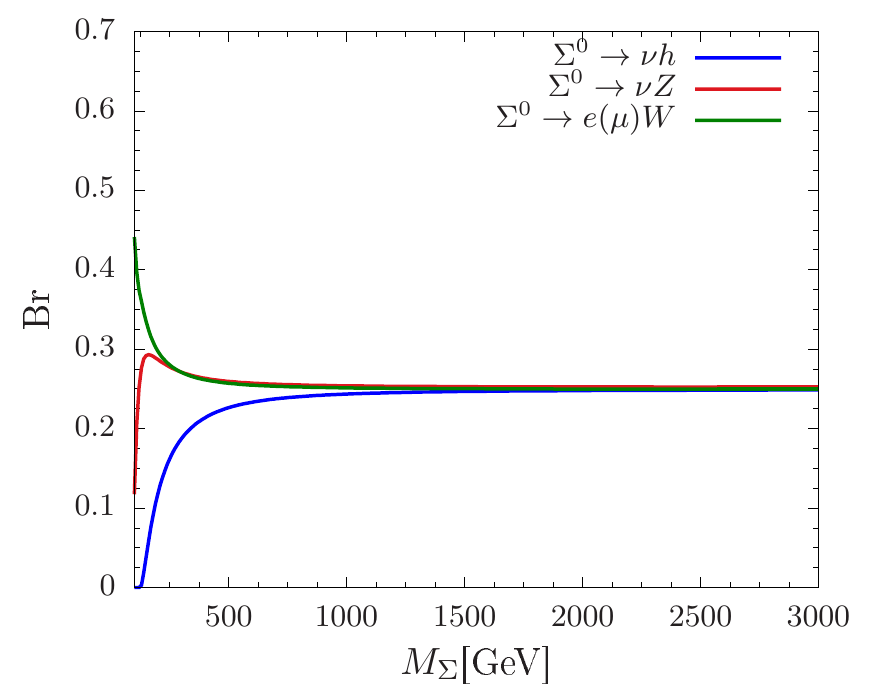}
\includegraphics[width=0.46\textwidth]{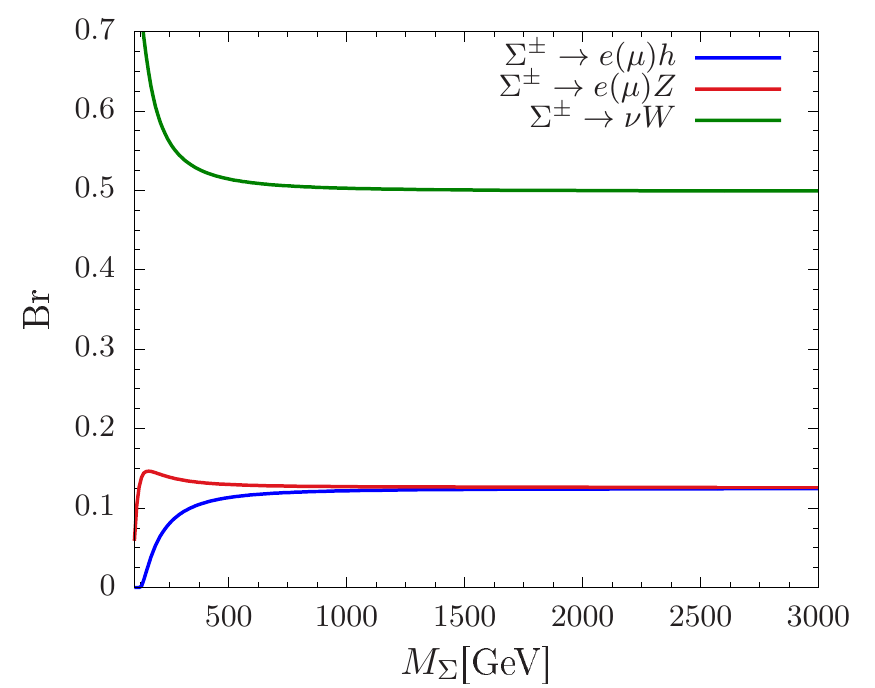}\\
\caption{Branching ratio (Br) of $\Sigma^0$~(left) and $\Sigma^{\pm}$~(right) into the SM particles as a function of $M_\Sigma$ for $V_e=$ $V_\mu=0.0001$ and $V_\tau=0$.}
\label{branching ratio2}
\end{figure}

Diagonalizing the neutrino mass matrix in Eq.~\ref{n1} we write the light neutrino mass eigenvalue in the following as 
\bea
m_\nu \simeq -\frac{v^2}{2} Y_D^T M^{-1} Y_D = M_D M^{-1} M_D^{T}.
\label{n3}
\eea
The mixing between the light and heavy mass eigenstates can be obtained as $\mathcal{O}(M_D M^{-1})$. The light neutrino flavor eigenstate can be written in terms of the light $(\nu_m)$ and heavy $(\Sigma_m)$ mass eigenstates as
\bea
\nu= \mathcal{A} \nu_m + V \Sigma_m ,
\label{n44}
\eea
where $V= M_D M^{-1}$ and $\mathcal{A}= \Big(1-\frac{1}{2} \tilde{\epsilon}\Big)U_{\text{PMNS}}$ respectively with $\tilde{\epsilon} = V^\ast V^T$ and $U_{\text{PMNS}}$ is the $3\times 3$ neutrino mixing matrix which diagonalizes the light neutrino mass matrix in the following way 
\bea
U_{\text{PMNS}}^T  m_\nu U_{\text{PMNS}}= \text{diag}(m_1, m_2, m_3).
\label{n4}
\eea
Due to the presence of $\tilde{\epsilon}$ the mixing matrix $(\mathcal{A})$ becomes non-unitary, $\mathcal{A}^\dagger\mathcal{A}\neq1$. Replacing the neutrino flavor eigenstates with the mass eigenstates from Eq.~\ref{n44} in the SM we express the charged current (CC) interactions in the following way 
\bea
 -\mathcal{L}_{\text{CC}}&=&\frac{g}{\sqrt{2}}
 \begin{pmatrix}
  \overline{e} & \overline{\Sigma} \\
 \end{pmatrix}
 \gamma^\mu W_\mu^- P_L  
 \begin{pmatrix}
 (1+\frac{\epsilon}{2}) U_{\text{PMNS}} &-\frac{Y_D^\dagger M^{-1} v}{\sqrt{2}}\\
 0&\sqrt{2}(1-\frac{\epsilon^\prime}{2})\\
 \end{pmatrix}
 \begin{pmatrix}
  \nu \\
  \Sigma^0 \\
 \end{pmatrix} \nonumber \\
 &+&\frac{g}{\sqrt{2}}
 \begin{pmatrix}
  \overline{e} & \overline{\Sigma} \\
 \end{pmatrix}
 \gamma^\mu W_\mu^- P_R  
  \begin{pmatrix}
 0&-\sqrt{2}m_{\ell} Y_D^\dagger M^{-2}v \\
 -\sqrt{2}M^{-1}Y_D(1-\frac{\epsilon^\ast}{2})V_{\text{PMNS}}^\ast&\sqrt{2}(1-\frac{\epsilon^{\prime^\ast}}{2})
 \end{pmatrix} 
 \begin{pmatrix}
  \nu \\
  \Sigma^0 \\
 \end{pmatrix}.
  \label{CC} 
   \eea
Similarly using Eq.~\ref{n44} in the SM neutral current (NC) interaction for the leptons we obtain the modified NC interaction as 
 \bea
- \mathcal{L}_{\text{NC}}&=&\frac{g}{\cos\theta_W}
 \begin{pmatrix}
  \overline{e} & \overline{\Sigma} \\
 \end{pmatrix}
 \gamma^\mu Z_\mu P_L  
 \begin{pmatrix}
  \frac{1}{2}-\cos^2\theta_W-\epsilon&\frac{Y_D^\dagger M^{-1} v}{2}\\
 \frac{M^{-1} Y_D v}{2}& \epsilon^\prime-\cos^2\theta_W
 \end{pmatrix}
 \begin{pmatrix}
  e \\
  \Sigma \\
 \end{pmatrix} \nonumber \\
 &+&\frac{g}{\cos\theta_W}
 \begin{pmatrix}
  \overline{e} & \overline{\Sigma} \\
 \end{pmatrix}
 \gamma^\mu Z_\mu P_R  
  \begin{pmatrix}
 1-\cos^2\theta_W&m_\ell Y_D^\dagger M^{-2} v \\
 M^{-2} Y_D m_{\ell} v&-\cos^2\theta_W
 \end{pmatrix} 
 \begin{pmatrix}
  e \\
  \Sigma \\
 \end{pmatrix} \nonumber \\
&+& \begin{pmatrix}
  \overline{\nu} & \overline{\Sigma^{0}} \\
 \end{pmatrix}
 \gamma^\mu Z_\mu P_L  
 \begin{pmatrix}
 1-U_{\text{PMNS}}^\dagger \epsilon U_{\text{PMNS}} & \frac{U_{\text{PMNS}}^\dagger Y_D^\dagger M^{-1} v}{\sqrt{2}}\\
 \frac{ M^{-1} Y_D U_{\text{PMNS}} v}{\sqrt{2}}& \epsilon^\prime
 \end{pmatrix}
 \begin{pmatrix}
  \nu \\
  \Sigma^{0} \\
 \end{pmatrix} ,
 \label{NC} 
 \eea
where $\theta_W$ is the weak mixing angle. Now we write down the interaction Lagrangian among the triplet fermion, SM lepton doublet and SM Higgs doublet $(H)$ in the following
\bea
 -\mathcal{L}_{H}&=&\frac{g}{2M_W}
 \begin{pmatrix}
  \overline{e} & \overline{\Sigma} \\
 \end{pmatrix}
 h P_L
 \begin{pmatrix}
  -\frac{m_\ell}{v}(1-3\epsilon)& m_{\ell} Y_D^\dagger M^{-1}\\
Y_D(1-\epsilon)+M^{-2} Y_D m_\ell^2 & Y_D Y_D^\dagger M^{-1} v
 \end{pmatrix}
  \begin{pmatrix}
  e \\
  \Sigma \\
 \end{pmatrix}  \nonumber \\ 
 &+&
 \frac{g}{2M_W}
 \begin{pmatrix}
  \overline{e} & \overline{\Sigma} \\
 \end{pmatrix}
 h P_R 
  \begin{pmatrix}
  -\frac{m_\ell}{v}(1-3\epsilon^\ast)& M^{-1}Y_D^\dagger m_\ell\\
(1-\epsilon^\ast) Y_D^\dagger+ m_\ell^2 Y_D^\dagger M^{-2} &  M^{-1}Y_D Y_D^\dagger v
 \end{pmatrix} 
   \begin{pmatrix}
  e\\
  \Sigma \\
 \end{pmatrix} \nonumber \\
&+& \begin{pmatrix}
  \overline{\nu} & \overline{\Sigma^0} \\
 \end{pmatrix}
 h P_L
 \begin{pmatrix}
\frac{\sqrt{2} m_\nu}{v}& U_{\text{PMNS}}^T m_\nu Y_D^\dagger M^{-1} \\
(Y_D-\frac{Y_D \epsilon}{2} -\frac{\epsilon^{\prime T}Y_D}{2})U_{\text{PMNS}}&\frac{Y_D Y_D^\dagger M^{-1} v}{\sqrt{2}}
 \end{pmatrix}
  \begin{pmatrix}
  \nu \\
  \Sigma^0 \\
 \end{pmatrix}  \nonumber \\ 
 &+&
 \begin{pmatrix}
  \overline{e} & \overline{\Sigma^0} \\
 \end{pmatrix}
 h P_R 
  \begin{pmatrix}
\frac{\sqrt{2} m_\nu}{v}&  M^{-1} Y_D m_\nu U_{\text{PMNS}}^\ast \\
U_{\text{PMNS}}^\ast(Y_D^\dagger-\frac{\epsilon^\ast Y_D^\dagger}{2} -\frac{Y_D^\dagger \epsilon^{\prime \ast}Y_D}{2})&\frac{M^{-1} Y_D Y_D^\dagger v}{\sqrt{2}}
 \end{pmatrix}
 \begin{pmatrix}
  \nu\\
  \Sigma^0 \\
 \end{pmatrix} .
 \label{H1}
\eea
The charged multiplets of the triplet fermions also interact with photon $(A_\mu)$. The corresponding interaction Lagrangian can be derived from Eq.~\ref{L} as  
\bea
- \mathcal{L}_{\gamma \Sigma\Sigma}&=&g\sin\theta_W
 \begin{pmatrix}
  \overline{e} & \overline{\Sigma} \\
 \end{pmatrix}
 \gamma^\mu A_\mu P_L  
 \begin{pmatrix}
 1&0\\
 0&1
 \end{pmatrix}
 \begin{pmatrix}
  e \\
  \Sigma \\
 \end{pmatrix} 
 +g\sin\theta_W
 \begin{pmatrix}
  \overline{e} & \overline{\Sigma} \\
 \end{pmatrix}
 \gamma^\mu A_\mu P_R  
  \begin{pmatrix}
  1&0\\
  0&1
 \end{pmatrix} 
 \begin{pmatrix}
  e \\
  \Sigma \\
 \end{pmatrix}.
\label{Ph}
\eea
In the Eqs.~\ref{CC}-\ref{H1} the parameters $\epsilon= \frac{v^2}{2} Y_D^\dagger M^{-2} Y_D$ and $\epsilon^\prime= \frac{v^2}{2} M^{-1} Y_DY_D^\dagger M^{-1}$ are the small quantities according to Refs.~\cite{Abada:2007ux,Abada:2008ea,Biggio:2011ja}. We neglect the effects of the higher powers (above $1$) of $\epsilon$ and $\epsilon^\prime$ in the calculations. Using Eq.~\ref{CC} to Eq.~\ref{H1}, we calculate the partial decay widths of $(\Sigma^0)$ in the following way
\bea
\Gamma(\Sigma^0 \to \ell^+ W)&=&\Gamma(\Sigma^0 \to \ell^- W)=\frac{g^2 |V_{\ell \Sigma}|^2}{64 \pi} \Big(\frac{M^3}{M_W^2}\Big) \Big(1-\frac{M_W^2}{M^2}\Big)^2 \Big(1+2\frac{M_W^2}{M^2}\Big), \nonumber \\
\Gamma(\Sigma^0 \to \nu Z)&=&\Gamma(\Sigma^0 \to \overline{\nu} Z)=\frac{g^2 |V_{\ell \Sigma}|^2}{128 \pi \cos^2\theta_W} \Big(\frac{M^3}{M_Z^2}\Big) \Big(1-\frac{M_Z^2}{M^2}\Big)^2 \Big(1+2\frac{M_Z^2}{M^2}\Big), \nonumber \\
\Gamma(\Sigma^0 \to \nu h)&=&\Gamma(\Sigma^0 \to \overline{\nu} h)=\frac{g^2 |V_{\ell \Sigma}|^2}{128 \pi} \Big(\frac{M^3}{M_W^2}\Big) \Big(1-\frac{M_h^2}{M^2}\Big)^2 .
\label{decay1}
\eea
Similarly the partial decay widths of $(\Sigma^\pm)$ can be calculated in terms of the mixing $(V_{\ell \Sigma})$ as
\bea
\Gamma(\Sigma^\pm \to \nu W)&=&\frac{g^2 |V_{\ell \Sigma}|^2}{32 \pi} \Big(\frac{M^3}{M_W^2}\Big) \Big(1-\frac{M_W^2}{M^2}\Big)^2 \Big(1+2\frac{M_W^2}{M^2}\Big), \nonumber \\
\Gamma(\Sigma^\pm \to \ell Z)&=&\frac{g^2 |V_{\ell \Sigma}|^2}{64 \pi \cos^2\theta_W} \Big(\frac{M^3}{M_Z^2}\Big) \Big(1-\frac{M_Z^2}{M^2}\Big)^2 \Big(1+2\frac{M_Z^2}{M^2}\Big), \nonumber \\
\Gamma(\Sigma^\pm \to \ell h)&=&\frac{g^2|V_{\ell \Sigma}|^2}{64 \pi} \Big(\frac{M^3}{M_W^2}\Big) \Big(1-\frac{M_h^2}{M^2}\Big)^2,
\label{decay2}
\eea
where $M_W$, $M_Z$ and $M_h$ are the SM $W$, $Z$ and Higgs boson masses respectively. The charged and neutral multiplets $\Sigma^{\pm, 0}$ are degenerate in mass at the tree-level, however, radiative corrections induced by the SM gauge boson in the loop can help in lifting such degeneracy. The estimation of this mass difference $\Delta M$ can be found in \cite{Cirelli:2005uq} and that is given by
\begin{align}
\Delta M=\frac{\alpha_2 M}{4\pi}\Big(f\big(\frac{M_W}{M}\big)-\cos^2\theta_W f\big(\frac{M_Z}{M}\big)\Big),
\end{align}
where the function $f$ and $A$ are defined as $f(r)=\frac{r}{2}\big(2r^3\text{ln}\,r-2r+\sqrt{r^2-4}(r^2+2)\text{ln}\,A\big)$ and $A=\big(r^2-2-r\sqrt{r^2-4}\big)/2$ respectively. This mass splitting saturates at $\Delta M\approx 170$ MeV for mass $M > 500$ GeV. If this mass splitting $\Delta M$ is larger than pion mass, then $\Sigma^\pm$ will have additional decay modes according to \cite{Cirelli:2005uq} and the corresponding partial decay widths can be written as 
\bea
\Gamma(\Sigma^\pm \to \Sigma^0 \pi^\pm)&=& \frac{2 G_F^2 V_{ud}^2 \Delta M^3 f_\pi^2}{\pi} \sqrt{1-\frac{m_\pi^2}{\Delta M^2}} ,\nonumber \\
\Gamma(\Sigma^\pm \to \Sigma^0 e \nu_e) &=& \frac{2 G_F^2 \Delta M^5}{15 \pi} ,\nonumber \\
\Gamma(\Sigma^\pm \to \Sigma^0 \mu \nu_\mu) &=&0.12 \Gamma(\Sigma^\pm \to \Sigma^0 e \nu_e) ,
\label{decay3}
\eea
which are independent of the free parameters.  The value of the Fermi constant, $G_F$, is $1.1663787 \times10^{-5}$ GeV$^{-2}$, the value of the CKM parameter $(V_{ud})$ is $0.97420 \pm 0.00021$ and the decay constant of the $\pi$ meson, $f_\pi$, is $0.13$ GeV~\cite{Tanabashi:2018oca}. Notice that for vanishing mixing angles $V_{\ell\Sigma}$, the $\Sigma^{\pm}$ dominantly decay into $\Sigma^0$, hence the decay width or the decay length is determined by $\Delta M$. Hence it is a constant. On the contrary, for very large mixing angles, $\Sigma^0$ decay width (decay length) is very large (very small). The Branching ratios (Br) of the neutral and charged multiplets $(\Sigma^{0, \pm})$ of the SU$(2)_L$ triplet fermion into SM particles are shown Fig.~\ref{branching ratio1} as a function of $M$ for $V_e=0.019$, $V_\mu=0$ and $V_\tau=0$. The same for $V_e=V_\mu=0.0001$ and $V_\tau=0$ are shown in Fig.~\ref{branching ratio2}. 
Note that for large values of $M_{\Sigma}$ with $|V_{\ell}|\neq 0$, the branching ratios can be obtained as
\begin{align*}
& \text{BR}(\Sigma^0\to\ell W):\text{BR}(\Sigma^0\to\nu_{\ell}Z):\text{BR}(\Sigma^0\to\nu_{\ell}h)=2:1:1 ,\\
& \text{BR}(\Sigma^\pm\to\nu_\ell W^\pm):\text{BR}(\Sigma^\pm\to\ell^\pm Z):\text{BR}(\Sigma^\pm\to\ell^\pm h)=2:1:1.
\end{align*} 
\section{Electron photon and photon photon colliders}
\label{egm}
\begin{figure}[h]
\centering
\includegraphics[width=0.9\textwidth]{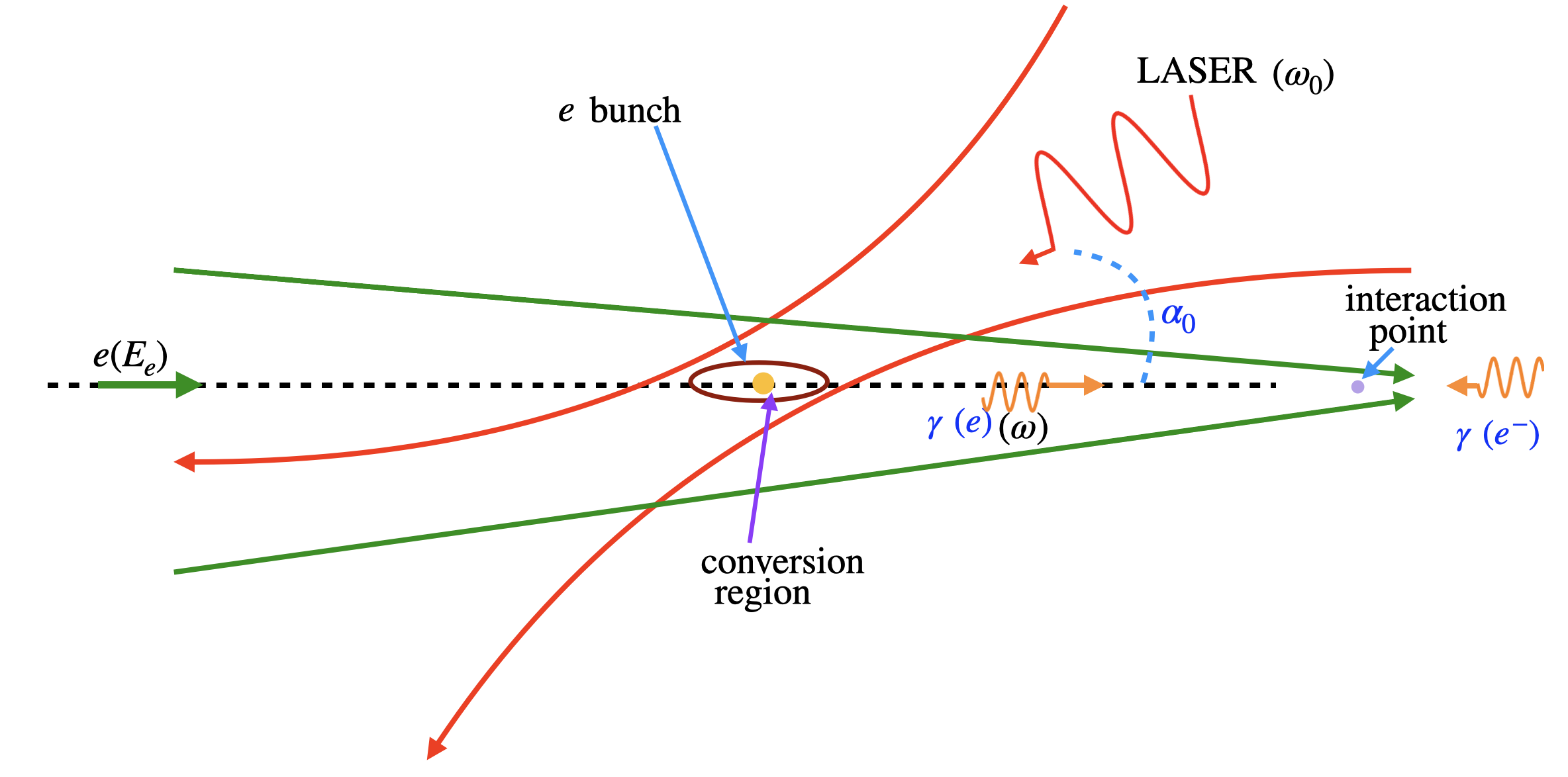}
\caption{Schematic presentation of the proposed $e^-\gamma$ and $\gamma \gamma$ colliders in $e^-e^+$ collider facilities where positron can be converted into a beam of high energy photons by Compton back scattering with LASER beams. Similar procedure can be performed with electrons. In this notation $e$ represents both electron and positron. A pictorial presentation from \href{https://www-zeuthen.desy.de/ILC/gammagamma/}{DESY}.}
\label{fig:kinematics-compton}
\end{figure}
A linear collider with colliding electrons and positrons can be transformed into electron photon collider where positron can be replaced by a backward Compton scattered photon which comes from the collision between LASER beam with energy $\omega_0$ focused on a bunch of positron (it can be potentially electrons if selected) with energy $E_e$. The positrons beam focused towards the interaction point collide with the focused LASER beam at the conversion region and the scattered high energy photons with energy $\omega$ follow the direction of incident positron trajectories with a small angular spread. Hence they are also focused towards the interaction point. Thus produced photon beam from the Compton backward scattering can further collides with the oncoming electron beam or a similar photon beam giving rise to $e^-\gamma$ or $\gamma \gamma$ collisions \cite{Ginzburg:1982yr,Ginzburg:1982bs}. The distance between the conversion region and impact point could be $\mathcal{O}(\rm mm)$ and the schematic presentation of the $e^-\gamma$ and $\gamma \gamma$ collisions are shown in Fig.~\ref{fig:kinematics-compton}. 

\begin{figure}[h]
\includegraphics[width=0.6\textwidth]{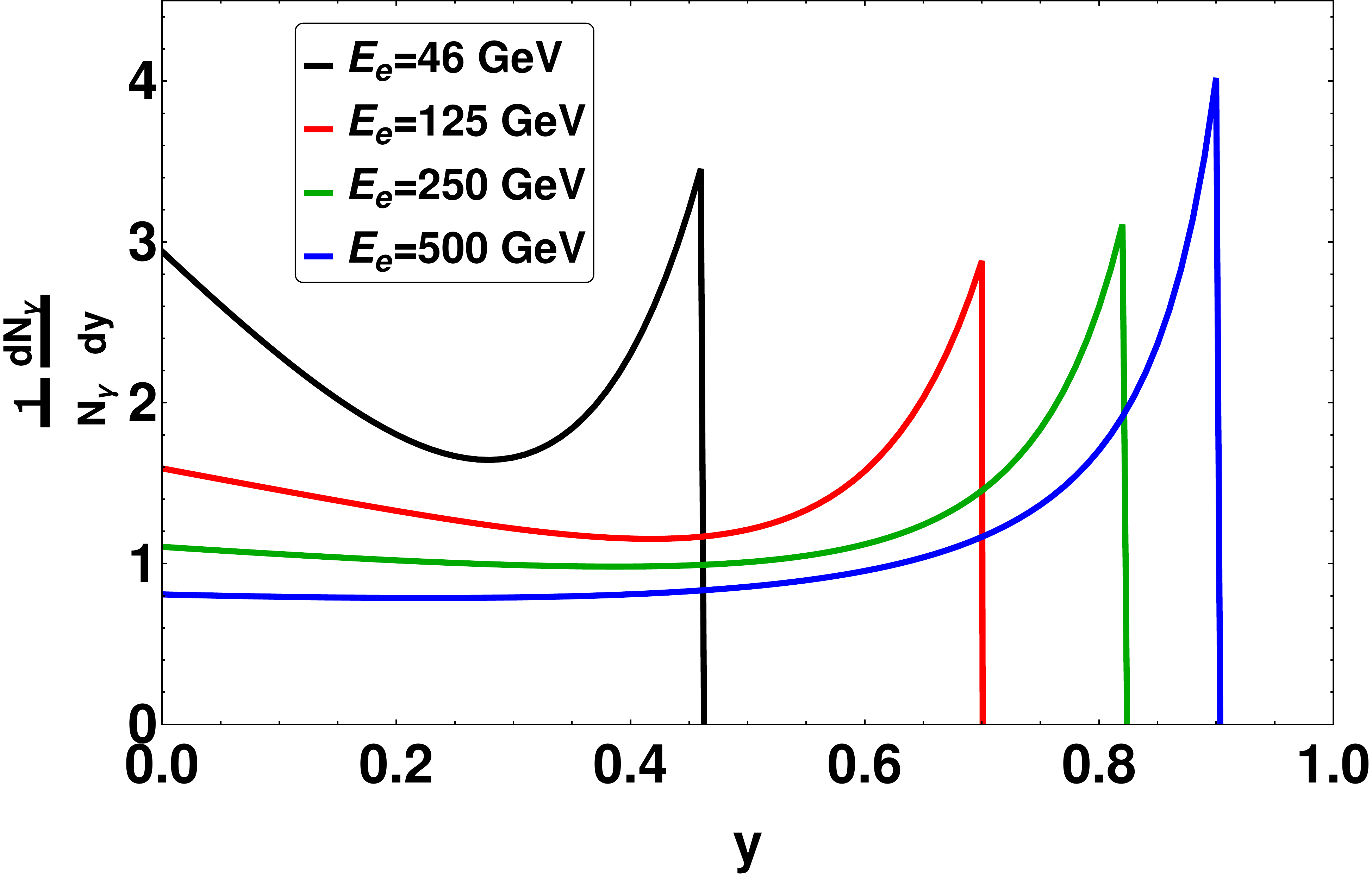}
\caption{Energy spectrum of photons. Four lines stand for three different electron beam energies $E_e=46$~GeV(black), 125 GeV(red), 250 GeV(green) and 500 GeV(blue), respectively.}
\label{fig:photon-distribution}
\end{figure}

Unlike electron and positron beams, the $e^- \gamma$ colliding beams will not be mono-chromatic. The full cross section should be evaluated from the sub-process by convoluting with the structure function of photon which is basically the energy spectrum of the photons described by the Compton distribution as
\begin{align}
\frac{1}{N_\gamma}\frac{dN_\gamma}{dy}\equiv F_{\gamma/e}(x,y)=\frac{1}{D(x)}\Big(\frac{1}{1-y}+1-y-\frac{4y}{x(1-y)}+\frac{4y^2}{x^2(1-y)^2}\Big),
\end{align}
with,
\begin{align}
D(x)&=\Big(1-\frac{4}{x}-\frac{8}{x^2}\Big)\text{ln}(1+x)+\frac{1}{2}+\frac{8}{x}-\frac{1}{2(1+x)^2},\,\,\,x=\frac{4E_e\omega_0}{m_e^2}\cos^2(\frac{\alpha_0}{2}),\nonumber \\
y&=\frac{\omega}{E_e}\leq y_m=\frac{\omega_m}{E_e}=\frac{x}{x+1}.
\label{eg0}
\end{align}
In the above $\omega_0$ is the energy of the incident photon which is scattered on the positron with energy $E_e$ at a small collision angle $\alpha_0$ which is shown in Fig.~\ref{fig:kinematics-compton}. Here $\omega_m$ is maximum achievable energy of the scattered photon and it can be large for large value of $x$. In Fig.~\ref{fig:photon-distribution}, we show the energy spectrum of the photon for different choices of electron or positron beam energies $E_e=46$ GeV, 125 GeV, 250 GeV and 500 GeV. We see that photon energy may reach up to $80\%(90\%)$ of the energy of the initial electrons $E_e=250$ GeV~(500 GeV) with LASER energy $\omega_0=1.17$ eV. Hence for larger $x$, the energy spectrum for the photon beam will be more mono-chromatic and will be peaked at high energy. 
\hspace{0.5cm} The total luminosity of the $e^-\gamma$ beam is $\mathcal{L}_{\gamma e}=\kappa \mathcal{L}_{ee}$, where $\kappa=\frac{N_\gamma}{N_e}$ is the photon conversion coefficient. This is basically the average number of high energy photons per one electron and is defined as~\cite{Telnov:2006cj}
\begin{align}
\kappa=\frac{N_\gamma}{N_e}=\frac{2|\vec{v}_e-\vec{v}_\gamma|\sigma_C}{N_e}\int n_e n_\gamma dV dt
\end{align}
$\vec{v}_e$ and $\vec{v}_\gamma$ is the velocity of electrons/ positrons and incoming photons. $\sigma_C$ is the cross section of the Compton scattering.
\subsection{Heavy Majorana neutrino production modes}
The heavy neutrinos can be produced at the electron positron colliders from a variety of production modes. An interesting production of the heavy neutrinos can be considered as $e^-e^+ \to N \nu$ through $t$ and $s-$channels exchanging the $W$ and $Z$ bosons, respectively. We mention these modes because of complementarity with the study of the heavy neutrino production at the electron photon collider. In electron photon colliders we can produce heavy neutrinos from $e^- \gamma \to N W^-$ involving the $t-$ channel and $s-$ channel processes. In both these case interference between these $s$ and $t-$channels take place. In both these colliders heavy neutrino production cross sections are suppressed by the square of the light-heavy mixing. For the process $e^-e^+ \to N \nu$, due to the interaction with the electron we dominantly produce the first generation heavy neutrino $(N=N_1)$, however, in the case of $Z$ mediated $s-$ channel process $e^-e^+ \to N \nu$ process, second or third generation RHNs can be produced. In the following we discuss about the heavy neutrino production cross sections at the $e^-e^+$ and $e^- \gamma$ colliders respectively.
\begin{figure}[h]
\centering
\includegraphics[width=0.75\textwidth]{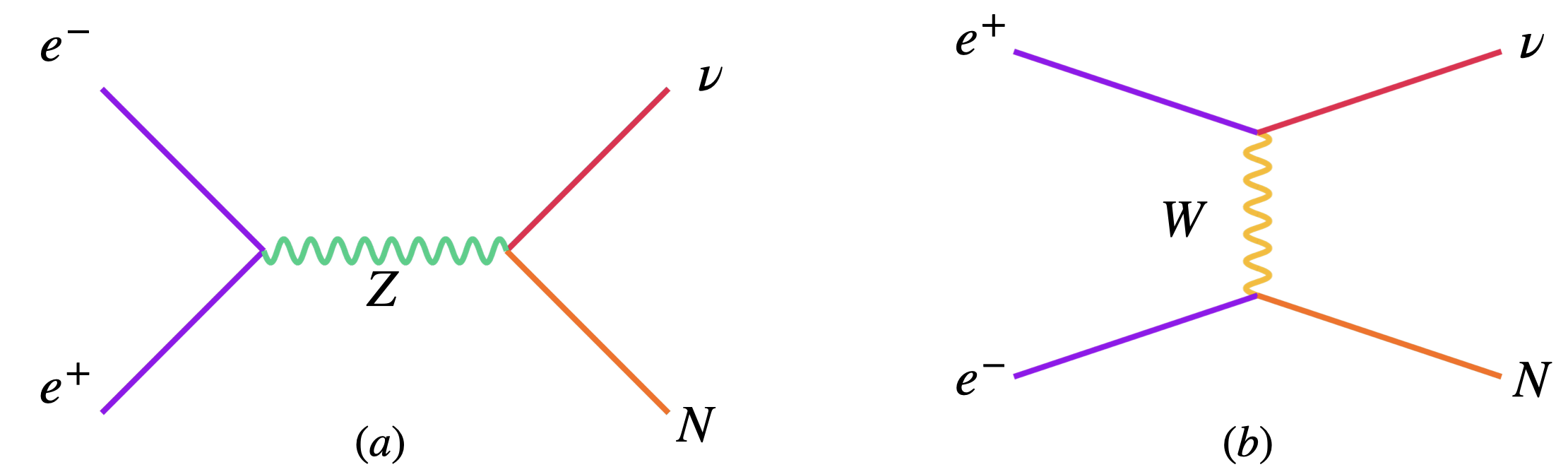}
\caption{Heavy neutrino production processes at $e^- e^+$ colliders in association with neutrino.}
\label{fig:FD-1}
\end{figure}

\begin{figure}[h]
\centering
\includegraphics[width=0.49\textwidth]{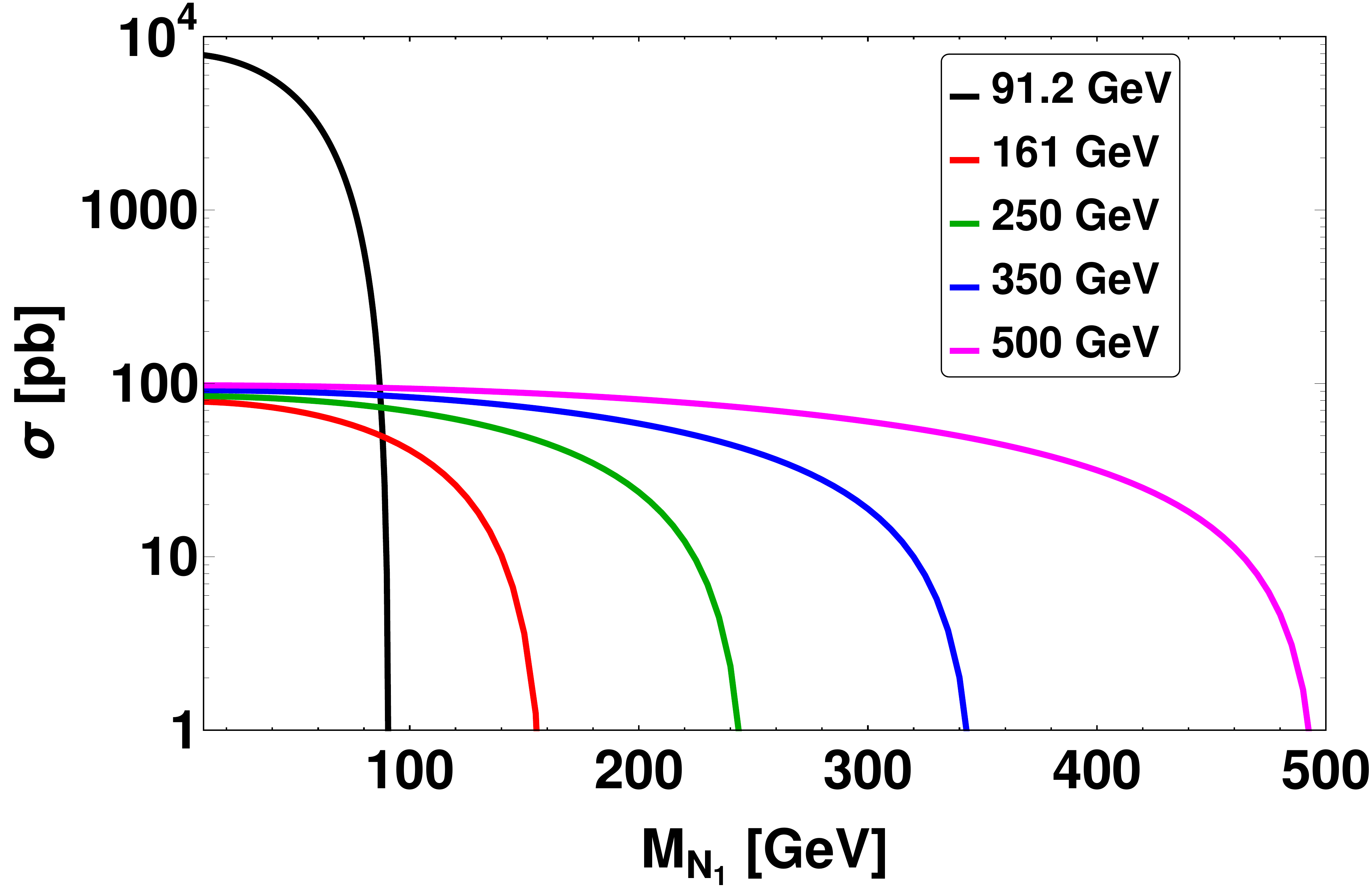}
\includegraphics[width=0.49\textwidth]{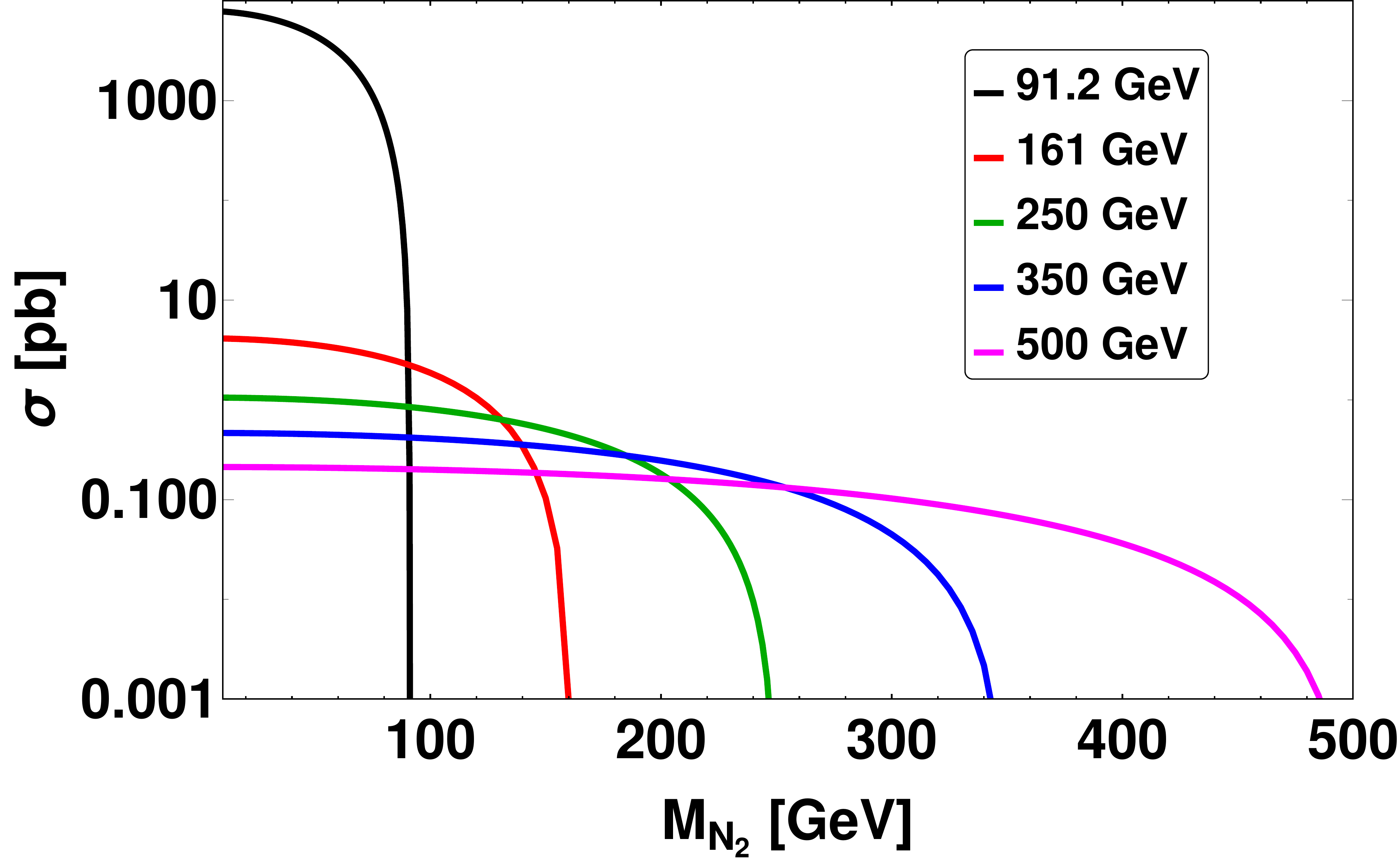}
\includegraphics[width=0.49\textwidth]{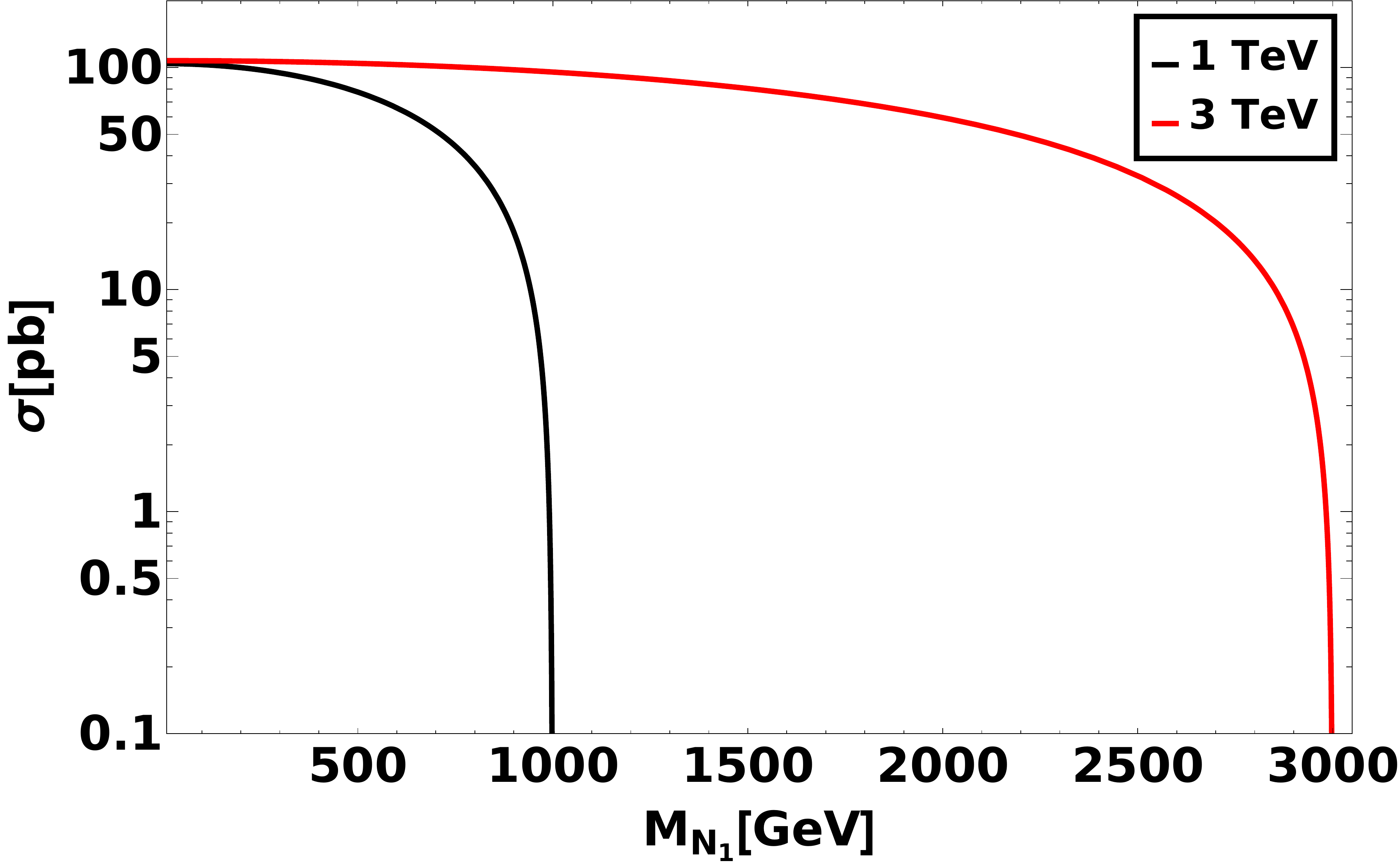}
\includegraphics[width=0.49\textwidth]{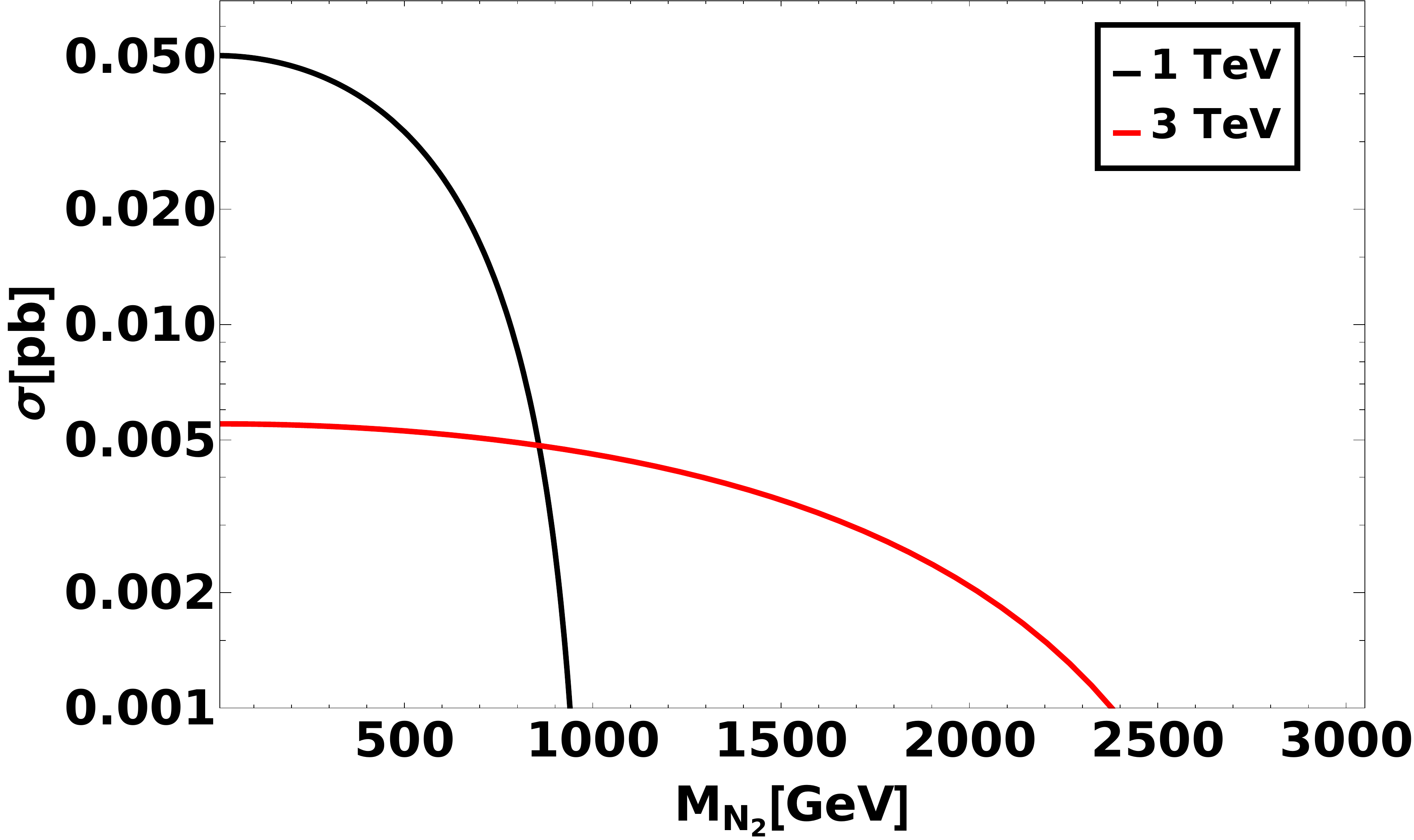}
\caption{RHN production cross section normalized by the square of the light-heavy neutrino mixing at the linear collider considering $e^+e^-\to \nu N_1$ (left panel) and $e^+e^-\to \nu N_2$ (right panel) process at the different center of mass energies.}
\label{fig:nuN1-and-nuN2-production}
\end{figure}

\hspace{0.5cm} Let us first consider the $e^- e^+\to\nu N_1$ process. Corresponding Feynman diagrams are shown in Fig.~\ref{fig:FD-1}. We calculate the differential scattering cross section including the $s-$ channel, $t-$ channel processes and the interference term as 
\begin{align}
d\sigma=\frac{1}{32\pi s}\left(1-\frac{M_N^2}{s}\right) \left(|\mathcal{M}_s|^2+|\mathcal{M}_t|^2+2\text{Re}\left(\mathcal{M}_s^{\dagger}\mathcal{M}_t\right)\right),
\end{align}
The differential cross-sections are with respect to angle $\cos\theta$ in the center of mass~(CM) frame. The corresponding scattering amplitude modulus squared are given below as 
\begin{align}
|\mathcal{M}_s|^2&=\frac{g_Z^4 |V_{eN_1}|^2}{16\left((s-m_Z^2)^2+m_Z^2\Gamma_Z^2\right)}\left((1-2x_w)^2 (s+t)(s+t-M_{N_1}^2)+4x_w^2 t (t-M_{N_1}^2)\right),\\
|\mathcal{M}_t|^2&=\frac{g_2^4 |V_{eN_1}|^2}{4\left((t-m_W^2)^2+m_W^2\Gamma_W^2\right)}(s+t)(s+t-M_{N_1}^2),
\end{align}
and the interference term can be written as
\begin{align}
2\text{Re}\left(\mathcal{M}_s^{\dagger}\mathcal{M}_t\right)&=-\frac{g_2^2 g_Z^2 |V_{eN_1}|^2 (1-2x_w)}{4\left((s-m_Z^2)^2+m_Z^2\Gamma_Z^2\right) \left((t-m_W^2)^2+m_W^2\Gamma_W^2\right)}\nonumber \\
& (s+t)(s+t-M_{N_1}^2)\left((s-m_Z^2)(t-m_W^2)+m_W m_Z\Gamma_W\Gamma_Z\right),
\end{align}
with $g_Z=\frac{e}{\sin\theta_W\cos\theta_W}$, $g_2=\frac{e}{\sin\theta_W}$, $x_w=\sin^2\theta_W$, $t=-\frac{1}{2}(s-M_N^2)(1-\cos\theta)$ and assuming initial massless states and neglecting the light neutrino masses respectively. The total cross sections normalized by the square of the mixing for the $N_1 \nu$ mode with respect to $M_{N_1}$ at different center of mass energies $(\sqrt{s})$ are shown in the left panel of Fig.~\ref{fig:nuN1-and-nuN2-production} and those for the $N_2 \nu$ are shown in the right panel of Fig.~\ref{fig:nuN1-and-nuN2-production}. In the second case only the $Z$ mediated $s-$ channel process participates. We find that the cross section can reach at a maximum value for $\sqrt{s}=91.2$ GeV which close to the $Z$ pole enhancing the effect of the $Z$ mediation for heavy neutrino production of any generation. The cross section decreases with the increase in $\sqrt{s}$ while going away from the $Z$ pole.
\begin{figure}[h]
\centering
\includegraphics[width=0.75\textwidth]{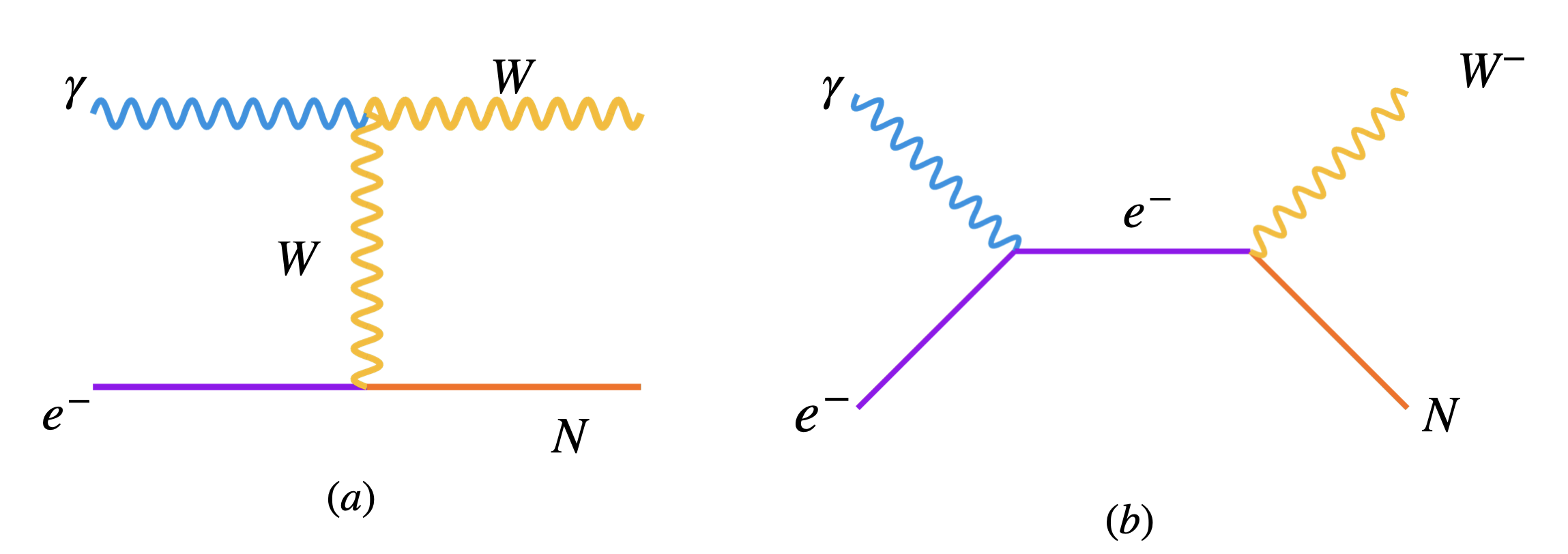}
\caption{Heavy neutrino production processes at $e^- \gamma$ colliders in association with $W^-$.}
\label{fig:FD-2}
\end{figure}

\begin{figure}[h]
\centering
\includegraphics[width=0.49\textwidth]{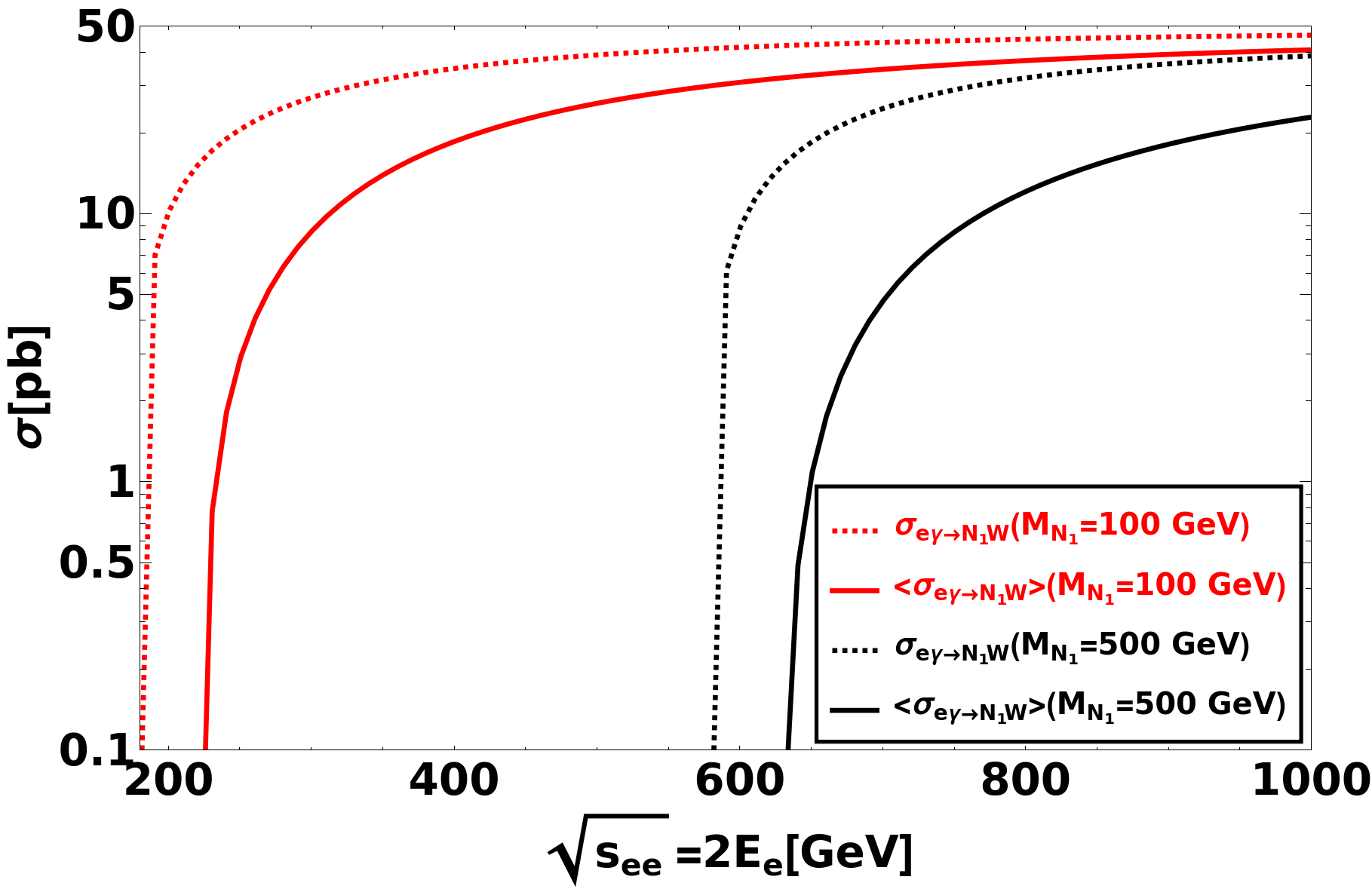}
\includegraphics[width=0.49\textwidth]{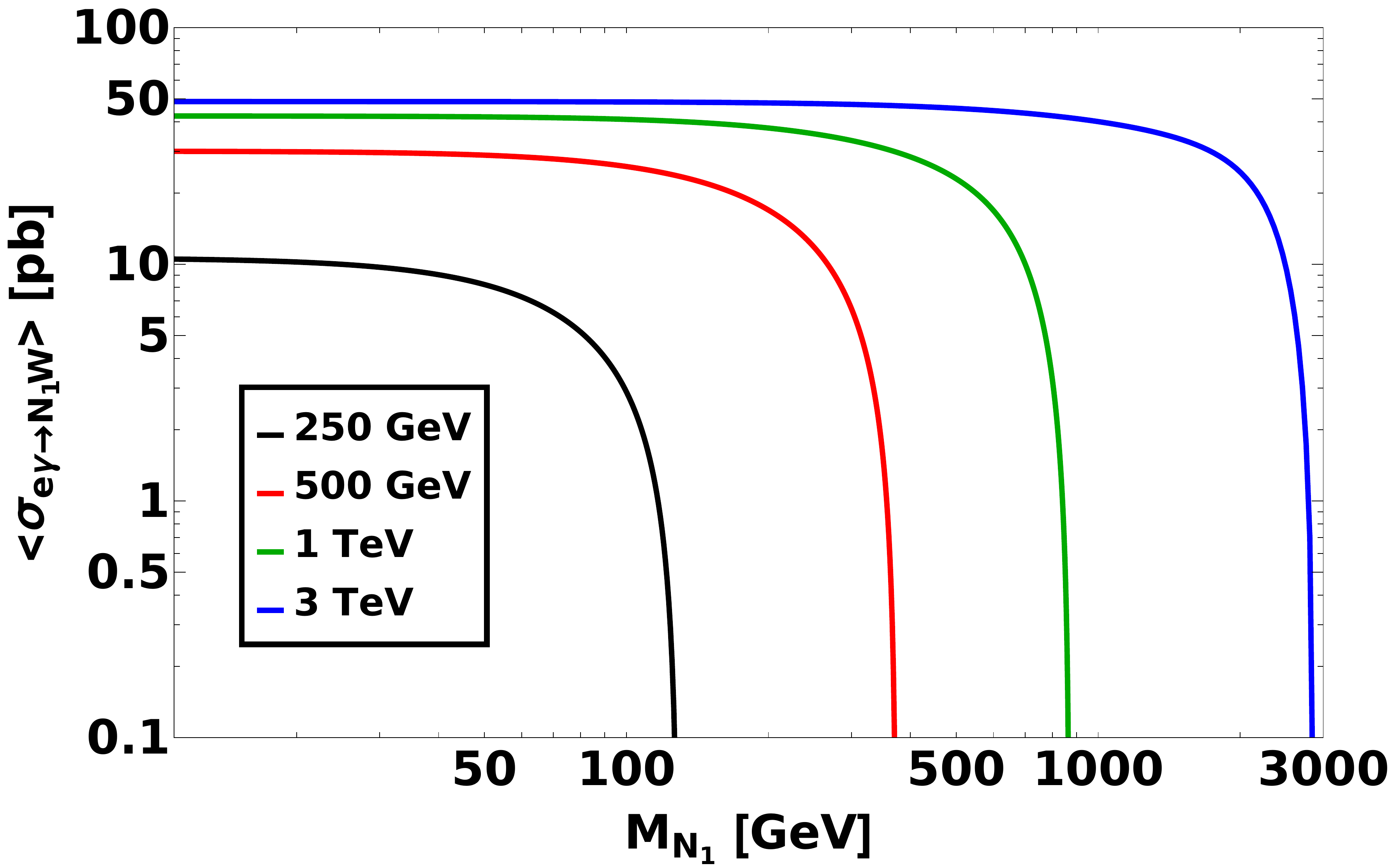}
\caption{Left panel: Comparison of $N_1W$ production cross section at $e^- \gamma$ collider for $M_{N_1}=100$ GeV and $M_{N_1}=500$ GeV. The red and black dashed line represents the bare cross section and the red, black solid line represents the total corss section averaged over the photon spectrum normalized by the square of the mixing. Right panel: Total cross section of the $N_1W$ production mode at $e^- \gamma$ colliders for different $\sqrt{s}$ as a function of the heavy neutrino mass after averaged over photon spectrum normalized by the square of the mixing.}
\label{fig:NW-production-cm}
\end{figure}

\hspace{0.5cm} We now calculate the $N_1 W$ production cross section from $e^-\gamma$ process. 
Corresponding Feynman diagrams are given in Fig.~\ref{fig:FD-2}. 
We estimate the differential scattering cross section of the process $\gamma e^-\to W^- N_1$ including the $s-$ channel and $t-$ channel processes and their interference as
\begin{align}
d\sigma=\frac{1}{32\pi s}\lambda^{\frac{1}{2}}\Big(1,\frac{M_{N_1}^2}{s},\frac{m_W^2}{s}\Big) \left(|\mathcal{M}_s|^2+|\mathcal{M}_t|^2+2\text{Re}\left(\mathcal{M}_s^{\dagger}\mathcal{M}_t\right)\right),
\label{eg1}
\end{align}
and the corresponding scattering amplitude modulus squared are given below as
\begin{align}
|\mathcal{M}_s|^2&=\frac{g_2^4 x_w V_{eN_1}^2}{4sm_W^2}\left(t(M_{N_1}^2-s)-m_W^2(2m_W^2+M_{N_1}^2-2(s+t))\right),\\
|\mathcal{M}_t|^2&=\frac{g_2^4 x_w V_{eN_1}^2}{4m_W^2\left((t-m_W^2)^2+m_W^2\Gamma_W^2\right)}\Big(-m_W^4(M_{N_1}^2+5s-t)\nonumber \\
&+m_W^2(-3M_{N_1}^4-M_{N_1}^2 t+5s^2 +6 s t+ 4t^2)\nonumber \\
&-t(-M_{N_1}^4+s^2 + s t + t^2)\Big),
\end{align}
and the interference term can be written as 
\begin{align}
2\text{Re}\left(\mathcal{M}_s^\dagger\mathcal{M}_t\right)&=\frac{g_2^4 x_w V_{eN_1}^2}{4 s m_W^2\left((t-m_W^2)^2+m_W^2\Gamma_W^2\right)}(t-m_W^2)\Big(-4 m_W^2(M_{N_1}^2+s-t)+m_W^2(2M_{N_1}^4-2M_{N_1}^2(s+t)\nonumber\\
&+3s^2)+2M_{N_1}^6-M_{N_1}^4(3s+2t)+M_{N_1}^2 s t + s(s^2+st+2t^2)\Big),
\end{align}
where $t=m_W^2-\frac{s}{2}\Big(\sqrt{\lambda\Big(1,\frac{M_{N_1}^2}{s},\frac{m_W^2}{s}\Big)+\frac{4m_W^2}{s}}-\lambda^{\frac{1}{2}}\Big(1,\frac{M_{N_1}^2}{s},\frac{m_W^2}{s}\Big)\cos\theta\Big)$. The total cross section can be obtained being averaged over the photon spectrum and that can be written as
\begin{align}
\braket{\sigma_{\gamma e\to N_1 W}(\sqrt{s_{ee}})}=\int_{y_\text{min}}^{y_{\text{max}}}dy F_{\gamma/e}(x,y)\sigma_{\gamma e\to N_1 W}(\sqrt{s_{e\gamma}}),
\label{eg2}
\end{align}
where $s_{ee}=4E_e^2$, $y_{\text{max}}=\frac{x}{1+x}$, $y_{\text{min}}=\frac{(M_{N_1}+M_W)^2}{s_{ee}}$ and $s_{e\gamma}=4yE_e^2$. The analytical expression for bare process $\sigma_{\gamma e\to N_1 W}(\sqrt{s_{e\gamma}})$ can be directly obtained from the differential scattering cross section in Eq.~\ref{eg1}. In the right panel of 
Fig.~\ref{fig:NW-production-cm}, we show the averaged cross section along with bare cross section for two choices of heavy neutrino masses $M_{N_1}=100$ GeV and $M_{N_1}=500$ GeV respectively. The total cross sections averaged over the photon spectrum as a function of the heavy neutrino mass are shown in the right panel of Fig.~\ref{fig:NW-production-cm} for different $\sqrt{s}$.
\subsection{Triplet scalar production modes}
\begin{figure}[h]
\centering
\includegraphics[width=1.0\textwidth]{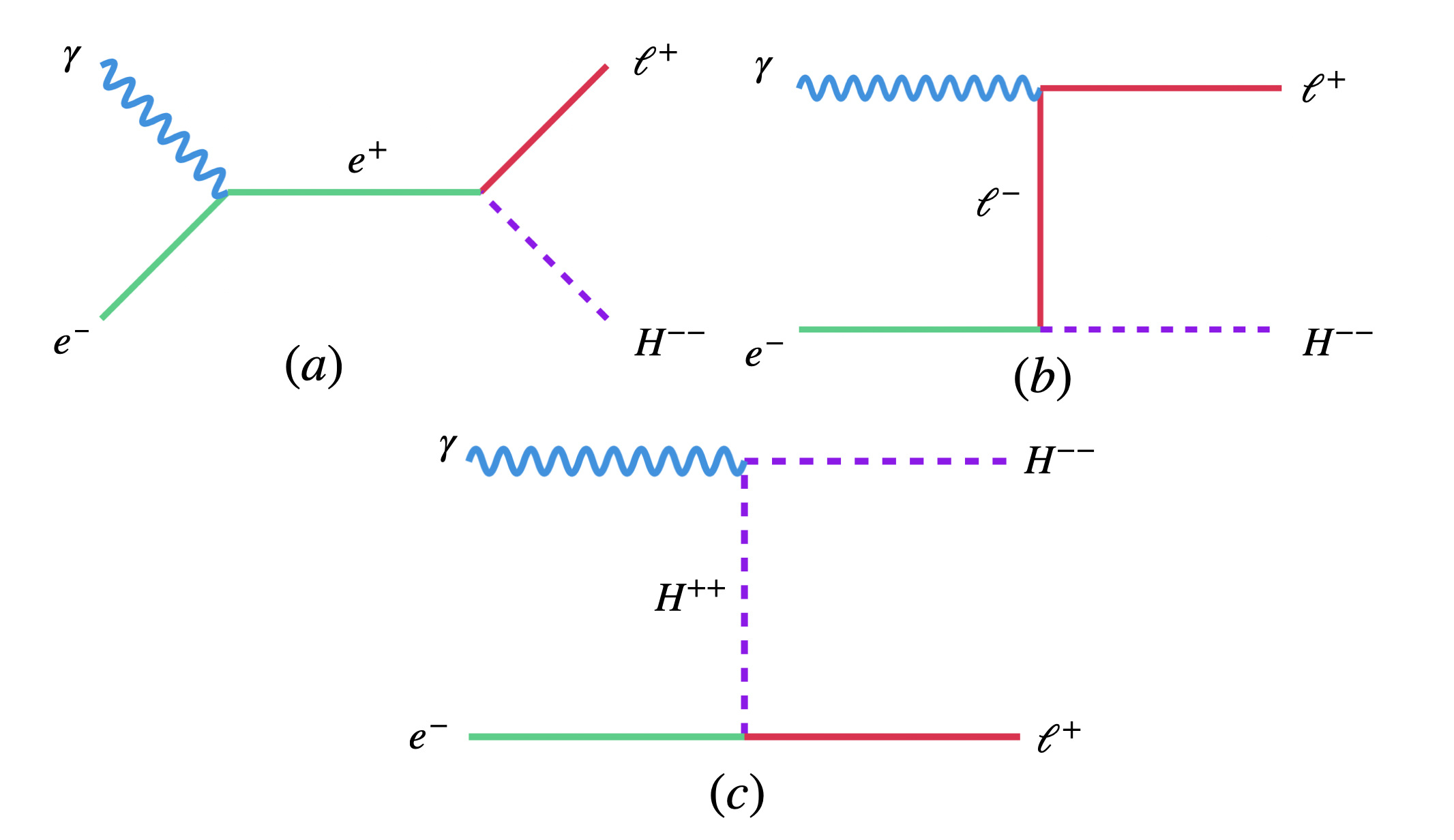}
\caption{Feynman diagrams for the production processes $\ell^+ H^{--}$ at the $e^- \gamma$ colliders where $\ell^+= e^+, \mu^+, \tau^+$.}
\label{fig:LFV}
\end{figure}

Triplet scalar has doubly charged multiplet $(H^{\pm \pm})$ which can be tested at the $e^- \gamma$ colliders. The Feynman diagrams of the doubly charged scalars in association with a charged lepton~($\ell^\pm$) are shown in Fig.~\ref{fig:LFV}.
In this case $\ell^\pm$ could be $e^\pm$, $\mu^\pm$ or $\tau^\pm$ which could help in testing the Lepton Flavor Violating (LFV) modes from the type-II seesaw mechanism at the $e^-\gamma$ coilliders. We shown the $s-$ channel and $t-$ channel processes in Fig.~\ref{fig:LFV}. Finally we estimate the production cross section of the $\ell^+ H^{--}$ process and the corresponding differential scattering cross section can be written as 
\begin{align}
d\sigma=\frac{1}{32\pi s}\bigg(1-\frac{M_{H^{\pm\pm}}^2}{s}\bigg) \Big(|\mathcal{M}_a|^2+|\mathcal{M}_b|^2+|\mathcal{M}_c|^2+2\text{Re}\left(\mathcal{M}_a^{\dagger}\mathcal{M}_b\right) + 2\text{Re}\left(\mathcal{M}_a^{\dagger}\mathcal{M}_c\right) + 2\text{Re}\left(\mathcal{M}_b^{\dagger}\mathcal{M}_c\right) \Big),
\label{eq:dsigegammatolHmm}
\end{align}
where $s-$ channel and two $t-$ channel scattering amplitude modulus squared terms can be written as 
\begin{align}
& |\mathcal{M}_a|^2=-g_2^2 x_w |Y_{\Delta}^{e\ell}|^2\frac{t}{s},\,\,\,|\mathcal{M}_b|^2=-g_2^2 x_w |Y_{\Delta}^{e\ell}|^2\frac{s}{t},\,\,|\mathcal{M}_c|^2=4g_2^2 x_w |Y_{\Delta}^{e\ell}|^2\frac{(M_{H^{\pm\pm}}^2-s-t) (2 M_{H^{\pm\pm}}^2-s-t)}{M_{H^{\pm\pm}}^2\Gamma_{H^{\pm\pm}}^2+(s+t)^2},
\end{align}
and the corresponding interference terms can be written as 
\begin{align}
& 2\text{Re}(\mathcal{M}_a^\dagger \mathcal{M}_b)=2 g_2^2 x_w |Y_{\Delta}^{e\ell}|^2\frac{(M_{H^{\pm\pm}}^2-s) (M_{H^{\pm\pm}}^2-t)}{st}, \nonumber \\
&2\text{Re}(\mathcal{M}_a^\dagger \mathcal{M}_c)=-2 g_2^2 x_w |Y_{\Delta}^{e\ell}|^2\frac{(2M_{H^{\pm\pm}}^2-s)(s+t) (M_{H^{\pm\pm}}^2-s-t)}{s(M_{H^{\pm\pm}}^2\Gamma_{H^{\pm\pm}}^2+(s+t)^2)}, \nonumber \\
& 2\text{Re}(\mathcal{M}_b^\dagger \mathcal{M}_c)=-2 g_2^2 x_w |Y_{\Delta}^{e\ell}|^2\frac{(2M_{H^{\pm\pm}}^2-t)(s+t) (M_{H^{\pm\pm}}^2-s-t)}{t(M_{H^{\pm\pm}}^2\Gamma_{H^{\pm\pm}}^2+(s+t)^2)}.
\end{align}

\begin{figure}[h]
\centering
\includegraphics[width=0.49\textwidth]{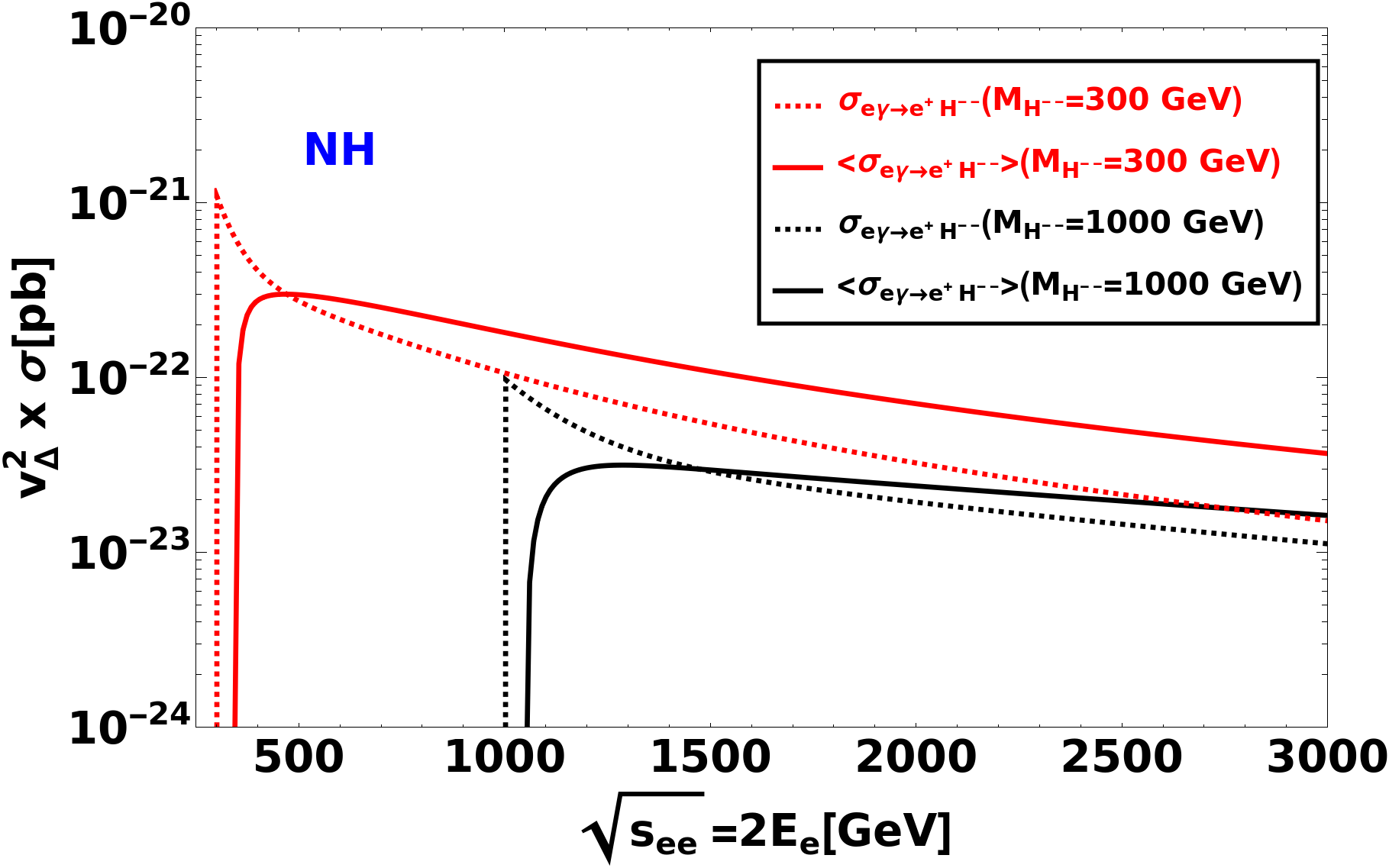}
\includegraphics[width=0.49\textwidth]{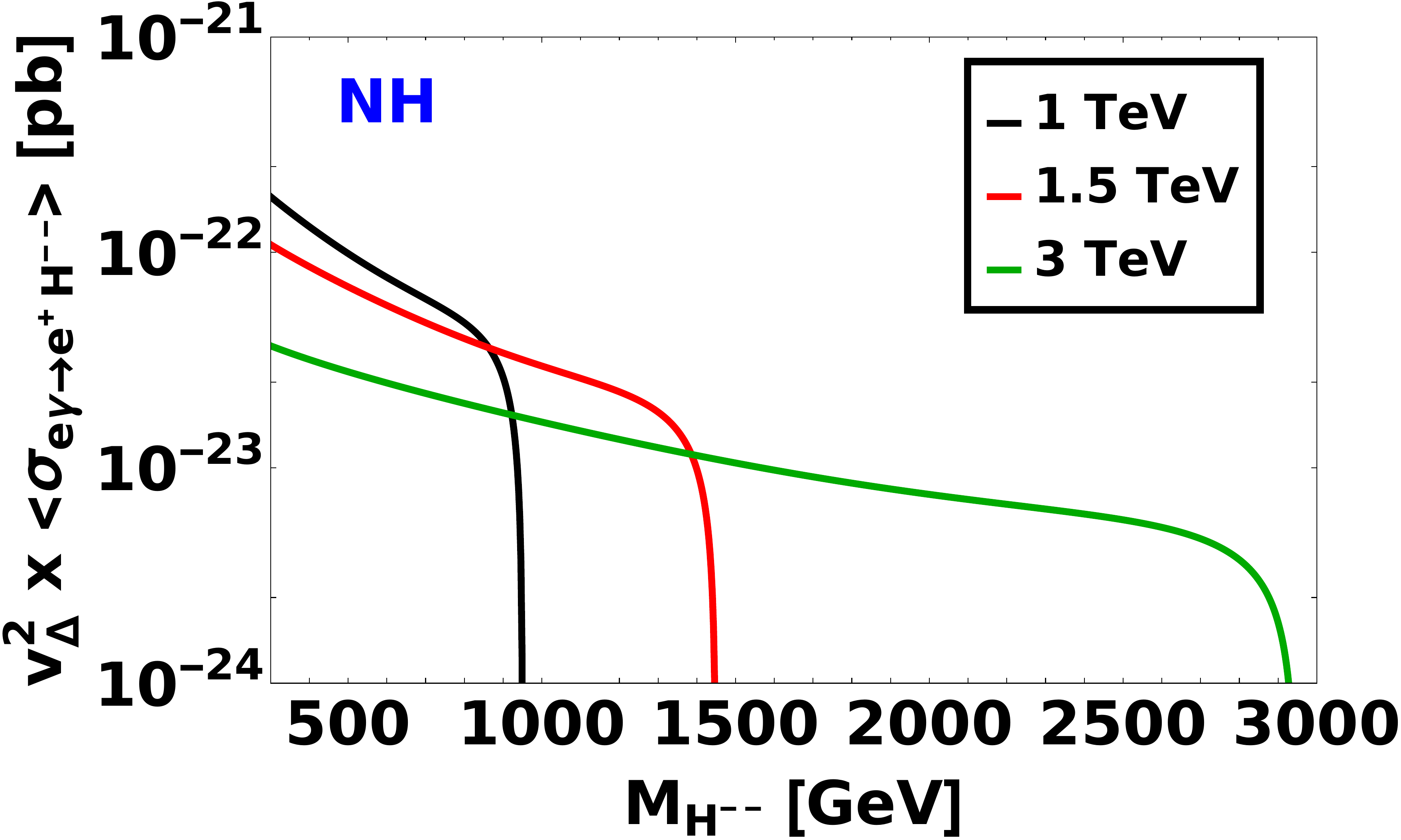}
\includegraphics[width=0.49\textwidth]{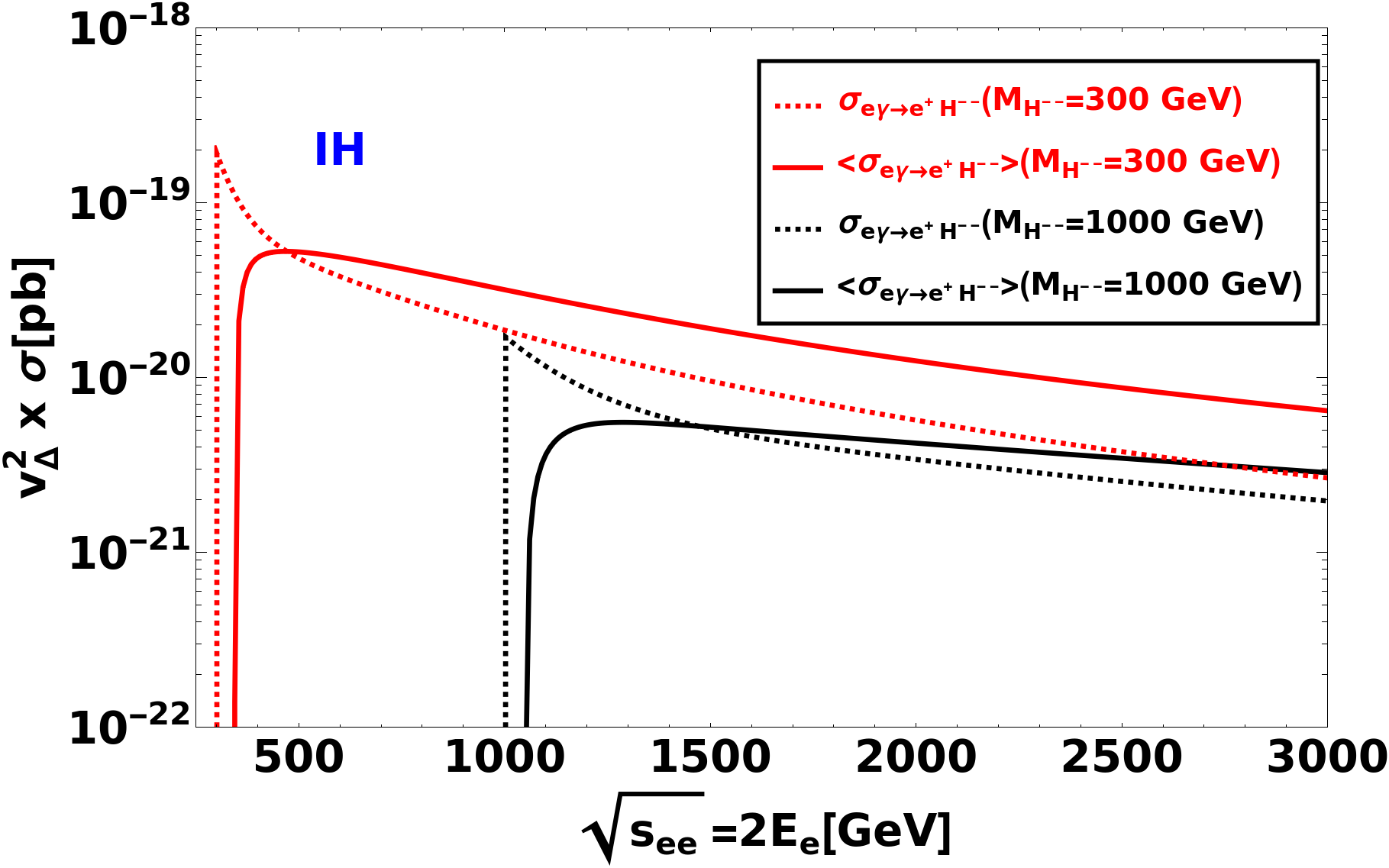}
\includegraphics[width=0.49\textwidth]{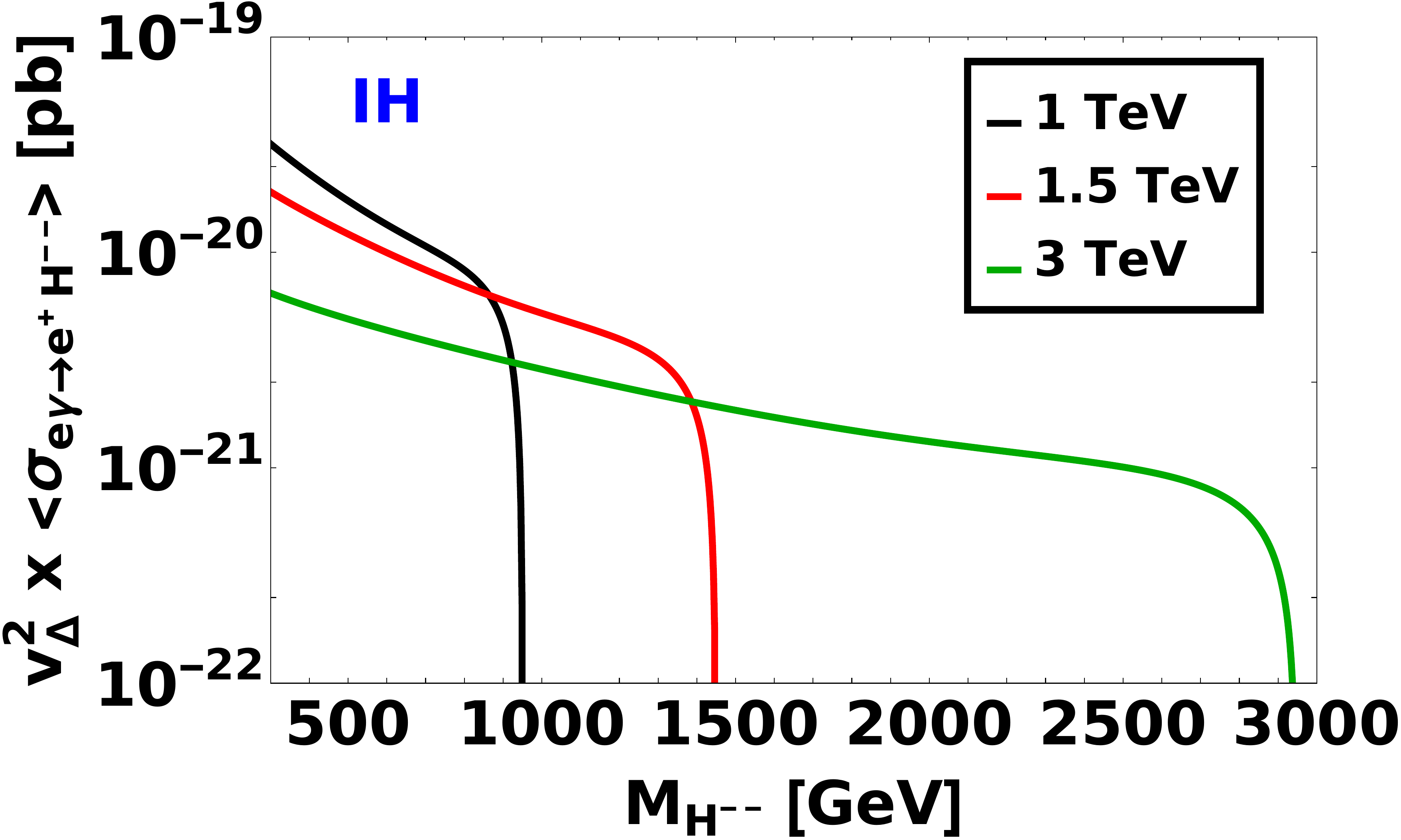}
\caption{Production cross section of $H^{--}\ell^+$ process at the $e^- \gamma$ colliders. The upper and lower panel are for normal and inverted hierarchy. Left panel: comparison of bare and averaged cross section for charged Higgs mass $M_{H^{\pm\pm}}=300$ GeV and $M_{H^{\pm\pm}}=1000$ GeV. The red and black dashed line represents the bare cross section and the red, black solid line represents the total cross section averaged over the photon spectrum. Right panel: cross section for different $\sqrt{s}$ as a function of the charged Higgs mass after averaged over photon spectrum.}
\label{fig:lHmm-production}
\end{figure}

In the upper and bottom left panel of Fig.~\ref{fig:lHmm-production}, we showed the averaged cross section along with bare cross section for two choices of charged Higgs mass $M_{H^{--}}=$ 300 GeV and 1000 GeV. The production
cross section of $H^{--} \ell^+$ as a function of $M_{H^{--}}$ for different $\sqrt{s}$ are shown in the upper right and bottom right panel. The production cross section drops with the increase in $M_{H^{--}}$ and sharply drops when $M_{H^{--}}\to \sqrt{s}$. Note that this production cross section is directly proportional to the Yukawa coupling $Y_\Delta$. Yukawa coupling $Y_\Delta$ is determined by Eq.~\ref{eq:YDelta} and is inversely proportional to $v_\Delta$. Also current measurements of neutrino oscillations parameter~\cite{10.5281/zenodo.4726908,deSalas:2020pgw} determines the pattern of Yukawa coupling $Y_\Delta$ and is different for normal hierarchy~(NH) and inverted hierarchy~(IH). This fact is clearly visible when comparing upper and bottom panel of Fig.~\ref{fig:lHmm-production}. Fig.~\ref{fig:lHmm-production} is also suggests that cross section will only be sizable if one consider very small value of VEV $v_\Delta$. As the Yukawa coupling is inversely proportional to $v_\Delta$, smaller the $v_\Delta$, larger the Yukawa coupling. But Yukawa coupling $Y_\Delta$ can not be arbitrarily large as there can be potential problem with constraints coming from charged lepton flavour violating observables such as $\mu\to e\gamma$ and $\mu\to \bar{e}ee$. This actually exclude the region with $v_\Delta\le 10^{-8}$ GeV~\cite{Antusch:2018svb}. With this kind of $v_\Delta$, the cross section will be very small and hence we decide not to discuss this production mode further.

\begin{figure}[h]
\centering
\includegraphics[width=1\textwidth]{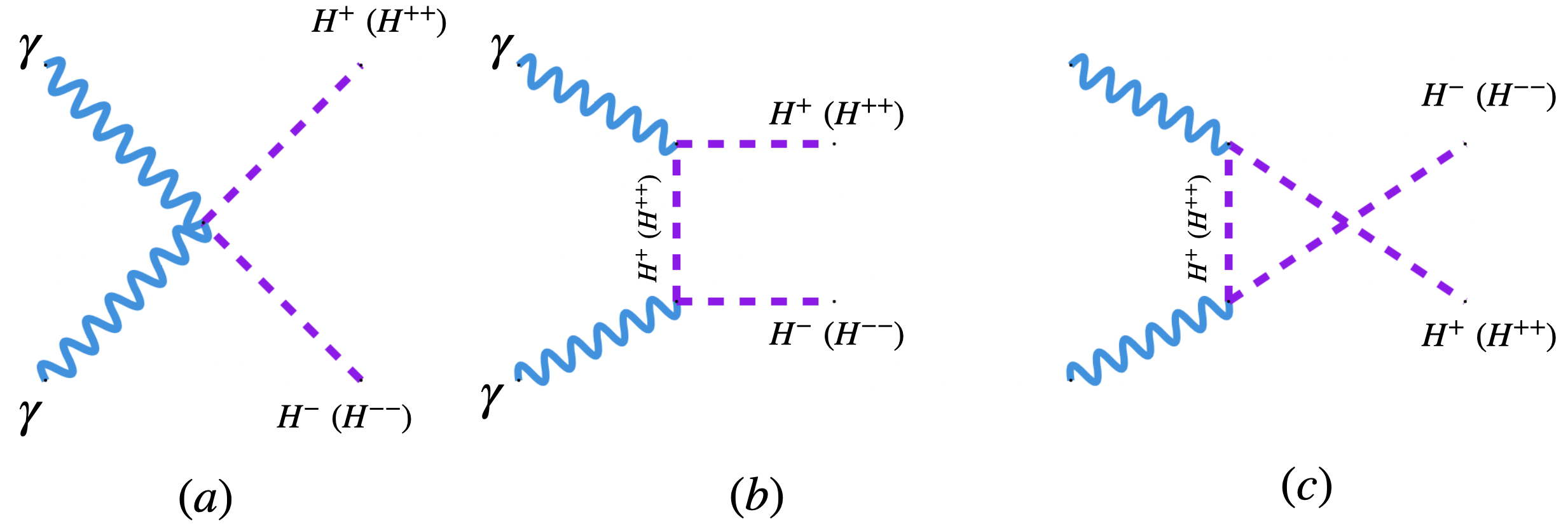}
\caption{Production processes of the singly $(H^\pm H^\mp)$ and doubly $(H^{\pm \pm} H^{\mp \mp})$ charged scalars at the $\gamma \gamma$ colliders.}
\label{fig:triplet scalar}
\end{figure}

\begin{figure}[h]
\centering
\includegraphics[width=0.49\textwidth]{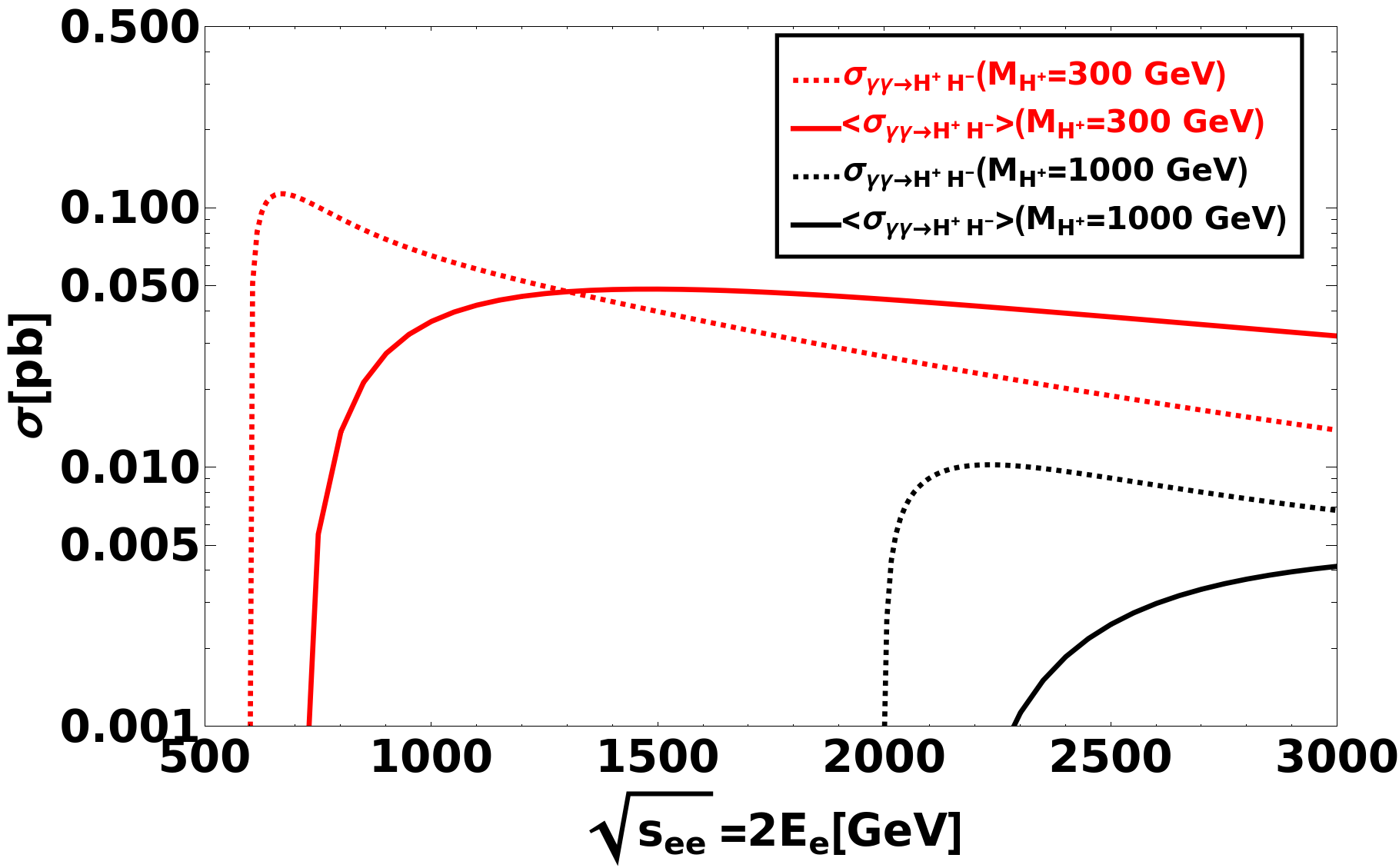}
\includegraphics[width=0.49\textwidth]{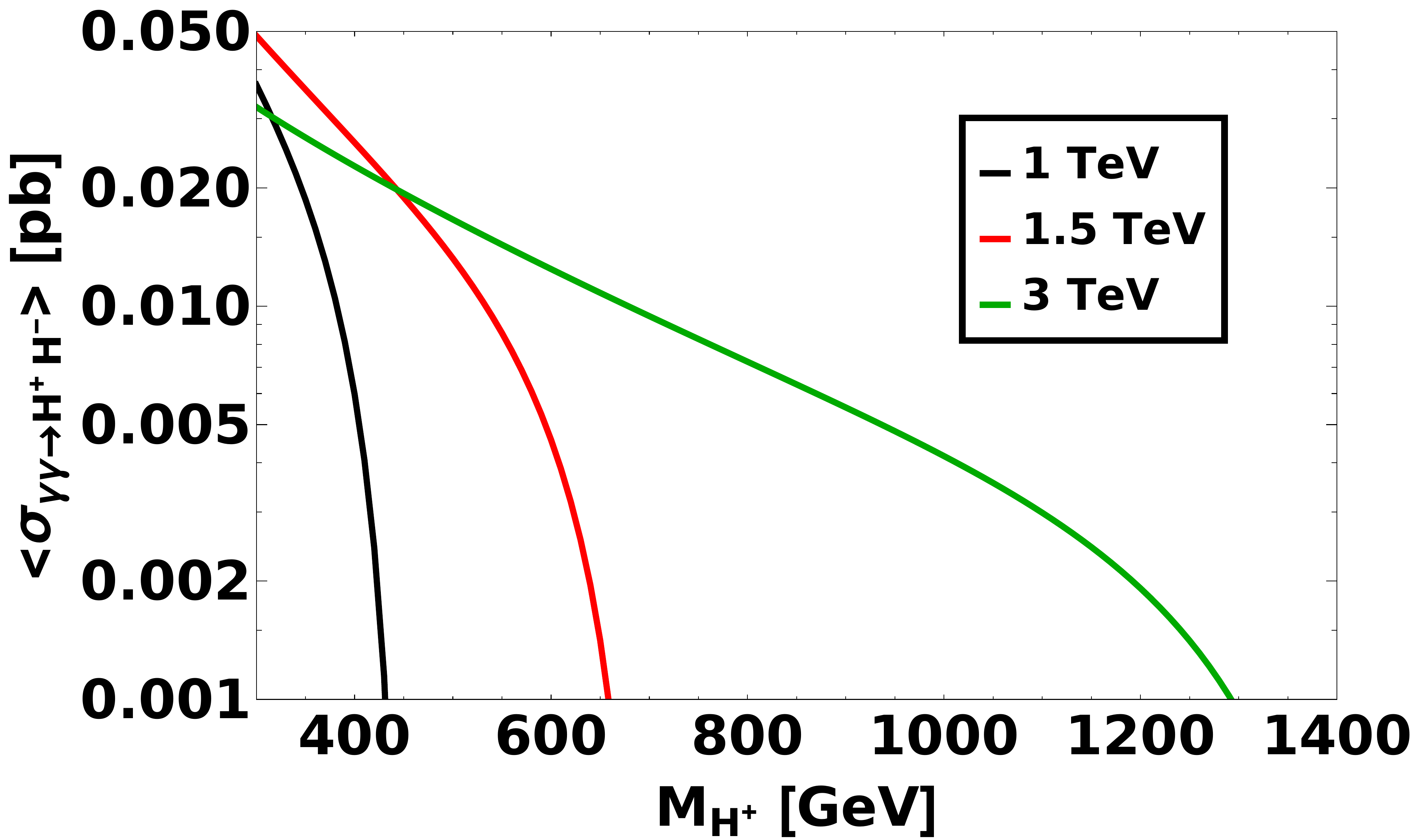}
\caption{Production cross section of $H^\pm$ pair production process at the $\gamma \gamma$ colliders. Left panel: comparison of bare and averaged cross section for charged Higgs mass $M_{H^{\pm}}=300$ GeV and $M_{H^{\pm}}=1000$ GeV. The red and black dashed line represents the bare cross section and the red, black solid line
represents the total corss section averaged over the photon spectrum. Right panel: cross section for different $\sqrt{s}$ as a function of the charged Higgs mass after averaging over photon spectrum.}
\label{fig:Hp-production}
\end{figure}

\begin{figure}[h]
\centering
\includegraphics[width=0.49\textwidth]{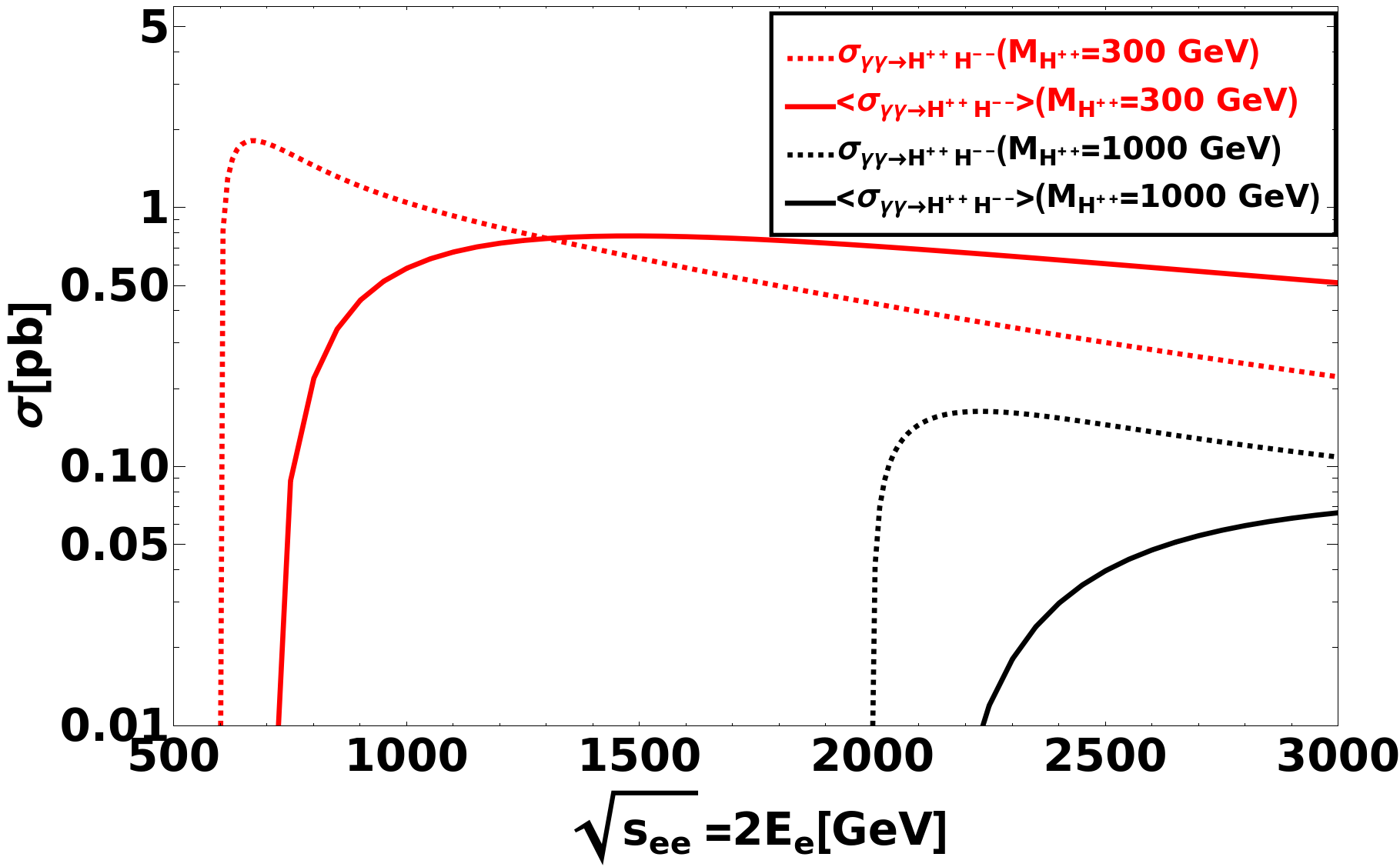}
\includegraphics[width=0.49\textwidth]{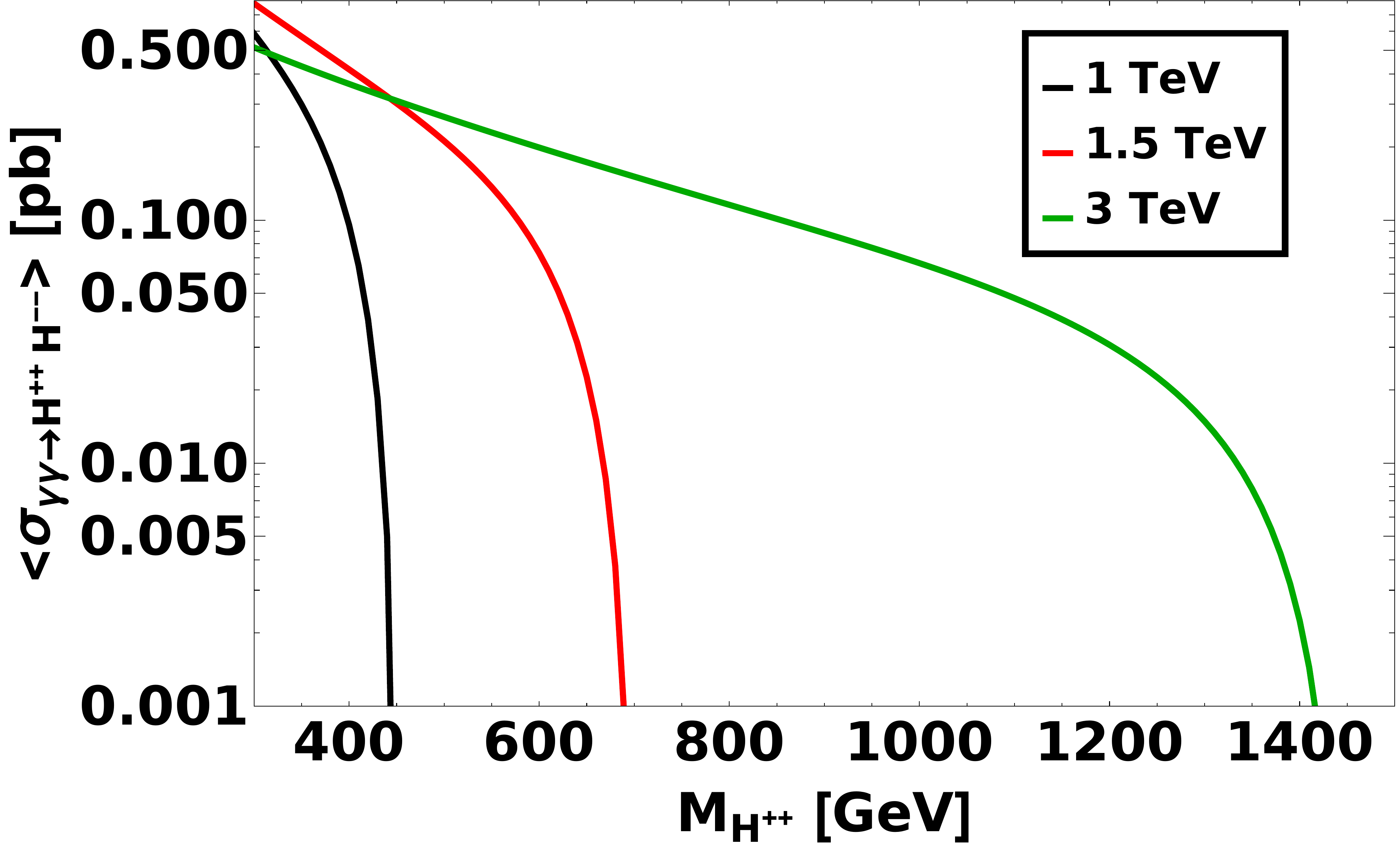}
\caption{Same as Fig.~\ref{fig:Hp-production} but now for process $\gamma\gamma\to H^{\pm\pm}H^{\mp\mp}$.}
\label{fig:Hpp-production}
\end{figure}

\hspace{0.5cm} Singly and doubly charged Higgs bosons can be produced in pair at $\gamma \gamma$ colliders. The corresponding Feynman diagrams are shown in Fig.~\ref{fig:triplet scalar}. 
The differential scattering cross section  for the process $\gamma \gamma \to H^\pm H^\mp (H^{\pm \pm}H^{\mp \mp})$ is can be written as
\begin{align}
d\sigma=\frac{1}{32\pi s}\sqrt{1-\frac{4 M_k^2}{s}} \Big(|\mathcal{M}_a|^2+|\mathcal{M}_b|^2+|\mathcal{M}_c|^2+2\text{Re}\left(\mathcal{M}_a^{\dagger}\mathcal{M}_b\right) + 2\text{Re}\left(\mathcal{M}_a^{\dagger}\mathcal{M}_c\right) + 2\text{Re}\left(\mathcal{M}_b^{\dagger}\mathcal{M}_c\right) \Big).
\label{dsig}
\end{align}
The respective terms in Eq.~\ref{dsig} are given below
\begin{align}
|\mathcal{M}_a|^2=4Q_k^4 ,\,\,\,\, |\mathcal{M}_b|^2=\frac{Q_k^4 (M_k^2+t)^2}{M_k^2\Gamma_k^2+(M_k^2-t)^2},\,\,\,\, |\mathcal{M}_c|^2=\frac{Q_k^4 (s+t-3M_k^2)^2}{M_k^2\Gamma_k^2+(s+t-M_k^2)^2},
\end{align}
and the interference terms can be written as
\begin{align}
& 2 \text{Re}(\mathcal{M}_a^\dagger \mathcal{M}_b)=\frac{Q_k^4 (M_k^2-t) (4M_k^2-s+4t)}{2\Big(M_k^2\Gamma_k^2+(M_k^2-t)^2\Big)},\nonumber \\
& 2 \text{Re}(\mathcal{M}_a^\dagger \mathcal{M}_c)=\frac{Q_k^4 (12 M_k^2 -5s -4t) (s+t-M_k^2)}{2\Big(M_k^2\Gamma_k^2 + (s+t-M_k^2)^2\Big)}, \nonumber\\
& 2 \text{Re}(\mathcal{M}_b^\dagger \mathcal{M}_c) = \frac{Q_k^4 (s-4 M_k^2)^2 \Big(M_k^2\Gamma_k^2+(M_k^2-t)(s+t-M_k^2)\Big)}{2\Big(M_k^2\Gamma_k^2+(M_k^2-t)^2\Big) \Big(M_k^2\Gamma_k^2+(s+t-M_k^2)^2\Big)},
\end{align}
where $M_k=\{M_{H^\pm},M_{H^{\pm\pm}}\}$, $t=M_k^2+\frac{s}{2}\Big(\cos\theta\sqrt{1-\frac{4M_k^2}{s}}-1\Big)$ and $Q_k=e(2e)$ for $H^\pm(H^{\pm\pm})$. 
We show the $H^\pm$ and $H^{\pm\pm}$ pair production cross section in Fig.~\ref{fig:Hp-production} and Fig.~\ref{fig:Hpp-production}, respectively. In the left panel of Fig.~\ref{fig:Hp-production} and \ref{fig:Hpp-production}, we showed the cross section averaged over photon spectrum along with bare cross section for two choices of charged Higgs mass $M_{H^{\pm},H^{\pm\pm}}= 300$ GeV and 1000 GeV. We also show the averaged cross section  as a function of charged Higgs mass $M_{H^\pm}$ and $M_{H^{\pm\pm}}$ for different center of mass energies in the right panel of Fig.~\ref{fig:Hp-production} and \ref{fig:Hpp-production}. The cross section decreases with the increase in mass and falls sharply at the vicinity of $\frac{\sqrt{s}}{2}$.
\subsection{Triplet fermion production modes}
\begin{figure}[h]
\centering
\includegraphics[width=0.9\textwidth]{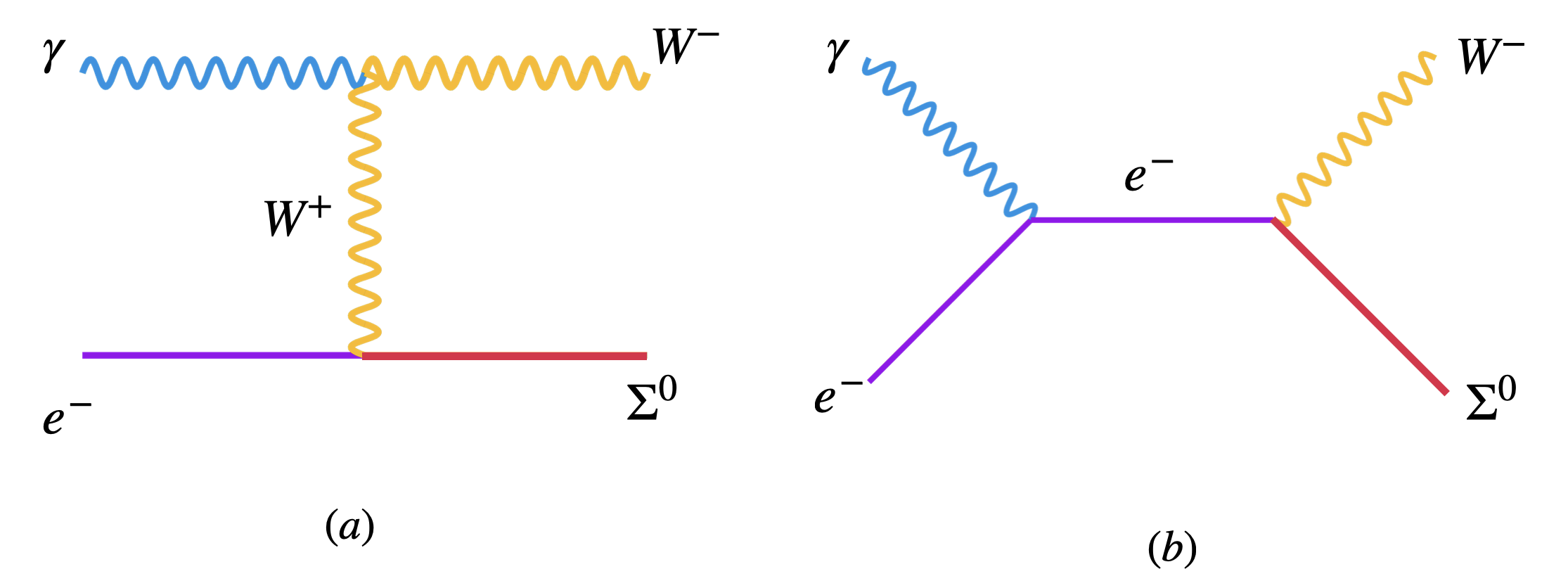}
\caption{Neutral component of the triplet fermion production at the $e^- \gamma$ colliders in association with $W^-$.}
\label{fig:FD-3}
\end{figure}

\begin{figure}[h]
\centering
\includegraphics[width=0.49\textwidth]{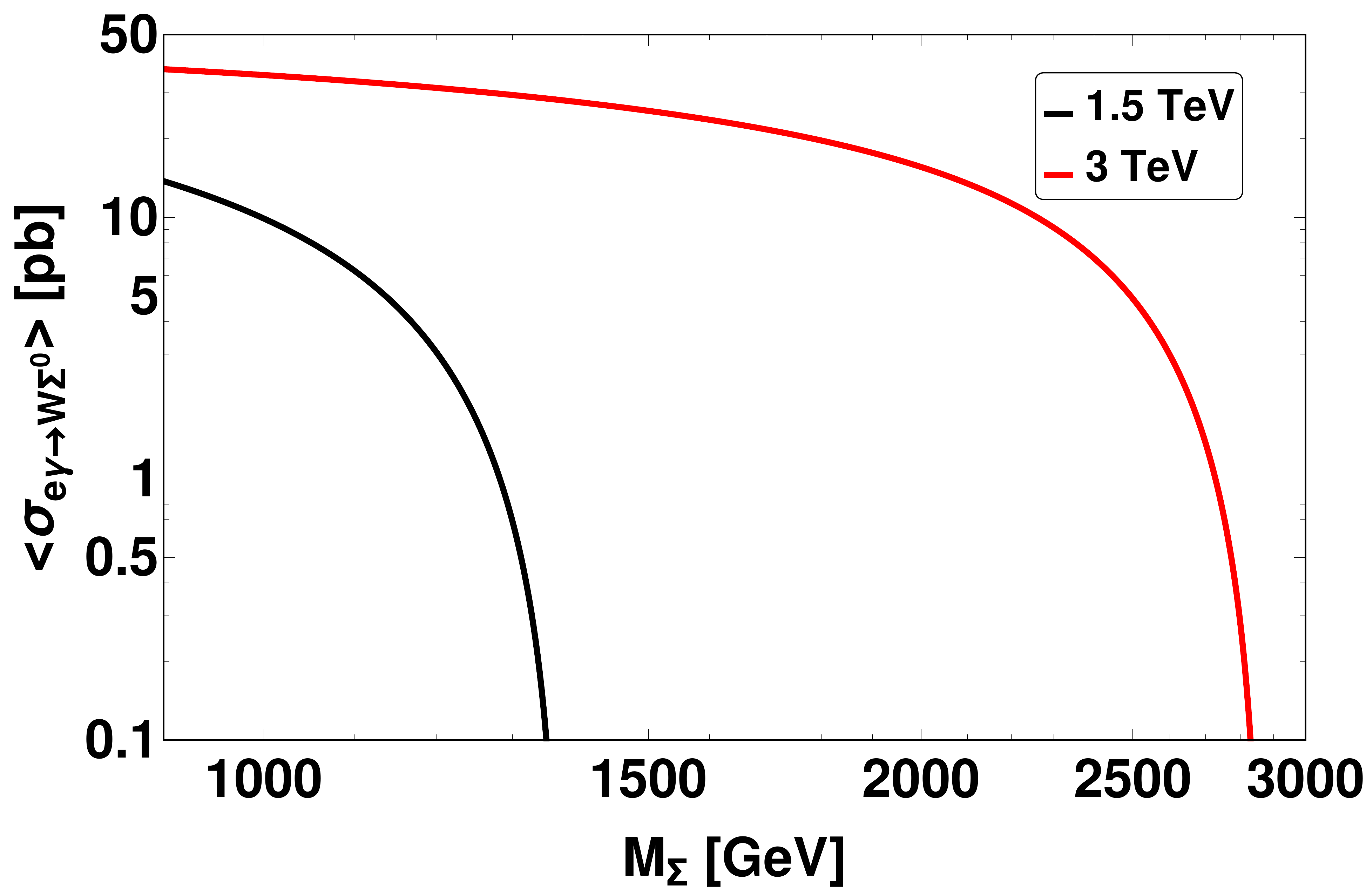}
\caption{Total cross section of the $W\Sigma^0$ process in $e^- \gamma$ colliders as a function of $M_\Sigma$ for different $\sqrt{s}$ normalized by the square of the mixing.}
\label{fig:Sigma-production}
\end{figure}

The triplet fermions can also be tested at the $e^- \gamma$ collider. The triplet fermion has neutral and charged multiplets which can be produced at the $e^-\gamma$ colliders in association with $W^-$, $Z$ and $h$ bosons respectively. The Feynman diagrams of $\Sigma^0$ production is shown in Fig.~\ref{fig:FD-3}. 
The analytical expressions for this process is same as $e^-\gamma\to N_1 W^-$, see Eqs.~\ref{eg1} and \ref{eg2}. We have to replace $M_{N_1}$ and $V_{eN_1}$ by $M_{\Sigma}$ and $V_e$, respectively. The total cross sections for different center of mass energies are shown in Fig.~\ref{fig:Sigma-production}. The production cross section increases with the increase in $\sqrt{s}$. In case of type-III seesaw the triplet mass is roughly ruled out below 1 TeV from the LHC searches at 139 fb$^{-1}$ luminosity \cite{ATLAS:2022yhd,CMS:2022nty}. Hence we consider lepton colliders with $\sqrt{s}=1.5$ TeV and 3 TeV respectively. 
\begin{figure}[h]
\centering
\includegraphics[width=0.9\textwidth]{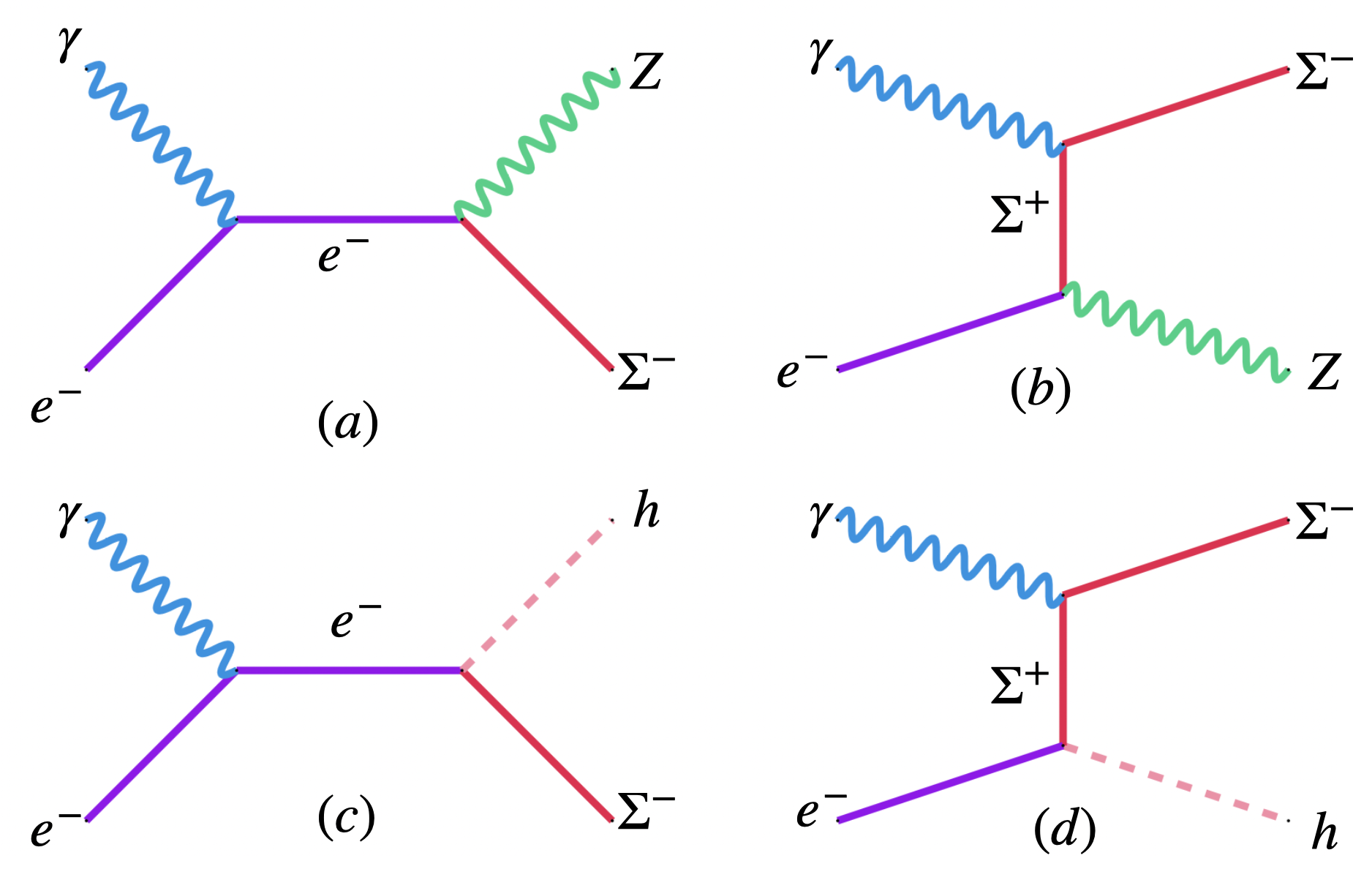}
\caption{Charged components of the triplet fermion production at the $e^- \gamma$ colliders in association with $Z$ (upper panel) and $h$ (lower panel).}
\label{fig:FD-4}
\end{figure}

\begin{figure}[h]
\centering
\includegraphics[width=0.49\textwidth]{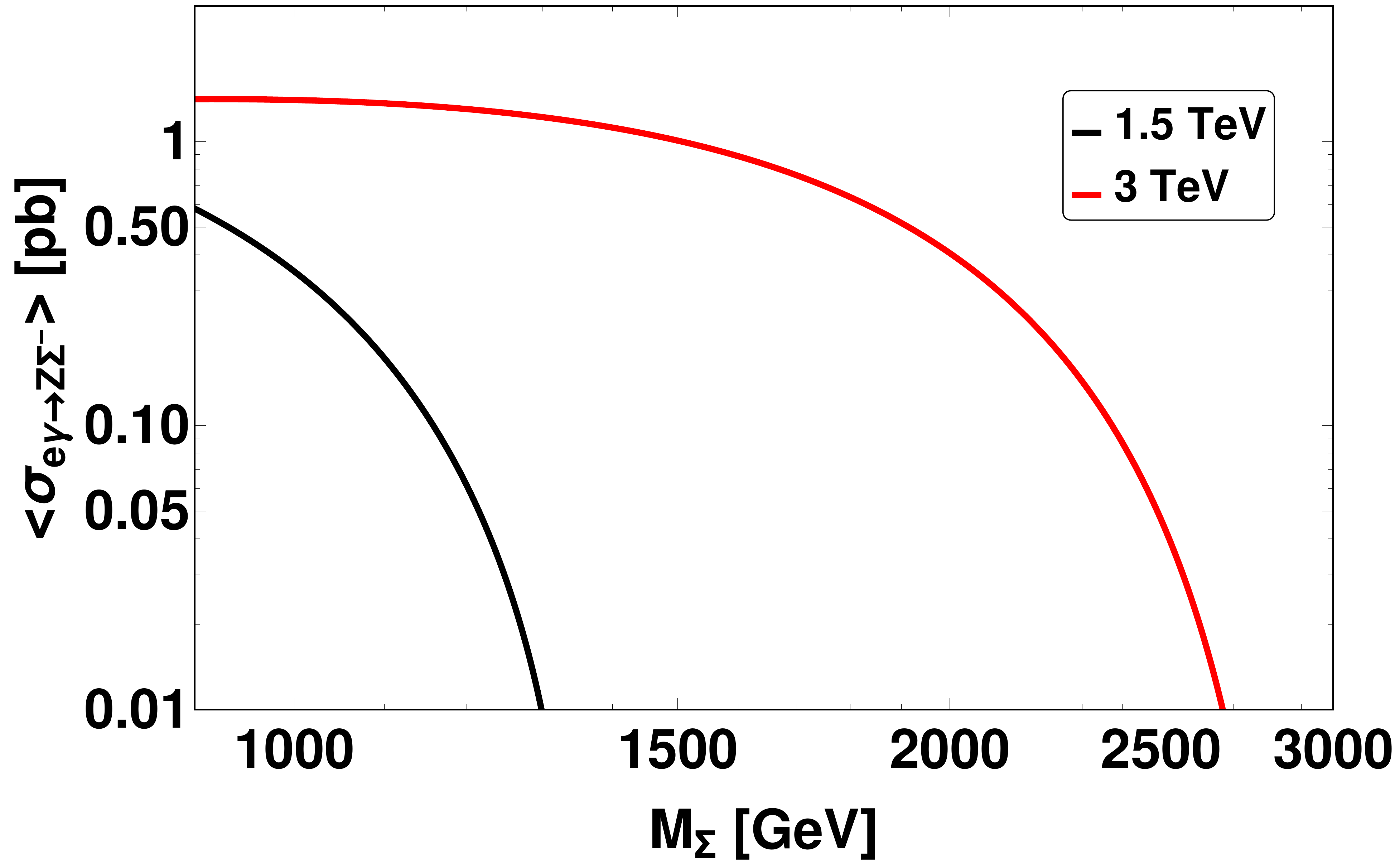}
\includegraphics[width=0.49\textwidth]{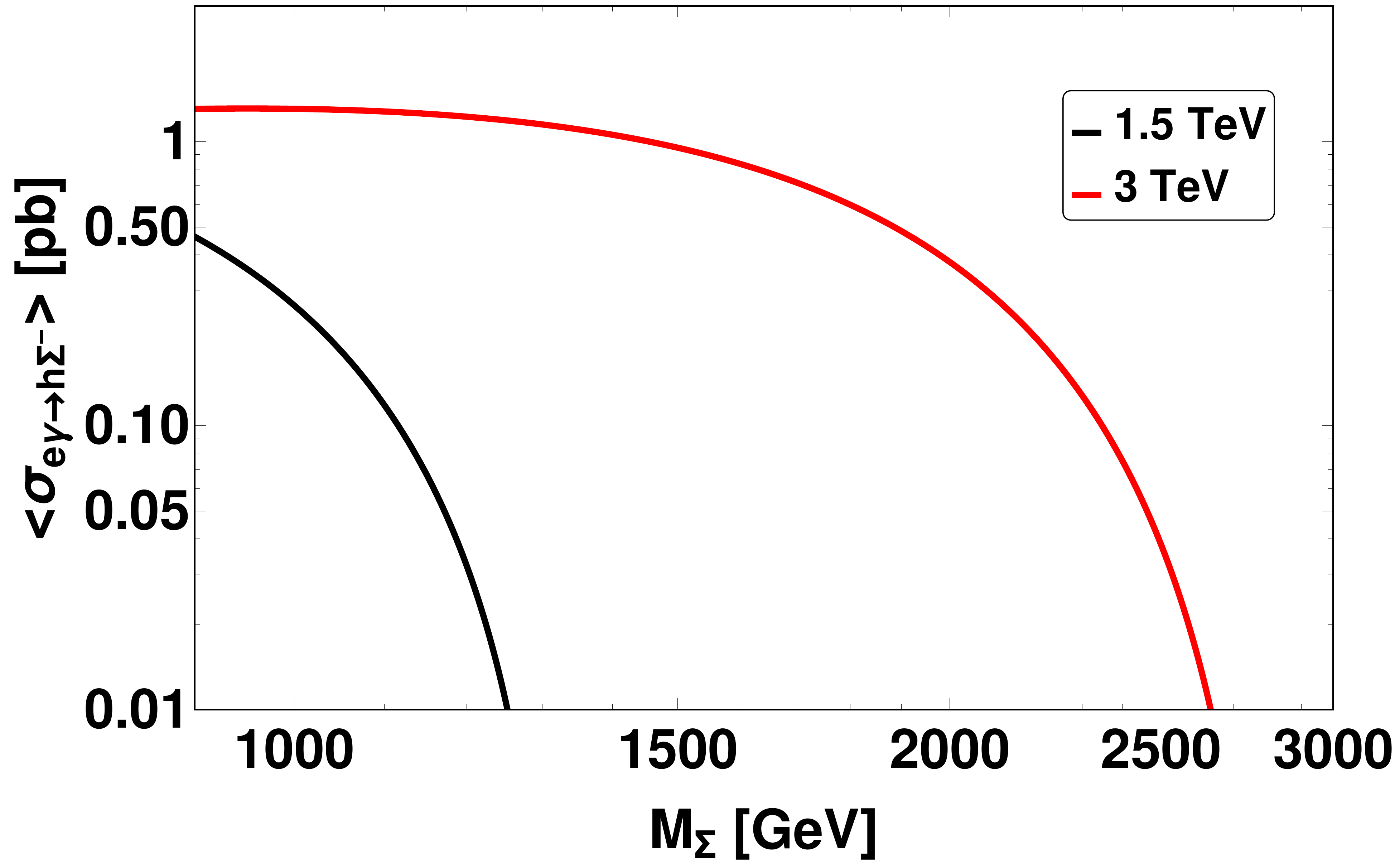}
\caption{Total production cross section of $Z\Sigma^-$ and $h\Sigma^-$ production modes at $\gamma e$ colliders as a function of $M_{\Sigma}$ for different $\sqrt{s}$ normalized by the square of the mixing.}
\label{fig:Sigma-production-1}
\end{figure}

\hspace{0.5cm} The charged multiplet of the triplet fermion $\Sigma^{\pm}$ can be produced in association with $Z$ and $h$ in the $e^-\gamma$ collider. The corresponding Feynman diagrams are shown in Fig.~\ref{fig:FD-4}. We first describe the $e^- \gamma \to \Sigma^- Z$ process and calculate the differential scattering cross section as
\begin{align}
d\sigma=\frac{1}{32\pi s}\lambda^{\frac{1}{2}}\Big(1,\frac{M_{\Sigma^-}^2}{s},\frac{m_Z^2}{s}\Big) \left(|\mathcal{M}_s|^2+|\mathcal{M}_t|^2+2\text{Re}\left(\mathcal{M}_s^{\dagger}\mathcal{M}_t\right)\right),
\end{align}
considering the $s-$ channel, $t-$ channel process and their interference. The corresponding scattering amplitude modulus squared are given below as 
\begin{align}
|\mathcal{M}_s|^2&=\frac{V_e^2 e^4}{4 s c_W^2 s_W^2 m_Z^2}\Big(m_Z^2(2M_{\Sigma^-}^2-s-2t)+(M_{\Sigma^-}^2-s)(M_{\Sigma^-}^2-s-t)\Big),\\
|\mathcal{M}_t|^2&=\frac{V_e^2 e^4}{4 c_W^2 s_W^2 m_Z^2 \Big(M_{\Sigma^-}^2\Gamma_{\Sigma^-}^2+(M_{\Sigma^-}^2-t)^2\Big)}\Big(4M_{\Sigma^-}^2 m_Z^4-m_Z^2(M_{\Sigma^-}^4-2sM_{\Sigma^-}^2+t(t+2s))\nonumber\\&+t(M_{\Sigma^-}^4-M_{\Sigma^-}^2(s+4t)+t(s+t))\Big),
\end{align}
and the interference term can be written as 
\begin{align}
2\text{Re}\left(\mathcal{M}_s^{\dagger}\mathcal{M}_t\right) &=\frac{V_e^2 e^4}{2 s c_W^2 s_W^2 m_Z^2 \Big(M_{\Sigma^-}^2\Gamma_{\Sigma^-}^2+(M_{\Sigma^-}^2-t)^2\Big)} (M_{\Sigma^-}^2-t) \Big(-m_Z^4(M_{\Sigma^-}^2+2(s+t)) \nonumber \\ 
&+m_Z^2(M_{\Sigma^-}^2-s)(t-M_{\Sigma^-}^2)+2m_Z^6+t((M_{\Sigma^-}-s)^2+st)\Big),
\end{align}
where $m_Z$ is the $Z$ mass. Next we consider the $e^- \gamma \to \Sigma^- h$ process and calculate the differential scattering cross section as
\begin{align}
d\sigma=\frac{1}{32\pi s}\lambda^{\frac{1}{2}}\Big(1,\frac{M_{\Sigma^-}^2}{s},\frac{m_h^2}{s}\Big) \left(|\mathcal{M}_s|^2+|\mathcal{M}_t|^2+2\text{Re}\left(\mathcal{M}_s^{\dagger}\mathcal{M}_t\right)\right).
\end{align}
We consider the $s-$ channel, $t-$ channel processes and their interference. The corresponding scattering amplitude modulus squared are given below as 
\begin{align}
|\mathcal{M}_s|^2 &=\frac{V_e^2 (1-V_e^2-V_\mu^2-V_\tau^2)^2 e^2 M_{\Sigma^-}^2}{s v_H^2} \Big(M_{\Sigma^-}^2-t\Big), \\
|\mathcal{M}_t|^2 &=\frac{V_e^2 (1-V_e^2-V_\mu^2-V_\tau^2)^2 e^2 M_{\Sigma^-}^2}{v_H^2 \Big(M_{\Sigma^-}^2\Gamma_{\Sigma^-}^2+(M_{\Sigma^-}^2-t)^2\Big)} \Big(M_{\Sigma^-}^2 (s+2 m_h^2)-t(s+2M_{\Sigma^-}^2)\Big),
\end{align}
and the interference term can be written as
\begin{align}
2\text{Re}\left(\mathcal{M}_s^{\dagger}\mathcal{M}_t\right) &=\frac{2 V_e^2 (1-V_e^2-V_\mu^2-V_\tau^2)^2 e^2 M_{\Sigma^-}^2}{s v_H^2 \Big(M_{\Sigma^-}^2\Gamma_{\Sigma^-}^2+(M_{\Sigma^-}^2-t)^2\Big)}
\Big(M_{\Sigma^-}^2-t\Big) \Big(m_h^4-m_h^2(M_{\Sigma^-}^2+s+t)+M_{\Sigma^-}^2(t-s)+st\Big),
\end{align}
where $m_h$ is the SM Higgs mass. The $Z \Sigma^-$ and $h \Sigma^-$ production cross section as a function of triplet mass for different center of mass energies has been shown in the left and right panel of Fig.~\ref{fig:Sigma-production-1}.
In the context of type-III seesaw we will consider $e^- \gamma \to \Sigma^0 W^-$ and $Z \Sigma^-$ modes to probe the triplet fermions at the $e^- \gamma$ colliders. 
\section{Different signatures form the seesaw models at the colliders}
\label{Analysis}
In the following we consider in detail to study various final states of the singlet heavy neutrino from type-I seesaw, triplet scalar from the type-II seesaw and triplet fermion from type-III seesaw respectively. We use MadGraph \cite{Alwall:2011uj} to generate the signal and background events and hadronize through Pythia 8 \cite{Sjostrand:2014zea}. We perform the detector simulation using Delphes~\cite{deFavereau:2013fsa}. We used the IDEA card for $\sqrt{s}=$91.2 GeV $e^-e^+$ colldier, ILC-gen card for $\sqrt{s}=250$ GeV and ILD-card for $\sqrt{s}=500$ GeV, 1 TeV and 3 TeV  $e^-e^+$, $e^-\gamma$ and $\gamma \gamma$ colliders respectively. To simplify the analysis, we assumed that heavy neutrinos $N_{2,3}$ are too high to acces at the considered colliders or their couplings to SM particles are zero. Further we assumed that the heavy neutrino $N_1$ is only coupled to electron, i.e. $V_{eN_1}=V_{eN}\neq 0, V_{\mu N_1}=V_{\tau N_1}=0$. Note that our analysis at $e^-\gamma$ and $\gamma\gamma$ collider are done assuming unpolarized electron and photon beams, which is ofcourse not true in realistic scenario as the backward Compton scattered photons are in general will be polarized. In the case of polarized beam, the signal and background behave differently for different polarization and this might weaken our obtained bound discussed below.
\subsection{Different signatures from heavy neutrinos}
In our analysis we concentrate on the first generation RHN which dominantly couples with first generation of the leptons. We consider the decay mode of the heavy neutrino as $N \to e W$ followed by the hadronic decay of the $W$ boson. We study the heavy neutrino production at the $e^- \gamma$ colliders and we estimate the bounds on the light-heavy mixing after studying various possible signal and their SM backgrounds. We also study final states coming from heavy neutrino production at the $e^-e^+$ collider as a part of complementarity study. 
\subsubsection{Heavy neutrinos at $e^-\gamma$ colliders}
At the $e^-\gamma$ collider, we have considered the following set of signals after the production of the heavy neutrinos in association with $W^-$:
\begin{itemize}
\item[(i)] $e^-\gamma \to N W^-\to e^\pm W^\mp W^-\to e^\pm + 4j$, where $W^\mp$ boson coming from $N$ decays and associated $W^-$ boson both decays into jets. The corresponding Feynman diagram is shown in the left panel of Fig.~\ref{fig1x}. We do not show the Feynman diagram for the $s$-channel process, however, included in the analysis. Note that the final state $e^- + 4j$ is lepton number conserving~(LNC) where as $e^+ + 4j$ final state is lepton number violating~(LNV). Hence one expect no SM background for $e^+ + 4j$ final state. In our analysis we combine both the LNV and LNC final state.
\item[(ii)] $e^-\gamma \to N W^-\to e^\pm W^\mp W^-\to e^- jj \ell^{-}\nu$, where in this case associated $W^{-}$ decays leptonically. The corresponding Feynman diagram is shown in the right panel of Fig.~\ref{fig1x}. We will refer to this final state $e^- jj \ell^{-}\nu$ as same sign dilepton~(SSDL). If $\ell^-=e^-$, we call it Same Sign Same Flavor (SSSF) scenario and if $\ell^-=\mu^-$, we call it Same Sign Different Flavor (SSDF) signal. The SSSF or SSDF signal is very interesting which may help to probe the Majorana nature of the heavy neutrino involved in this process. We study the SSDL mode combining SSSF and SSDF modes respectively. 
\begin{figure}[h]
\centering
\includegraphics[width=1\textwidth]{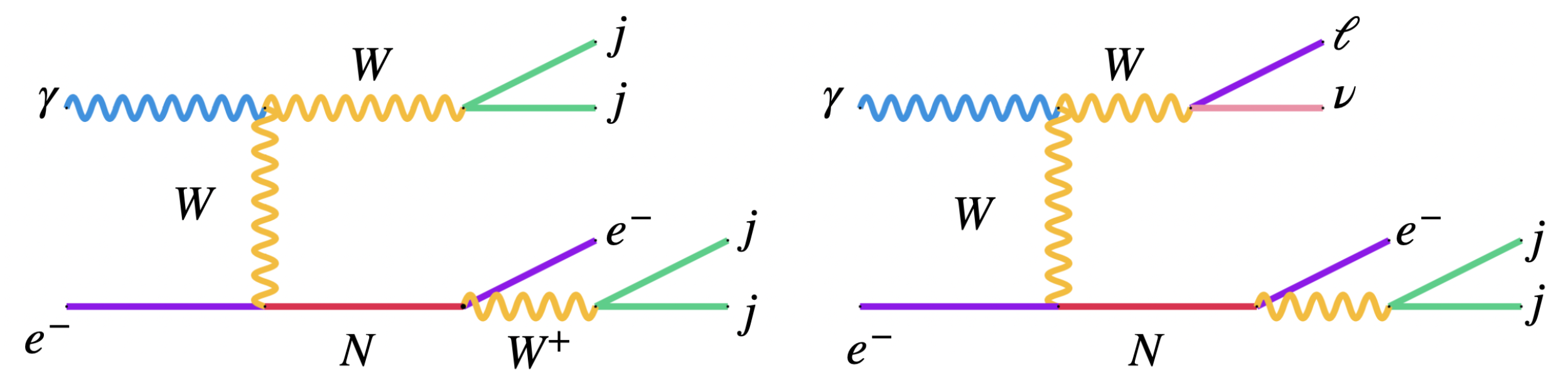}
\caption{$e^\pm+4j$ (left) and SSDL (right, $\ell^-=e^-, \mu^-$) final states in the context of $e^-\gamma$ colliders. The SSDL signal is obtained combining SSSF and SSDF signals. The corresponding $s$-channel processes has been included in the analyses.}
\label{fig1x}
\end{figure}
\begin{figure}[h]
\centering
\includegraphics[width=1\textwidth]{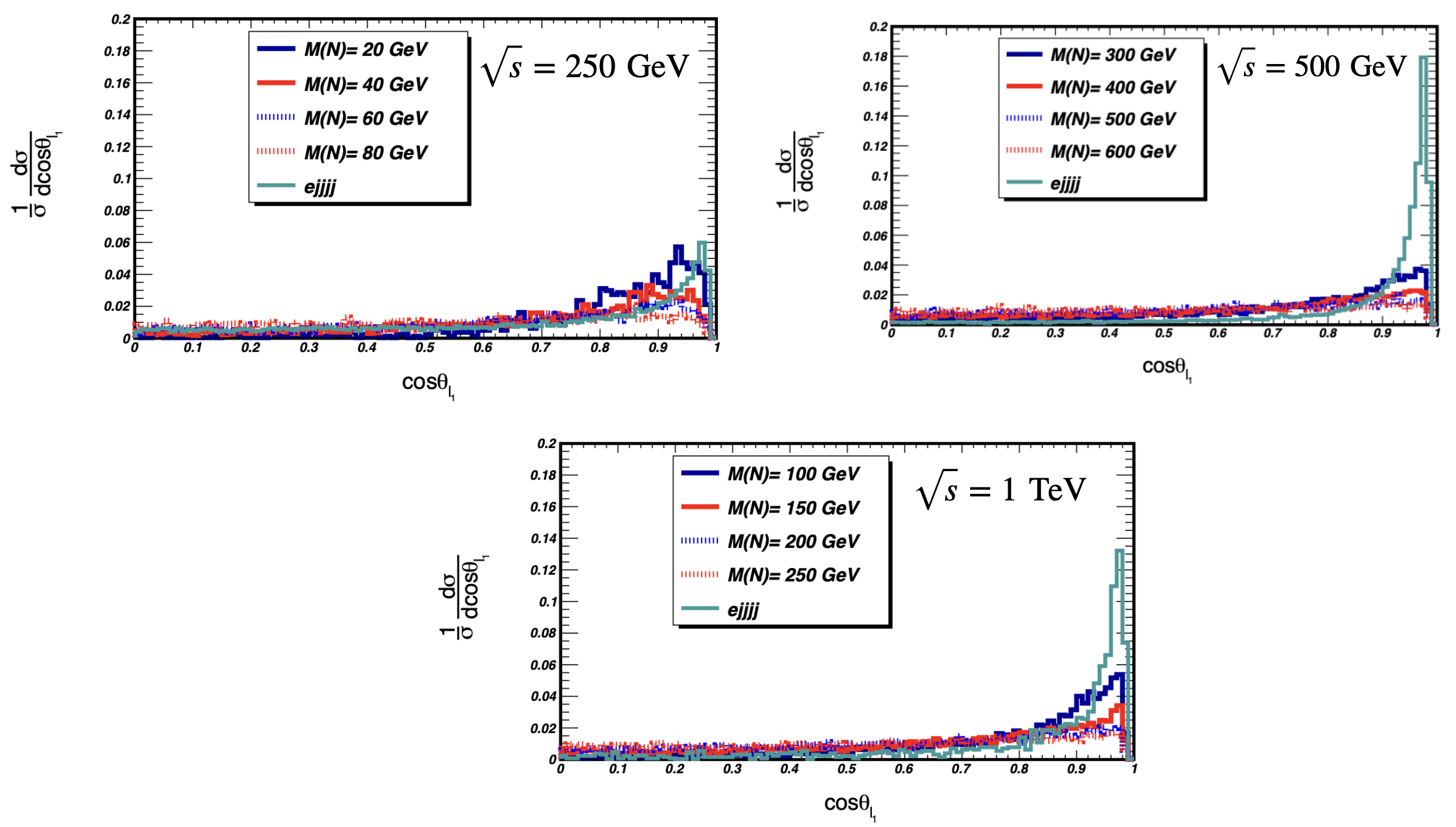}
\caption{The polar angle $(\cos\theta_{\ell_1})$ of the electron for the signal and generic background processes for $e^- \gamma$ collider at $\sqrt{s}=$250 GeV, 500 GeV and 1 TeV respectively for the $e^\pm +4j$ process. The cyan line stands for the leading SM background where as other lines stand for signal with different mass of RHN.}
\label{figx2}
\end{figure}
In our analysis, we combine LNV and LNC channels to obtain the final states as $e^\pm + 4j$. The LNV signal $e^+ + 4j$ is almost background free untill some $e^++\text{jets}$ events appear from some radiations, however that effect will be negligible. For LNC channel $e^- + 4j$, the leading SM backgrounds come from $e^-jj, e^-jjj$ and $e^-jjjj$ including initial and final state radiations. Generating the $e^\pm +4j$ events we show the distributions of the polar angle of the lepton, $\cos\theta_{\ell_1}$, of the signal and the corresponding generic background in Fig.~\ref{figx2} for $\sqrt{s}=250$ GeV, 500 GeV and 1 TeV respectively. The polar angle variable for the lepton $\cos\theta_{\ell}$ in Fig.~\ref{figx2} is defined as $\theta_{\ell}=\text{tan}^{-1}(\frac{p_T^\ell}{p_z^\ell})$, where $p_z^\ell$ is the $z$ component of the three momentum of the lepton. This is a very effective cut which reduces the SM background significantly. Note that the invariant mass cut of jets $m_{jj}$ which is coming from $W$ boson is also very effective to reduce SM background. This is because the invariant mass distribution of jets has low energy peaks which come from the hadronic activity of the low energy jets. In addition to this one can also use the invariant mass cut of the RHN. As the RHN will decay as $N\to ejj$, the invariant mass of $ejj$ will peak at $M_{ejj}\sim M_N$. Although we dont need to use this cut as it becomes redundant after using the lepton polar angle cut and invariant mass $m_{jj}$ cut. 
\begin{table}[h]
\begin{center}
\begin{tabular}{|c|c|c|c|c|c|}
\hline
\multirow{3}{*}{$\sqrt{s}$ (GeV)} &\multirow{3}{*}{$M_N$ (GeV)} & \multicolumn{2}{c|}{Signal} & %
    \multicolumn{2}{c|}{Background} \\
\cline{3-6}
&& before cuts (fb) & after cuts (fb) & before cuts (fb) & after cuts (fb)\\
\hline
&20 & 4108 & 63.26 &102.86&4.21 \\
&40&  3629  & 290.3 &102.86&4.21 \\
250 & 60 & 2923 & 426.7 &102.86&4.21\\
&80 & 3460 & 477.8 & 102.86 &4.21\\
\hline
\end{tabular}
\caption{Cross sections of the signal (normalized by $|V_{eN}|^2$) and generic background before and after cuts for $e^- \gamma$ collider at $\sqrt{s}=$250 GeV for the final state $e^\pm +4j$. We have used the following cuts: 60 GeV $< m_{jj} <$ 100 GeV, $\cos\theta_{\ell_1} < 0.94$, $p_T^{j_{1, \rm leading}} > 25$ GeV, $p_T^{j_{2, \rm trailing}} > 15$ GeV, $p_T^{j_{3, \rm trailing}} > 10$ GeV, $p_T^{j_{4, \rm trailing}} > 7$ GeV and $p_T^\ell > 12$ GeV.}
\label{tab5}
\end{center}
\end{table}
\begin{table}[h]
\begin{center}
\begin{tabular}{|c|c|c|c|c|c|}
\hline
\multirow{3}{*}{$\sqrt{s}$ (GeV)} &\multirow{3}{*}{$M_N$ (GeV)} & \multicolumn{2}{c|}{Signal} & %
    \multicolumn{2}{c|}{Background} \\
\cline{3-6}
&& before cuts (fb) & after cuts (fb) & before cuts (fb) & after cuts (fb)\\
\hline
&100 & 15601 & 3744 & 126.32 & 7.94 \\
&150 & 7279 & 2984 & 126.32 & 7.94 \\
500&200& 5677 & 1446 & 126.32 & 1.96\\
&250& 3952 & 1288 & 126.32& 1.96 \\
\hline
\end{tabular}
\caption{Cross sections of the signal (normalized by $|V_{eN}|^2$) and generic background before and after cuts for $e^- \gamma$ collider at $\sqrt{s}=$500 GeV for the final state $e^\pm+4j$. We have used the following cuts: $\cos\theta_{\ell_1} < 0.94$, 60 GeV $< m_{jj} <$ 100 GeV, $p_T^{j_{1, \rm leading}} > 30$ GeV, $p_T^{j_{2, \rm trailing}} > 20$ GeV, $p_T^{j_{3, \rm trailing}} > 20$ GeV, $p_T^{j_{4, \rm trailing}} > 10$ GeV and $p_T^\ell > 10$ GeV for $M_{N_1}=100-150$~GeV, where as for $M_{N_1}>150$~GeV we used the following cuts: $\cos\theta_{\ell_1} < 0.92$, 60 GeV $< m_{jj} <$ 100 GeV, $p_T^{j_{1, \rm leading}} > 50$ GeV, $p_T^{j_{2, \rm trailing}} > 40$ GeV, $p_T^{j_{3, \rm trailing}} > 20$ GeV, $p_T^{j_{4, \rm trailing}} > 10$ GeV and $p_T^\ell > 40$ GeV.}
\label{tab6}
\end{center}
\end{table}
\begin{table}[h]
\begin{center}
\begin{tabular}{|c|c|c|c|c|c|}
\hline
\multirow{3}{*}{$\sqrt{s}$ (TeV)} &\multirow{3}{*}{$M_N$ (GeV)} & \multicolumn{2}{c|}{Signal} & %
    \multicolumn{2}{c|}{Background} \\
\cline{3-6}
&& before cuts (fb) & after cuts (fb) & before cuts (fb) & after cuts (fb)\\
\hline
&300 & 11161 & 1339 & 94.93 & 0.241 \\
&400& 9519 & 1561 & 94.93 & 0.241 \\
1&500 & 7696 & 1301 & 94.93 & 0.082 \\
&600 & 5648 & 1174 & 94.93 & 0.082\\
&700& 3368 & 720.7 & 94.93 & 0.082\\
\hline
\end{tabular}
\caption{Cross sections of the signal (normalized by $|V_{eN}|^2$) and generic background before and after cuts for $e^- \gamma$ collider at $\sqrt{s}=$1 TeV for the final state $e^\pm+4j$. We have used the following cuts: $\cos\theta_{\ell_1} < 0.92$, $M_{N_1}-40$ GeV $< m_{e jj} < M_{N_1}+40$, 60 GeV $< m_{jj} <$ 100 GeV, $p_T^{j_{1, \rm leading}} > 60$ GeV, $p_T^{j_{2, \rm trailing}} > 30$ GeV, $p_T^{j_{3, \rm trailing}} > 20$ GeV, $p_T^{j_{4, \rm trailing}} > 10$ GeV and $p_T^\ell > 60$ GeV for $M_{N_1}=300-400$~GeV, whereas for $M_{N_1}>400$~GeV, we have used the following cuts: $\cos\theta_{\ell_1} < 0.92$, $M_{N_1}-40$ GeV $< m_{e jj} < M_{N_1}+40$, 60 GeV $< m_{jj} <$ 100 GeV, $p_T^{j_{1, \rm leading}} > 100$ GeV, $p_T^{j_{2, \rm trailing}} > 50$ GeV, $p_T^{j_{3, \rm trailing}} > 30$ GeV, $p_T^{j_{4, \rm trailing}} > 10$ GeV and $p_T^\ell > 120$ GeV.}
\label{tab8}
\end{center}
\end{table}
We study the $e^\pm+4j$ signal from the RHN production in a $e^-\gamma$ collider at the $\sqrt{s}=250$ GeV, 500 GeV and 1 TeV. The signal normalized by the square of the mixing and corresponding generic background cross sections for the $e^\pm+4j$ final state at $\sqrt{s}=$250 GeV, 500 GeV and 1 TeV are given in Tab.~\ref{tab5}, \ref{tab6} and \ref{tab8}, respectively.
\begin{table}[h]
\begin{center}
\begin{tabular}{|c|c|c|c|c|c|}
\hline
\multirow{3}{*}{$\sqrt{s}$ (GeV)} &\multirow{3}{*}{$M_N$ (GeV)} & \multicolumn{2}{c|}{Signal} & %
    \multicolumn{2}{c|}{Background} \\
\cline{3-6}
&& before cuts (fb) & after cuts (fb) & before cuts (fb) & after cuts (fb)\\
\hline
&20 & 539.18 & 11.93 & 2.02 & 0.105 \\
&40 & 467.44 & 24.71 & 2.02 & 0.105 \\
250& 60 & 379.17 & 47.05 & 2.02 & 0.105\\
&80& 1676.1 & 223.91 & 2.02 & 0.105\\
\hline
\end{tabular}
\caption{Cross sections of the signal (normalized by $|V_{eN}|^2$) and generic background before and after cuts for $e^- \gamma$ collider at $\sqrt{s}=$250 GeV for the SSDL final state. We have used the following cuts: $p_T^{j_1, \rm leading}$, $p_T^{j_2, \rm trailing}$ $> 10$ GeV, $p_T^{\ell_1, \rm leading}, p_T^{\ell_2, \rm trailing} > 10$  GeV, 60 GeV $< m_{j_{1}j_{2}}<$ 100 GeV and $\cos\theta_{l_{1, \rm leading}} < 0.9$, $\cos\theta_{l_{2, \rm trailing}} < 0.92$.}
\label{tab9}
\end{center}
\end{table}
\begin{table}[h]
\begin{center}
\begin{tabular}{|c|c|c|c|c|c|}
\hline
\multirow{3}{*}{$\sqrt{s}$ (GeV)} &\multirow{3}{*}{$M_N$ (GeV)} & \multicolumn{2}{c|}{Signal} & %
    \multicolumn{2}{c|}{Background} \\
\cline{3-6}
&& before cuts (fb) & after cuts (fb) & before cuts (fb) & after cuts (fb)\\
\hline
&100 &1480.1 & 211.82 & 57.64 & 5.31 \\
&150 &931.49& 236.61 & 57.64 & 7.42 \\
500&200&643.48 & 189.58 & 57.64 & 6.09\\
&250&419.61 &135.61 & 57.64&4.68\\
\hline
\end{tabular}
\caption{Cross sections of the signal (normalized by $|V_{eN}|^2$) and generic background before and after cuts for $e^- \gamma$ collider at $\sqrt{s}=$500 GeV for the SSDL final state. We have used the following cuts: $p_T^{j_1, \rm leading}$, $p_T^{j_2, \rm trailing}$ $> 10$ GeV, $p_T^{\ell_1, \rm leading}, p_T^{\ell_2, \rm trailing} > 10$  GeV, 60 GeV $< m_{j_{1}j_{2}}<$ 100 GeV, $M_{N_1}-40$ GeV $< m_{\ell jj} < M_{N_1}+40$ GeV and $\cos\theta_{l_{1, \rm leading}} < 0.9$, $\cos\theta_{l_{2, \rm trailing}} < 0.92$.}
\label{tab10}
\end{center}
\end{table}
\begin{table}[h]
\begin{center}
\begin{tabular}{|c|c|c|c|c|c|}
\hline
\multirow{3}{*}{$\sqrt{s}$ (TeV)} &\multirow{3}{*}{$M_N$ (GeV)} & \multicolumn{2}{c|}{Signal} & %
    \multicolumn{2}{c|}{Background} \\
\cline{3-6}
&& before cuts (fb) & after cuts (fb) & before cuts (fb) & after cuts (fb)\\
\hline
&300 &2596.01&123.3&128.82&1.76 \\
&400&2137.9&307.5&128.82&1.68 \\
1&500&1699.8&109.5&128.82&0.22\\
&600&1236.02&190.9&128.82&0.37\\
&700&733.0&139.5&128.82&0.25\\
\hline
\end{tabular}
\caption{Cross sections of the signal (normalized by $|V_{eN}|^2$) and generic background before and after cuts for $e^- \gamma$ collider at $\sqrt{s}=$1 TeV for the SSDL final state. We have used the following cuts: $p_T^{j_1, \rm leading} > 80$ GeV, $p_T^{j_2, \rm trailing} > 20$ GeV, $p_T^{\ell_1, \rm leading} > 120$  GeV,  $p_T^{\ell_2, \rm trailing} > 20$ GeV, $ E_{\ell_1, \rm leading} > 120$ GeV, 60 GeV $< m_{j_{1}j_{2}}<$ 100 GeV, $M_{N_1}-40$ GeV $< m_{\ell jj} < M_{N_1}+40$ GeV and $\cos\theta_{l_{1, \rm leading}} < 0.92$.}
\label{tab11}
\end{center}
\end{table}
\begin{table}[h]
\begin{center}
\begin{tabular}{|c|c|c|c|c|c|}
\hline
\multirow{3}{*}{$\sqrt{s}$ (TeV)} &\multirow{3}{*}{$M_N$ (GeV)} & \multicolumn{2}{c|}{Signal} & %
    \multicolumn{2}{c|}{Background} \\
\cline{3-6}
&& before cuts (fb) & after cuts (fb) & before cuts (fb) & after cuts (fb)\\
\hline
&1000 & 1342.1 & 167.94 & 85.47 & 0.14 \\
3 & 1500 & 1128.67 & 269.75 & 85.47 & 0.13 \\
& 2000 & 821.51 & 130.49 & 85.47 & 0.012 \\
& 2500 & 325.73 & 77.05 & 85.47 & 0.008 \\
\hline
\end{tabular}
\caption{Cross sections of the signal (normalized by $|V_{eN}|^2$) and generic background before and after cuts for $e^- \gamma$ collider at $\sqrt{s}=$3 TeV for the SSDL final state. We have used the following cuts:  $p_T^{J}>400$ GeV, $p_T^{\ell_1, \rm leading} > 400$  GeV for $M_N=1000$ GeV, 1500 GeV and  $p_T^{J}>800$ GeV, $p_T^{\ell_1, \rm leading} > 800$  GeV for $M_N=2000$ GeV, 2500 GeV. We have used the following additional cuts irrespective of heavy neutrino masses: 60 GeV $< m_{J}<$ 100 GeV, $M_{N_1}-40$ GeV $< m_{\ell J} < M_{N_1}+40$ GeV and $\cos\theta_{l_{1, \rm leading}} < 0.92$.}
\label{tab111}
\end{center}
\end{table}

For the case of SSDL signal we again use the polar angle cut of lepton coming from RHN. We have not shown polar angle distribution as it's behaviour almost same as in Fig.~\ref{figx2}. In addition to this we used the leading lepton energy cut and the invariant mass of the jets $(m_{j_{1}j_{2}})$ are considered to be  60 GeV $< m_{j_{1}j_{2}}<$ 100 GeV for the $W$ boson being originated from the RHN decay. The signal normalized by the square of the mixing and corresponding generic background cross sections at $\sqrt{s}=$250 GeV, 500 GeV and 1 TeV before and after our applied cuts are given in Table.~\ref{tab9}, \ref{tab10} and \ref{tab11}, respectively. When necessary we also use the invariant mass of $ejj$ which peaks at RHN mass. The specific cuts we have used for each center of mass energies are mentioned in the caption of Table.~\ref{tab9}, \ref{tab10} and \ref{tab11}, respectively. 

We have also considered center of mass energy $\sqrt{s}=3$ TeV for the SSDL signal. As, in this case one can consider heavy RHN mass, $W$ boson from the RHN decay can be boosted so that the hadronic jets could be collimated to produce a boosted jet called fatjet~($J$). We have shown the corresponding results in Table.~\ref{tab111}. We have used high $p_T$ cut for fatjet and the invariant mass cut $60\text{ GeV}\leq m_J \leq 100\text{ GeV}$, $M_{N_1}-40\text{ GeV}\leq m_{eJ}\leq M_{N_1}+40$~GeV. 

\item[(iii)] We study another interesting signal in the context of $e^- \gamma$ collider in the form of opposite sign dilepton (OSDL) final state considering $e^-\gamma \to N W^-\to e^\pm W^\mp W^-\to e^- \ell^{+}jj\nu/e^+\mu^-jj\nu$, where in this case associated $W^{-}$ decays either leptonically or hadronically. In this case we consider $\ell=e,\mu$ in the following:
\begin{itemize}
\item[(a)] Note that the final state $e^+\mu^-jj\nu$ is almost background free and we call it opposite sign different flavor signal. In Table.~\ref{tab:OSDL}, we have shown the corresponding results for $\sqrt{s}=250$ GeV, 500 GeV, 1 TeV and 3 TeV, respectively. As this final state is almost background free, we have used very basic cuts such as $p_T^{j}>10$~GeV, $p_T^{\ell}>10$~GeV and $|\eta_{\ell,j}|<2.5$ to study the events. 
\begin{table}[h]
\begin{center}
\begin{tabular}{|c|c|c|c|}
\hline
\multirow{3}{*}{$\sqrt{s}$ (GeV)} &\multirow{3}{*}{$M_N$ (GeV)} & \multicolumn{2}{c|}{Signal}  \\
\cline{3-4}
&& before cuts (fb) & after cuts (fb) \\
\hline
& 20 & 386.6 & 77.33  \\
& 40 & 341.6 & 102.5   \\
250 & 60 & 275.2 & 110.1  \\
& 80 & 195.6 & 68.46   \\
\hline
\hline
& 100 & 1481 & 698.3  \\
& 150 & 928.5 & 601.4   \\
500 & 200 & 641.1 & 432.9  \\
& 250 & 418.8 & 283.1   \\
\hline
\hline
& 300 & 1161 & 678.4  \\
& 400 & 954.1 & 584.1   \\
1000 & 500 & 756.9 & 471.2  \\
& 600 & 547.1 & 339.1   \\
& 700 & 322.9  & 190.2   \\
\hline
\hline
& 1000 & 1299 & 879.3  \\
& 1500 & 1085 & 804.6   \\
3000 & 2000 & 783.6 & 625.4  \\
& 2500 & 309.5 & 262.9   \\
\hline
\end{tabular}
\caption{Cross sections of the signal (normalized by $|V_{eN}|^2$) before and after cuts for $e^- \gamma$ collider at different center of mass energies for the final state $ e^+\mu^- jj p_{T}^{\rm miss}$. We have used the following basic cuts: $p_T^{j}>10$~GeV, $p_T^{\ell}>10$~GeV and $|\eta_{\ell,j}|<2.5$. For $\sqrt{s}=$250 GeV, 500 GeV and 1 TeV we have demanded at least two jets, whereas for $\sqrt{s}=$3 TeV we have considered final states with one or more jets as for this one has the possibility of a fat-jet. The criteria for fat-jet are same the SSDL scenario.}
\label{tab:OSDL}
\end{center}
\end{table}

\item[(b)] On the other hand the final state  $e^-\ell^+ j j \nu$ with $\ell^+=e^+, \mu^+$has sizable SM background from the process $e^-\gamma\to e^-W^+ W^-$. In Table.~\ref{tab:OSSF}, we show the signal and corresponding backgrounds for $\sqrt{s}=250$ GeV, 500 GeV, 1 TeV and 3 TeV, respectively. The specific cuts we have used for each center of mass energies are mentioned in the caption of Table.~\ref{tab:OSSF}. 

\begin{table}[h]
\begin{center}
\begin{tabular}{|c|c|c|c|c|c|}
\hline
\multirow{3}{*}{$\sqrt{s}$ (GeV)} &\multirow{3}{*}{$M_N$ (GeV)} & \multicolumn{2}{c|}{Signal} & %
    \multicolumn{2}{c|}{Background} \\
\cline{3-6}
&& before cuts (fb) & after cuts (fb) & before cuts (fb) & after cuts (fb)\\
\hline
 & 20 & 866.9 & 83.69 & 2.08 & 0.79 \\
 & 40 & 752.7 & 121.65 & 2.08 & 0.79 \\
250 & 60 & 618.2 & 153.97 & 2.08 & 0.79 \\
 & 80 & 2953 & 670.28 & 2.08 & 0.79 \\
\hline
\hline
    & 100 & 2962  &  490.53  &  59.78  & 10.35 \\
    & 150 & 1860 &   455.06  &  59.78  & 10.35 \\
500 & 200 & 1287 &  371.46   &  59.78  & 10.35  \\
    & 250 &  838.75 & 253.79  & 59.78  & 10.35 \\    
\hline
\hline
    & 300 & 2318 &  248.91  & 141.75 &  7.38 \\
    & 400 & 1908 &  361.26  & 141.75 &  7.38 \\
1000 & 500 & 1513 & 375.15 & 141.75  & 7.38 \\
    & 600 & 1095  & 240.59  & 141.75  & 2.58 \\
    & 700 & 644.5 & 151.98  & 141.75   & 2.58 \\
\hline
\hline
     & 1000 &  2591  &  91.66  &  120.56  &  0.34 \\
     & 1500 &  2168  & 242.91  &   120.56  &  0.34 \\
3000 & 2000 & 1572   & 133.91  &   120.56  &  0.017 \\
     & 2500 &  616.5 & 42.46   &   120.56  &  0.017 \\
\hline                  
\end{tabular}
\caption{Cross sections of the signal (normalized by $|V_{eN}|^2$) before and after cuts for $e^- \gamma$ collider at different center of mass energies for the final state $ e^-\ell^{+} jj p_{T}^{\rm miss}$. For $\sqrt{s}=$250 GeV, 500 GeV and 1 TeV we have demanded at least two jets, whereas for $\sqrt{s}=$3 TeV we have considered final states with one or more jets as for this one has the possibility of a fat-jet. The criteria for fat-jet are same the SSDL scenario. For all of considered center of mass energies we used the polar angle cuts for leptons as $|\cos\theta_{\ell_1}|<0.9$, $|\cos\theta_{\ell_2}|<0.92$ and invariant mass cut $60 < m_{j_1 j_2} < 100$~GeV. For $\sqrt{s}=250$~GeV and 500 GeV, we have used the following basic cuts: $p_T^{j_i}>10$~GeV, $p_T^{\ell_i}>10$~GeV. For $\sqrt{s}=1000$~GeV, we used the following cuts: $p_T^{j_1}>20$~GeV, $p_T^{j_2}>10$~GeV, $p_T^{\ell_1}>90$~GeV, $p_T^{\ell_2}>20$~GeV for $M_N=300-500$~GeV whereas for $M_N>500$~GeV, we used $p_T^{j_1}>20$~GeV, $p_T^{j_2}>10$~GeV, $p_T^{\ell_1}>150$~GeV, $p_T^{\ell_2}>20$~GeV. For $\sqrt{s}=3000$~GeV, we used the following cuts: $p_T^{J}>180$~GeV, $p_T^{\ell_1}>450$~GeV, $p_T^{\ell_2}>30$~GeV, $60 < m_J < 100$~GeV in mass range $M_N=1000-1500$~GeV, where as for $M_N>1500$~GeV we used: $p_T^{J}>180$~GeV, $p_T^{\ell_1}>800$~GeV, $p_T^{\ell_2}>40$~GeV, $60 < m_J < 100$~GeV.}
\label{tab:OSSF}
\end{center}
\end{table}

\end{itemize} 
\end{itemize}

Using the normalized signal events from Tabs.~\ref{tab5}-\ref{tab:OSSF} and the corresponding backgrounds we estimate a $2\sigma$ contour on the $M_{N_1}-|V_{eN}|^2$ plane solving the following equation
\bea
\sigma = \frac{\rm S \times |V_{eN}|^2}{\sqrt{\rm S \times |V_{eN}|^2+\rm B}},
\label{signi}
\eea
where $S$ and $B$ stands for signal and corresponding backgrounds, respectively. We consider 5.6 ab$^{-1}$, 4 ab$^{-1}$, 8 ab$^{-1}$ and 4 ab $^{-1}$ luminosities as benchmarks for the colliders with center of mass energies 250 GeV, 500 GeV, 1 TeV and 3 TeV respectively. The 2$\sigma$ limits are shown in Fig.~\ref{fig:mix1} by the thick solid (combined $e^\pm+4j$ and SSDL, `with SSDL') blue line. Combining the results given in Tab.~\ref{tab:OSDL} and \ref{tab:OSSF} and applying Eq.~\ref{signi} we estimate bounds on the mixing angle and it is shown by the dashed (combined $e^\pm+4j$ and OSDL, `with OSDL') orange line.
\subsubsection{Heavy neutrinos at $e^+e^-$ colliders for complementarity}
We consider $e^+ e^- \to N \nu$ process followed by $N \to e^\pm W^\mp$ and hadronic decay of the $W$ boson showing $e^\pm +2j+p_T^{\rm miss}$ final state. To study this final state we consider $\sqrt{s}=91.2$ GeV, 250 GeV and 500 GeV. The distributions of the polar angle of the electron for the signal and corresponding backgrounds are same as in Fig.~\ref{figx2}.  To study this final state at $\sqrt{s}=91.2$ GeV, we consider a generic background. The signal normalized by the square of the mixing and generic background cross sections before and after cuts at $\sqrt{s}=91.2$ GeV are given in in Tab.~\ref{tab1}. The selection cuts for the signal and generic background are mentioned in the caption of Table.~\ref{tab1}.
\begin{table}[h]
\begin{center}
\begin{tabular}{|c|c|c|c|c|c|}
\hline
\multirow{3}{*}{$\sqrt{s}$ (GeV)} &\multirow{3}{*}{$M_N$ (GeV)} & \multicolumn{2}{c|}{Signal} & %
    \multicolumn{2}{c|}{Background} \\
\cline{3-6}
&& before cuts (fb) & after cuts (fb) & before cuts (fb) & after cuts (fb)\\
\hline
&20 &370.21 $\times 10^{4}$&93.24$\times 10^{4}$&&0.163 \\
&40&282.10$\times 10^{4}$&105.53$\times 10^{4}$&2.52&0.163 \\
91.2&60 &156.23$\times 10^{4}$&63.16$\times 10^{4}$&&0.320\\
&80&183.47$\times 10^{4}$&69.63$\times 10^{4}$&&0.564\\
\hline
\end{tabular}
\caption{Cross sections of the signal (normalized by $|V_{eN}|^2$) and generic background before and after cuts for $e^-e^+$ collider at $\sqrt{s}=$91.2 GeV for the final state $e^\pm+2j+p_T^{\rm miss}$. We have used the following cuts: $\cos\theta_\ell < 0.92$, $p_T^{j_1}> 15$ GeV, $p_T^{j_2}> 10$ GeV, $p_T^{\rm miss}> 15$ GeV and $p_T^{\ell} > 10$ GeV.}
\label{tab1}
\end{center}
\end{table}
To study the $e^\pm+2j+p_T^{\rm miss}$ final state at $\sqrt{s}=250$ GeV $e^-e^+$ collider, in addition to the generic background we also consider the background coming from the $ZZ$ final state where one $Z$ boson decays leptocically and the other hadronically. For relatively larger center of mass energy such as $\sqrt{s}=500$ GeV one need to also add another important background in the form of $t\overline{t}$ process. The signal normalized by the square of the mixing and generic background cross sections before and after cuts at $\sqrt{s}=250$ GeV and $\sqrt{s}=500$ GeV for $e^\pm+2j+p_T^{\rm miss}$  final state are shown in Tab.~\ref{tab2} and Tab.~\ref{tab3}, respectively.
\begin{table}[h]
\begin{center}
\begin{tabular}{|c|c|c|c|c|c|}
\hline
\multirow{3}{*}{$\sqrt{s}$ (GeV)} &\multirow{3}{*}{$M_N$ (GeV)} & \multicolumn{2}{c|}{Signal} & %
    \multicolumn{2}{c|}{Backgrounds} \\
\cline{3-6}
&& before cuts (fb) & after cuts (fb) & before cuts (fb) & after cuts (fb)\\
\hline
&100&40510&5502&$e\nu jj:$ 4536&267 \\
\cline{5-6}
&&&&$ZZ:$ 101&4.6\\
\cline{2-6}
&250&38200&3097&$e\nu jj:$ 4536&99 \\
\cline{5-6}
&&&&$ZZ:$ 101&1.3\\
\cline{2-6}
250&150&20300&3813&$e\nu jj:$ 4536&92.4\\
\cline{5-6}
&&&&$ZZ:$ 101&1.3\\
\cline{2-6}
&450&8600&2320&$e\nu jj:$ 4536&60\\
\cline{5-6}
&&&&$ZZ:$ 101&0.63\\
\hline
\end{tabular}
\caption{Cross sections of the signal (normalized by $|V_{eN}|^2$) and backgrounds before and after cuts for $e^-e^+$ collider at $\sqrt{s}=$250 GeV for the final state $e^\pm+2j+p_T^{\rm miss}$. We have used the following cuts: $\cos\theta_\ell < 0.92$, $p_T^{j_1}> 30$ GeV, $p_T^{j_2}> 20$ GeV, $p_T^{\ell} > 10$ GeV and $p_T^{\rm miss}> 20$ GeV.}
\label{tab2}
\end{center}
\end{table}
\begin{table}[h]
\begin{center}
\begin{tabular}{|c|c|c|c|c|c|}
\hline
\multirow{3}{*}{$\sqrt{s}$ (GeV)} &\multirow{3}{*}{$M_N$ (GeV)} & \multicolumn{2}{c|}{Signal} & %
    \multicolumn{2}{c|}{Backgrounds} \\
\cline{3-6}
&& before cuts (fb) & after cuts (fb) & before cuts (fb) & after cuts (fb)\\
\hline
&100&51028&16409&$e\nu jj:$ 2251&841 \\
\cline{5-6}
&&&&$ZZ:$ 38&1.53\\
\cline{5-6}
&&&&$t\overline{t}:$ 198.5&8.8\\
\cline{2-6}
&250&24506&11023&$e\nu jj:$ 2251&392.4 \\
\cline{5-6}
&&&&$ZZ:$ 38&0.35\\
\cline{5-6}
&&&&$t\overline{t}:$ 198.5&2.54\\
\cline{2-6}
500&350&15212&6303&$e\nu jj:$ 2251&146.3\\
\cline{5-6}
&&&&$ZZ:$ 38&0.2\\
\cline{5-6}
&&&&$t\overline{t}:$ 198.5&0.6\\
\cline{2-6}
&450&4701&2017&$e\nu jj:$ 2251&51.0\\
\cline{5-6}
&&&&$ZZ:$ 38&0.1\\
\cline{5-6}
&&&&$t\overline{t}:$ 198.5&0.06\\
\hline
\end{tabular}
\caption{Cross sections of the signal (normalized by $|V_{eN}|^2$) and backgrounds before and after cuts for $e^-e^+$ collider at $\sqrt{s}=$500 GeV for the final state $e^\pm+2j+p_T^{\rm miss}$. We have used the following cuts: $\cos\theta_\ell < 0.92$, $p_T^{j_1}> 45$ GeV, $p_T^{j_2}> 10$ GeV, $p_T^{\rm miss}> 15$ and $p_T^{\ell} > 50$.}
\label{tab3}
\end{center}
\end{table}
\hspace{0.5cm} Using the normalized signal events $(S)$ from Tabs.~\ref{tab1}-\ref{tab3} and the corresponding backgrounds $(B)$, we estimate the significance of a $2\sigma$ contour on the $M_{N_1}-|V_{eN}|^2$ plane solving Eq.~\ref{signi} with luminosities 150 ab$^{-1}$, 5.6 ab$^{-1}$ and 4 ab$^{-1}$ for $\sqrt{s}=91.2$ GeV \cite{Blondel:2021ema}, 250 GeV and 500 GeV \cite{CEPCStudyGroup:2018rmc,LCCPhysicsWorkingGroup:2019fvj,CLICdp:2018cto} respectively. In this context we have scaled the previous $\sqrt{s}=$1 TeV \cite{LCCPhysicsWorkingGroup:2019fvj} and 3 TeV \cite{CLICdp:2018cto} results using fat-jet signatures (hadronic decay of the $W$ boson) from \cite{Das:2018usr} at 8 ab$^{-1}$ and 5 ab$^{-1}$ luminosities respectively. These are represented by thick, darker green dot dashed contour in Fig.~\ref{fig:mix1} as `$e^- e^+$'. We show these bounds as a part of complementary study.
\subsubsection{Different limits on light-heavy mixing angle}
\begin{figure}[h]
\centering
\includegraphics[width=0.9\textwidth]{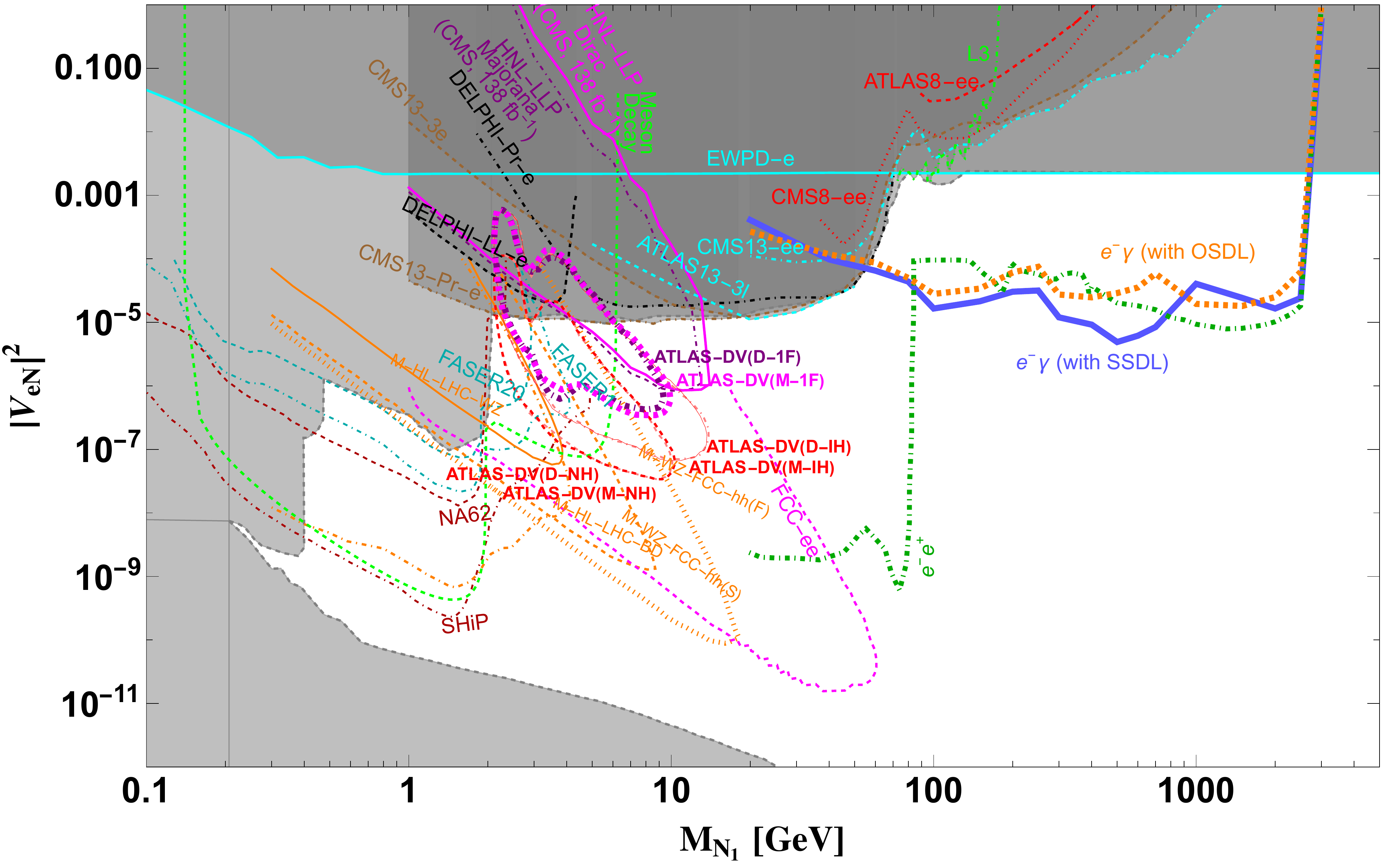}
\caption{Limits on mixing angles of the first generation heavy neutrino compared with existing and some prospective bounds. Shaded regions are already ruled out from different searches. The closed ATLAS contours represent limits from different displaced vertex searches. The prospective limits obtained from $e^-\gamma$ colliders at $\sqrt{s}=250$ GeV, 500 GeV, 1 TeV and 3 TeV are shown by light blue and orange thick dashed lines. In this case `with SSDL' (blue, thick solid) stands for mixings coming from combined $e^\pm+4j$ and SSDL modes. On the other hand `with OSDL' (orange, thick dashed) stands for  mixings coming from combined $e^\pm+4j$ and OSDL modes. The thick light green dot-dashed line represents limits for the $e^-e^+$ colliders for $\sqrt{s}=91.2$ GeV, 250 GeV, 500 GeV, 1 TeV and 3 TeV respectively as complementarity.}
\label{fig:mix1}
\end{figure}
We compare our results with the existing bounds from a variety of experiments shown by the shaded gray region in Fig.~\ref{fig:mix1}. The ATLAS and CMS bounds from the SSDL signal at $\sqrt{s}=8$ TeV LHC is shown by the red dashed and dotted lines from \cite{ATLAS:2015gtp,CMS:2015qur} respectively. The bounds from the L3 detector of LEP are shown by light green dot dashed line from \cite{L3:2001zfe}. The bounds from $\sqrt{s}=13$ TeV LHC using the SSDL searches from CMS \cite{CMS:2018jxx} using cyan dot-dashed and trilepton searches from ATLAS \cite{ATLAS:2019kpx} using cyan dashed and CMS \cite{CMS:2018iaf} brown dot dashed lines respectively. The bounds obtained from the prompt (Pr) and long-lived (LL) heavy neutrinos searches from DELPHEI \cite{DELPHI:1996qcc} are shown by dot dashed and dashed lines respectively. The limits from the EWPD \cite{deBlas:2013gla,delAguila:2008pw,Akhmedov:2013hec} for the electron flavor are shown by the solid cyan line.  Bounds on the mixing using long-lived particle (LLP) searches from the CMS \cite{CMS:2022fut} are shown by the solid magenta line for Dirac type and dot dashed purple line for Majorana types heavy neutrinos respectively. The conservative limits on the heavy neutrino mixing from the meson decay are shown by light green dashed line from \cite{Chun:2019nwi}. The displaced vertex (DV) searches in ATLAS from the heavy neutrinos are taken from \cite{ATLAS:2022atq} where Dirac (D) and Majorana (M) types heavy neutrinos were considered for single flavor (1F) and different neutrino mass hierarchies such as normal hierarchy (NH) and inverted hierarchy (IH). The 1F case for the Dirac and Majorana types neutrinos are shown by purple and magenta dotted closed curves whereas other DV scenarios are shown by 
red solid (M-IH), light red dot dashed (D-IH), red dashed (D-NH) and light red dashed (M-NH) closed curves respectively. The prospective bounds obtained from the future circular collider for electron positron (FCC-ee) are shown by dashed Magenta line from \cite{Blondel:2014bra}. The limits on the light-heavy mixing from SHiP are shown by the darker red dot dashed line from \cite{Alekhin:2015byh,SHiP:2018xqw} and those from the NA62 experiments are shown by darker red dashed line \cite{NA62:2017rwk,NA62:2017qcd,Lanfranchi:2017wzl,Drewes:2018gkc}.
Prospective upper limits from MATHUSLA at FCC-hh for the $W/Z$ boson decays for MATHUSLA surface version are presented by M-WZ-FCC-hh(S) and forward version M-WZ-FCC-hh(F) and the lines are shown by orange dashed and orange dotted lines respectively.  The prospective limits for the heavy neutrinos produced from the $W/Z$ boson decays at the high luminosity LHC (HL LHC)are presented by  M-HL-LHC-WZ and shown by solid orange line. The bounds on the heavy neutrinos obtained from the $B/D$ meson decays are represented by M-HL-LHC-BD and shown by orange dot dashed line \cite{Curtin:2018mvb}. The prospective bounds obtained from the FASER collaboration with detector radius 20 cm and 1 m are shown by the darker cyan dashed and dot dashed lines being marked as FASER20 and FASER1 respectively \cite{Kling:2018wct,FASER:2022hcn}. For $M_{N_1} \geq 2$ GeV we find that recent bounds from LHC are stronger and rules out many existing results. Furthermore EWPD is stronger for heavier RHNs showing the gray shaded regions as ruled out parameter space. For $M_{N_1} \leq$ 2 GeV we find strong limits from CHARM\cite{CHARM:1985nku,CHARMII:1994jjr,Orloff:2002de}, JNIR \cite{Baranov:1992vq},  PS191 \cite{Bernardi:1987ek}, BBN \cite{Boyarsky:2009ix,Ruchayskiy:2012si} and theoretical bounds can be obtained from the seesaw scenario \cite{deGouvea:2009fp,deGouvea:2005er,Cirelli:2004cz} in the anti clockwise direction. We find that the prospective limits from the $e^-\gamma$ colliders could be compared with the bounds obtained from $e^-e^+$ colliders from $M_{N_{1}} \geq 80$ GeV. \\
Here we would like to mention that one can also obtain prospective bound on mixing angle $|V_{\mu N}|^2$ at futuristic muon collider, see Refs.~\cite{Mekala:2023diu,Kwok:2023dck,Li:2023tbx} for recent studies. They found that the 3 TeV/10 TeV muon collider can probe mixing angle as $|V_{\mu N}|^2\sim 10^{-5}/10^{-6}$ for $M_N\sim \mathcal{O}(100\,\,\text{GeV})/\mathcal{O}(1\,\text{TeV})$.
\subsection{Different signatures from triplet scalars}
At $\gamma \gamma$ collider, one can easily produce $H^\pm$ and $H^{\pm \pm}$ in pair. We found that $\gamma \gamma \to H^+ H^-$ mode has nearly one order of magnitude less cross section than $\gamma \gamma \to H^{++} H^{--}$ process. Therefore we consider only the pair production of doubly charged scalar multiplet at $\gamma \gamma$ collider. As shown in Fig.~\ref{fig:HppBR}, depending on the choices of $v_\Delta$ and $\Delta m$, $H^{\pm\pm}$ decays to either $\ell^\pm\ell^\pm$, $W^\pm W^\pm$ or $H^\pm W^{\pm *}$ with 100$\%$ branching ratio. In our analysis we will assume the mass splitting $\delta m\approx 0$ so that $H^{\pm\pm}$ either decays to $\ell^\pm\ell^\pm$~($v_\Delta < 10^{-4}\,\text{GeV}$) or $W^\pm W^\pm$~($v_\Delta > 10^{-4}\,\text{GeV}$). Hence depending on the values of $v_\Delta$, pair production of  $H^{\pm \pm}$ can produce either four leptons $\ell_i^\pm\ell_j^\pm\ell_k^\mp\ell_m^\mp$ or $W^\pm W^\pm W^\mp W^\mp$ in the final state. The $W^\pm W^\pm W^\mp W^\mp$ mode can produce several jets in the final state following the hadronic decay of the four $W$ bosons.

\begin{figure}
\centering
\includegraphics[width=1\textwidth]{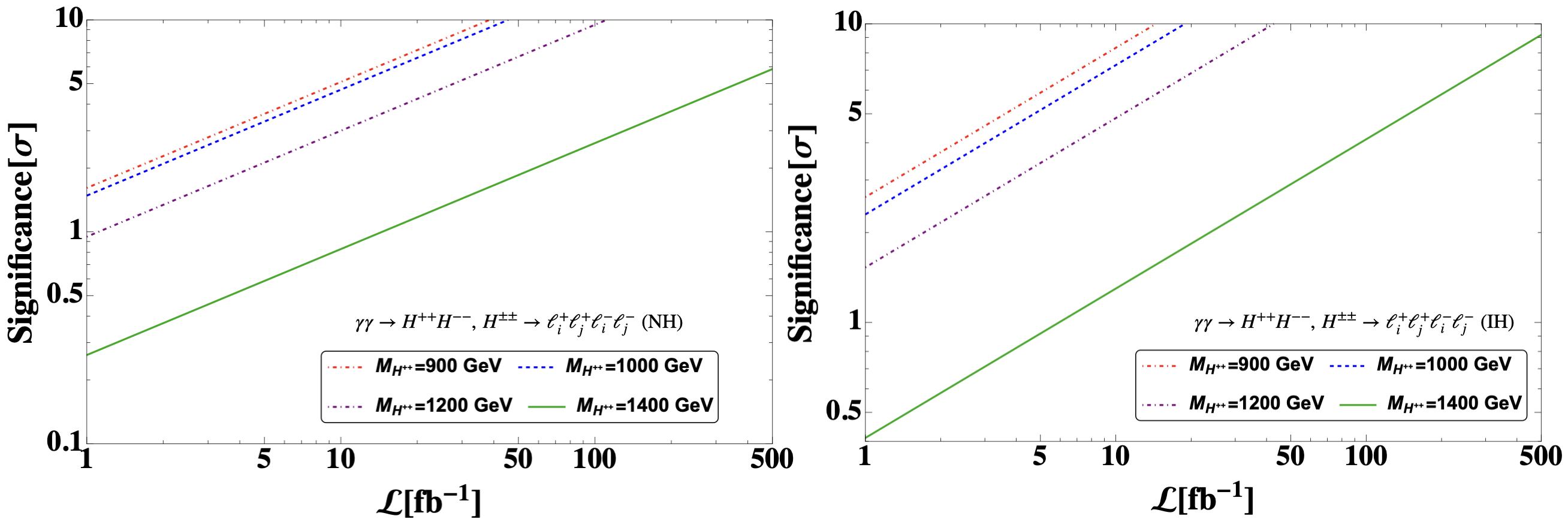}
\includegraphics[width=1\textwidth]{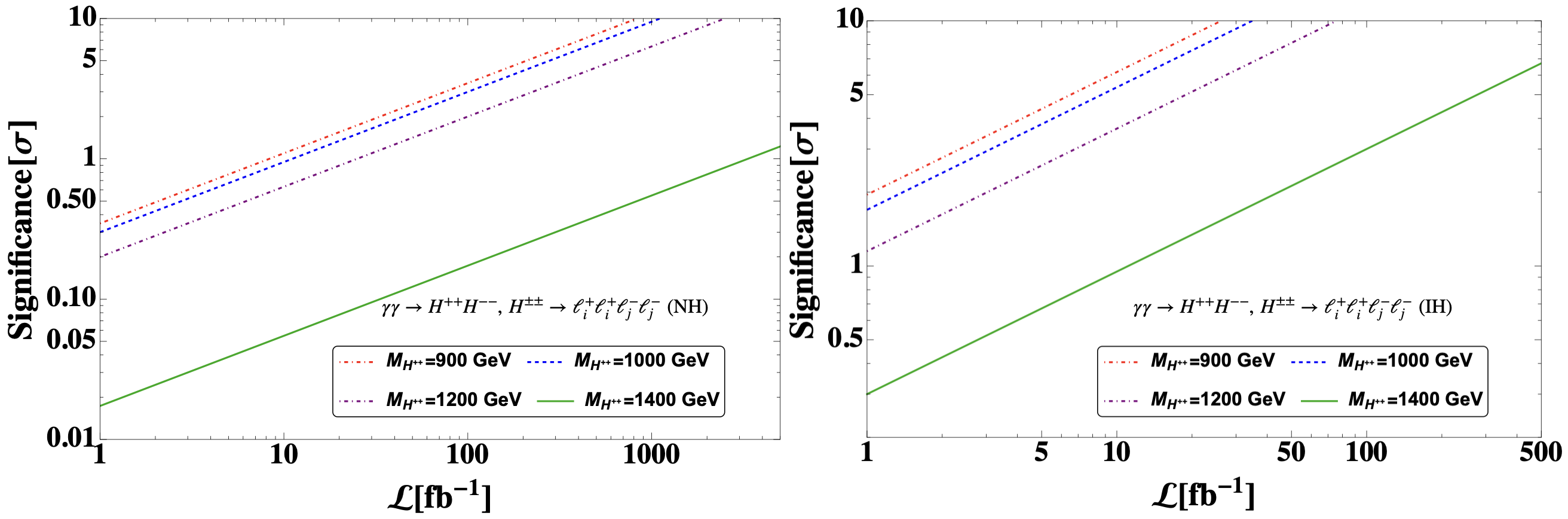}
\includegraphics[width=1\textwidth]{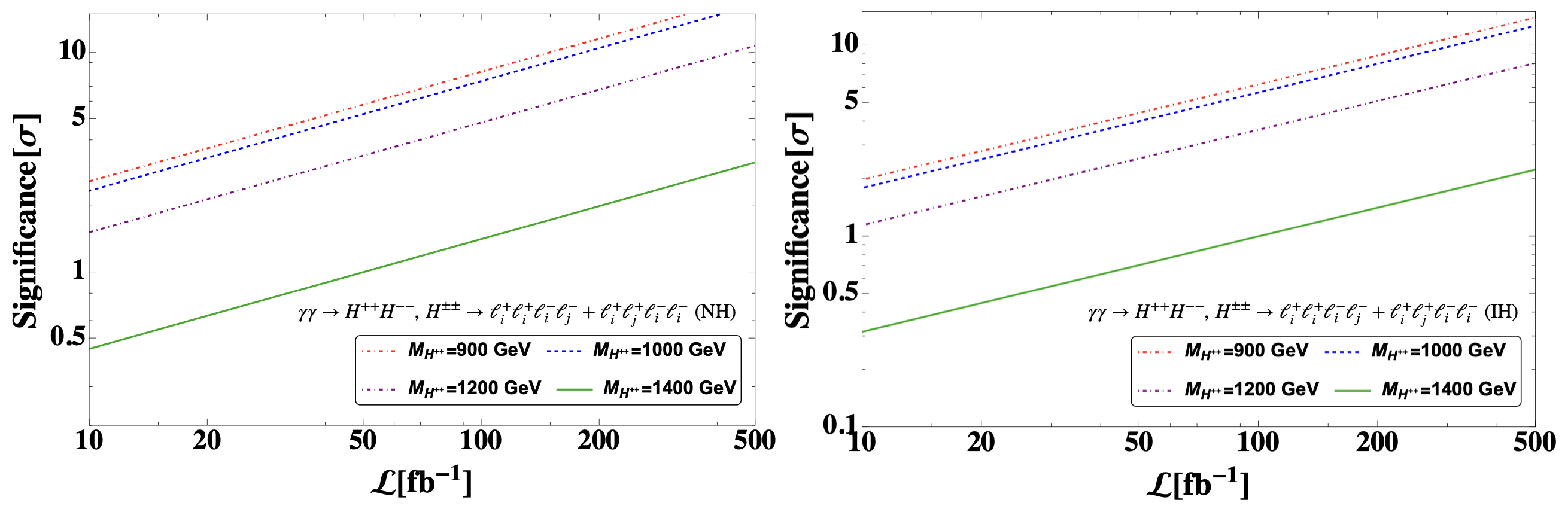}
\caption{Significance of different final sates from $\gamma \gamma \to H^{++} H^{--}$ process at $\sqrt{s}=3$ TeV as a function of integrated luminosity.}
\label{signi-2x}
\end{figure}

\hspace{0.5cm} Let us first discuss the leptonic mode $\gamma\gamma\to H^{\pm\pm}H^{\mp\mp}\to\ell_i^\pm \ell_j^\pm\ell_k^\mp\ell_m^\mp$ at $e^+e^-$ collider with $\sqrt{s}=3$ TeV. In this case, we assume $v_\Delta < 10^{-4}$ GeV such that the leptonic decay from the charged Higgs $\text{BR}(H^{\pm\pm} \to \ell^\pm \ell^\pm)$ can be dominant. Note that the doubly charged Higgs boson decay to leptonic final state goes as $\Gamma^{H^{\pm\pm}}_{\ell_i\ell_j}\propto |Y_{\Delta}^{ij}|^2$. Hence, the patterns of various leptonic channels will exactly follow the pattern of $Y_\Delta^{ij}$, which is determined by oscillation parameter. As a result the $\text{BR}(H^{\pm\pm}\to\ell_i^\pm \ell_j^\pm)$ will also depend on the ordering of light neutrino mass. Hence, one expects that collider observables such as the cross-section $\sigma(\gamma\gamma\to H^{\pm\pm}H^{\mp\mp}\to\ell_i^\pm \ell_j^\pm\ell_k^\mp\ell_m^\mp)$ may strongly correlate with neutrino mass ordering. To show that indeed this is the case, we consider three possible scenario:
\begin{itemize}
\item[(i)] $\gamma\gamma\to H^{++}H^{--}\to\ell_i^+\ell_j^+\ell_i^-\ell_j^-$, where the index $i,j$ is summed over. We are not considering $\tau$ channel as it is harder to detect compare to electron or muon. Note that this channel is lepton flavor conserving~(LFC). In this analysis we consider a generic background of four lepton $4\ell$ and two lepton with two $W$ bosons $2\ell 2W$ where the $W$ bosons decay leptonically. As for $v_\Delta < 10^{-4}$ GeV, there exist tight constraint on the mass of $M_{H^{\pm\pm}}$, we consider relatively high $M_{H^{\pm\pm}}$. Due to this reason one expect that lepton $p_T$ distribution will peak at high $p_T$. Also lepton polar angle distribution will be similar as Fig.~\ref{figx2}. The signal and backgrounds cross sections before and after the cuts are given in Tab.~\ref{tabchpp-1} for the NH and IH cases which affect the branching ratio of the charged scalars into leptons depending on the light neutrino mass hierarchy as the corresponding Yukawa couplings carry information from the neutrino oscillation data.
\begin{table}[h]
\begin{center}
\begin{tabular}{|c|c|c|c|c|c|c|c|}
\hline
\multirow{3}{*}{$\sqrt{s}$ (TeV)} &\multirow{3}{*}{$M_{H^{\pm\pm}}$ (GeV)} & \multicolumn{2}{c|}{Signal(NH)} & \multicolumn{2}{c|}{Signal(IH)}& %
    \multicolumn{2}{c|}{Background} \\
\cline{3-8}
&& before cuts (fb) & after cuts (fb) & before cuts (fb) & after cuts (fb) & before cuts (fb) & after cuts (fb)\\
\hline
&900 &5.5&2.6& 13.536 & 6.916 &4$\ell$: 1.4&0.005 \\
& &&&&&2$\ell$2$W$: 3.74&0.0008 \\
\hline
&1000 &4.2&2.2 & 10.81 & 5.29 &4$\ell$: 1.4&0.005 \\
3& &&&&&2$\ell$2$W$: 3.74&0.0008 \\
\hline
&1200 &1.91&0.9 & 4.69 & 2.34 &4$\ell$: 1.4&0.0011 \\
& &&&&&2$\ell$2$W$: 3.74&0.00022 \\
\hline
&1200 &0.14&0.07& 0.34 & 0.17 &4$\ell$: 1.4&0.0011 \\
& &&&&&2$\ell$2$W$: 3.74&0.00022 \\
\hline
\end{tabular}
\caption{Cross sections for the signal and SM backgrounds before and after cuts for $\ell_i^{+} \ell_j^{+} \ell_i^{-} \ell_j^{-}$ final state from the $\gamma \gamma \to H^{++} H^{--}$ process where both $i=j$ and $i \neq j$ are possible scenarios. We have used the following cuts: $\cos\theta_{l_{1, 3}} < 0.9$, $\cos\theta_{l_{2,4}} < 0.92$, $p_T^{\ell_{1, 3}} > 400$ GeV and $p_T^{\ell_{2, 4}} > 150$ GeV.}
\label{tabchpp-1}
\end{center}
\end{table}
We estimate the significance $(\sigma)$ of the process as a function of luminosity $(\mathcal{L})$ 
and it is shown in the top row for the NH (IH) case in the left (right) panel of Fig.~\ref{signi-2x}. For the case of NH, the signal can reach at a significance of 5$\sigma$ between 10 fb$^{-1}$- 12 fb$^{-1}$ luminosity for $M_{H^{\pm\pm}}=900$ GeV and 1000 GeV. That for $M_{H^{\pm\pm}}=1200$ GeV can reach at 5$\sigma$ at 30 fb$^{-1}$ luminosity. A significance of 3$\sigma$ can be reached around 150 fb$^{-1}$ luminosity for $M_{H^{\pm\pm}}=1400$ GeV. We find slight improvement in the IH case where more than 5$\sigma$ significance could be attained around 10 fb$^{-1}$ luminosity for $M_{H^{\pm\pm}}=900$ GeV, 1000 GeV and 1200 GeV respectively, however, a 5$\sigma$ significance could be attained around a luminosity slightly above 200 fb$^{-1}$ for $M_{H^{\pm\pm}}=$ 1400 GeV.
\item[(ii)] Second we consider the process $\gamma \gamma \to H^{++} H^{--} \to \ell_i^+ \ell_i^+ \ell_j^- \ell_j^-$ where $i\ne j$ and $i,j$ are summed over electron and muon channel. This final state violates lepton flavor by two units. In this case we use the similar mass points and cuts as shown in Tab.~\ref{tabchpp-1}. For this signal we find there is no irreducible backgrounds in this case. The signal cross sections before and after cuts are shown in Tab.~\ref{tabchpp-2} for the NH and IH case.
\begin{table}[h]
\begin{center}
\begin{tabular}{|c|c|c|c|c|c|}
\hline
\multirow{3}{*}{$\sqrt{s}$ (TeV)} &\multirow{3}{*}{$M_H^{\pm \pm}$ (GeV)} & \multicolumn{2}{c|}{Signal(NH)} & \multicolumn{2}{c|}{Signal(IH)} \\
\cline{3-6}
&& before cuts (fb) & after cuts (fb) & before cuts (fb) & after cuts (fb) \\
\hline
&900&0.162&0.12 & 5.31 & 3.82 \\
3&1000&0.122&0.09 & 3.99 & 2.87 \\
&1200&0.056&0.04 & 1.84 & 1.32\\
&1400&0.0041&0.003 & 0.13 & 0.09\\
\hline
\end{tabular}
\caption{Cross sections for the signal $\ell_i^{+} \ell_i^{+} \ell_j^{-} \ell_j^{-}$ before and after cuts,  where $i\ne j$ and $i, j$ are summed over electron and muon channel. The cuts are same as in Table.~\ref{tabchpp-1}.}
\label{tabchpp-2}
\end{center}
\end{table}
The estimated significance of this signal are shown in the middle row of Fig.~\ref{signi-2x} at $\sqrt{s}=3$ TeV for NH (IH) case in the left (right) panel. The NH cases of $M_{H^{++}}=900$ GeV, 1000 GeV and 1200 GeV could be observed at 5$\sigma$ significance around 100 fb$^{-1}$ luminosity. The case with $M_{H^{++}}=1400$ GeV will have low significance throughout the considered range of luminosity. We find significant improvement in the IH case where 5$\sigma$ significance could be achieved within 10 fb$^{-1}$ luminosity to 20 fb$^{-1}$ luminosity for $M_{H^{++}}=900$ GeV, 1000 GeV and 1200 GeV respectively, however, that could be achieved around 300 fb$^{-1}$ luminosity for $M_{H^{++}}=$ 1400 GeV.
\item[(iii)] Finally we consider $\gamma \gamma \to H^{++} H^{--} \to \ell_i^+ \ell_i^+ \ell_i^- \ell_j^- + \ell_i^+ \ell_j^+ \ell_i^- \ell_i^-$, where $i\ne j$ and $i,j$ is summed over electron and muon channel. This final state violates lepton flavor by one unit. In this case we use the similar mass points and cuts as shown in Tab.~\ref{tabchpp-1}. For this signal we find that the important background only comes from $2\ell 2W$ final state. The signal cross sections before and after cuts are shown in Tab.~\ref{tabchpp-3} for the NH and IH case.
\begin{table}[h]
\begin{center}
\begin{tabular}{|c|c|c|c|c|c|c|c|}
\hline
\multirow{3}{*}{$\sqrt{s}$ (TeV)} &\multirow{3}{*}{$M_H^{\pm\pm}$ (GeV)} & \multicolumn{2}{c|}{Signal(NH)} & \multicolumn{2}{c|}{Signal(IH)}& %
    \multicolumn{2}{c|}{Background $2\ell$2$W$} \\
\cline{3-8}
&& before cuts (fb) & after cuts (fb) & before cuts (fb) & after cuts (fb) & before cuts (fb) & after cuts (fb)\\
\hline
& 900 & 1.51 & 0.67 & 0.87 & 0.39 & 3.74 & $9.72\times 10^{-4}$ \\
3 & 1000 & 1.13 & 0.55 & 0.66 & 0.32 & 3.74 & $9.72\times 10^{-4}$ \\
 &  1200 & 0.52 & 0.23 & 0.31 & 0.13 & 3.74 & $7.48\times 10^{-5}$ \\
 & 1400 & 0.04 & 0.02 & 0.02 & 0.01 & 3.74 & $7.48\times 10^{-5}$ \\
\hline
\end{tabular}
\caption{Cross sections for the signal and SM backgrounds before and after cuts for $\ell_i^+ \ell_i^+ \ell_i^- \ell_j^- + \ell_i^+ \ell_j^+ \ell_i^- \ell_i^-$, where $i\ne j$ and $i,j$ is summed over electron and muon channel. The cuts are same as in Table.~\ref{tabchpp-1}.}
\label{tabchpp-3}
\end{center}
\end{table}
The estimated significance of this signal are shown in the bottom row of Fig.~\ref{signi-2x} at $\sqrt{s}=3$ TeV for NH (IH) case in the left (right) panel.
In NH case a 5$\sigma$ significance could be attained with 50 fb$^{-1}$ luminosity to 120 fb$^{-1}$ luminosity for $M_{H^{\pm\pm}}=900$ GeV, 1000 GeV and 1200 GeV respectively whereas this signal can be probed at 3$\sigma$ significance around 500 fb$^{-1}$ luminosity for $ M_{H^{\pm\pm}}=1400$ GeV. On the other hand this signal from the IH case can be observed with a significance of 5$\sigma$  within 60 fb$^{-1}$ luminosity to 150 fb$^{-1}$ luminosity for $M_{H^{\pm\pm}}=900$ GeV, 1000 GeV and 1200 GeV respectively whereas the case for $ M_{H^{\pm\pm}}=1400$ GeV could be observed at 2$\sigma$ significance with a luminosity of 500 fb$^{-1}$. 
\end{itemize}

\begin{figure}
\centering
\includegraphics[width=0.6\textwidth]{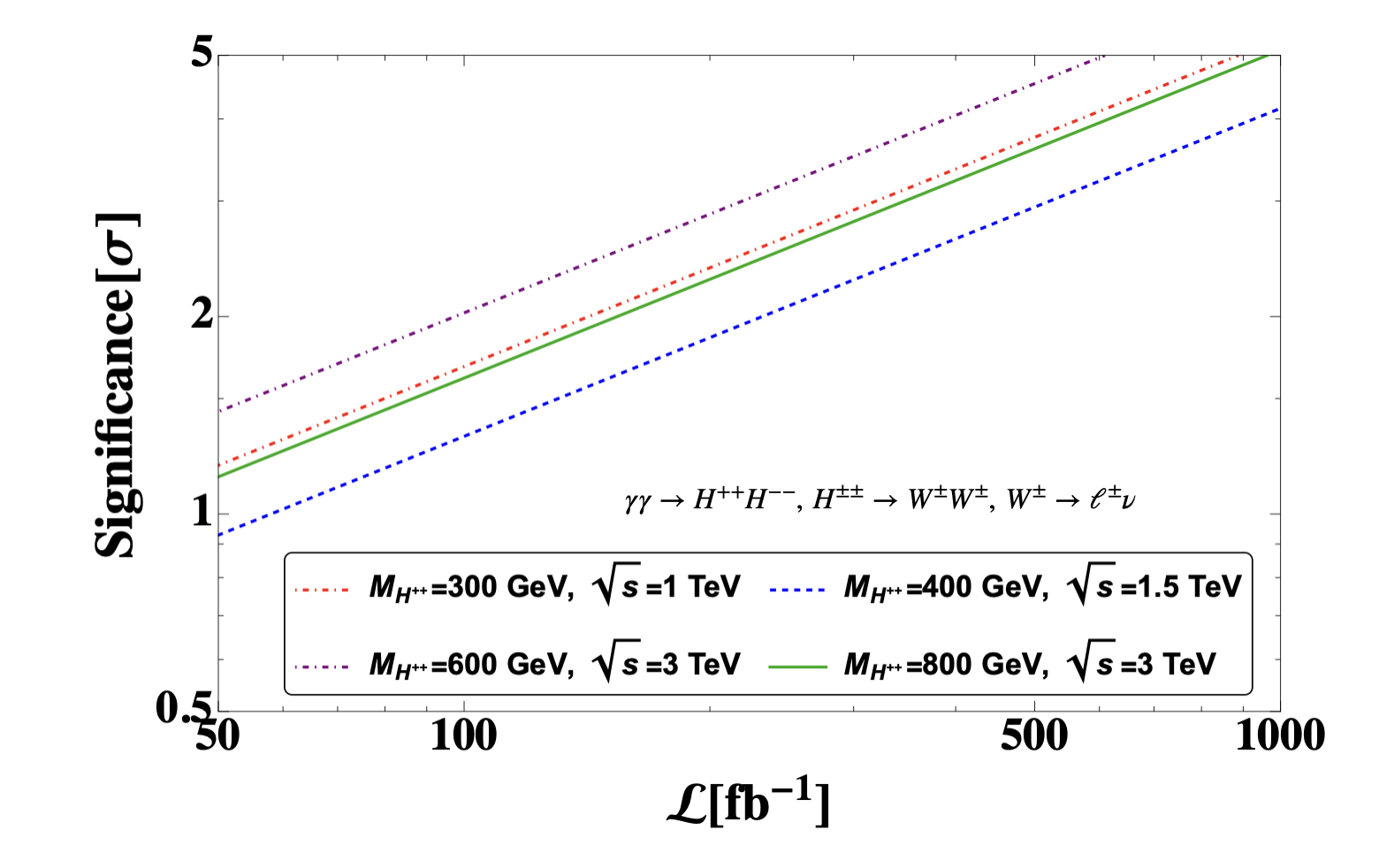}
\caption{Significance of different final sates from $\gamma \gamma \to H^{++} H^{--}$ process at different $\sqrt{s}$ as a function of integrated luminosity.}
\label{signi-2x-1}
\end{figure}

\hspace{0.5cm} Next we consider $v_{\Delta} > 10^{-3}$ GeV where $H^{\pm\pm} \to  W^{\pm} W^{\pm}$ is the dominant mode. We specifically focous on the leptonic mode coming from $\gamma \gamma \to H^{++} H^{--}$ process followed by $H^{\pm \pm} \to  W^{\pm} W^{\pm}$ and $W^{\pm} \to \ell^{\pm} \nu$ in association with missing momentum where $\ell= e, \mu$. We generate the SM backgrounds like 4$W$, $W^\pm Z$ and 4$\ell$. In this case $W$ and $Z$ bosons decay leptonically taking electrons and muons under consideration. We estimate the signal and background cross sections before and after the following selection cuts:
\begin{itemize}
\item[(i)] Transverse momentum for the $p_T$ ordered leptons follow: (1) $p_T^{\ell_{1,3}} > 80$ GeV, $p_T^{\ell_{2,4}}> 30$ GeV for $M_{H^{++}}=300$ GeV at $\sqrt{s}=1$ TeV, (2) $p_T^{\ell_{1,3}} > 100$ GeV, $p_T^{\ell_{2,4}}> 50$ GeV for $M_{H^{++}}=400$ GeV at $\sqrt{s}=1.5$ TeV and (3) $p_T^{\ell_{1,3}} > 150$ GeV, $p_T^{\ell_{2,4}}> 75$ GeV for $M_{H^{++}}=600$ GeV and 800 GeV at $\sqrt{s}=3$ TeV respectively.  
\item[(ii)]  The energy of the leptons are followed as: (1) $E_{\ell_{1,3}}> 100$ GeV and $E_{\ell_{2,4}}> 50$ GeV for $M_{H^{++}}=300$ GeV at $\sqrt{s}=1$ TeV, (2) $E_{\ell_{1,3}}> 150$ GeV and $E_{\ell_{2,4}}> 75$ GeV for $M_{H^{++}}=400$ GeV at $\sqrt{s}=1.5$ TeV and (3) $E_{\ell_{1,3}}> 200$ GeV and $E_{\ell_{2,4}}> 100$ GeV for $M_{H^{++}}=600$ GeV and $800$ GeV at $\sqrt{s}=3$ TeV respectively. 
\item[(iii)] The azimuthal angles of the leptons follow: (1) $\cos\theta_{\ell_{1,3}} < 0.9$, $\cos\theta_{\ell_{2,4}} < 0.92$ for $M_{H^{++}}=300$ GeV at $\sqrt{s}=1$ TeV, (2) $\cos\theta_{\ell_{1,3}} < 0.88$, $\cos\theta_{\ell_{2,4}} < 0.92$ for $M_{H^{++}}=400$ GeV at $\sqrt{s}=1.5$ TeV and (3) $\cos\theta_{\ell_{1,3}} < 0.88$, $\cos\theta_{\ell_{2,4}} < 0.92$ for $M_{H^{++}}=600$ GeV and 800 GeV at $\sqrt{s}=3$ TeV, respectively. 
\end{itemize}
The signal and backgrounds cross sections before and after the cuts are given in Tab.~\ref{tabchpp-4}.
\begin{table}[h]
\begin{center}
\begin{tabular}{|c|c|c|c|c|c|}
\hline
\multirow{3}{*}{$\sqrt{s}$ (TeV)} &\multirow{3}{*}{$M_H^{\pm\pm}$ (GeV)} & \multicolumn{2}{c|}{Signal} & %
    \multicolumn{2}{c|}{Background} \\
\cline{3-6}
&& before cuts (fb) & after cuts (fb) & before cuts (fb) & after cuts (fb)\\
\hline
1&300 &1.21&0.08&4$W$: 0.02&0.0004\\
& &&&4$\ell$: 8.0&0.112 \\
& &&&$W^{\pm} Z$: 1.45&0.035 \\
\hline
1.5&400 &0.88&0.052&4$W$: 0.092&0.0008 \\
& &&&4$\ell$: 4.45&0.066 \\
& &&&$W^{\pm}Z$: 3.21&0.038 \\
\hline
3&600 &0.42&0.054&4$W$: 1.0&0.0023 \\
& &&&4$\ell$: 1.4&0.00027 \\
& &&&$W^\pm Z$: 7.2&0.0145 \\
\hline
&800 &0.25&0.05&4$W$: 0.46&0.0023 \\
& &&&4$\ell$: 1.4&0.03 \\
& &&&4$\ell$: 7.2&0.014 \\
\hline
\end{tabular}
\caption{Cross sections for the signal and SM backgrounds before and after cuts for 4$W$ final state from the $\gamma \gamma \to H^{++} H^{--}$ process where $W$ bosons decay leptonically.}
\label{tabchpp-4}
\end{center}
\end{table}
We show the significance of this process as a function of the integrated luminosity in Fig.~\ref{signi-2x-1}. The significance can reach at 5$\sigma$ with a luminosity between 600 fb$^{-1}$ luminosity to 1 ab$^{-1}$ luminosity. 
\subsection{Different signatures from triplet fermions}
We study the production of triplet fermion at the $e^-\gamma$ collider. The triplet fermion has neutral and charged multiplets. The lower limit on the triplet fermion mass form the LHC is nearly 1 TeV. As a result we consider $M_{\Sigma} \geq 1.2$ TeV and center of mass energy of $e^- \gamma$ collider as $3$ TeV. The dominant decay mode of the neutral multiplet is $W^\mp \ell$ and the second dominant mode of the charged multiplet is $Z \ell$. Due to the heavy triplet fermion mass under consideration, these $W$ and $Z$ bosons will be boosted and produce fat-jets. In this scenario we consider two different aspects in the following:
\subsubsection{SSDL mode}
\begin{figure}[h]
\centering
\includegraphics[width=0.65\textwidth]{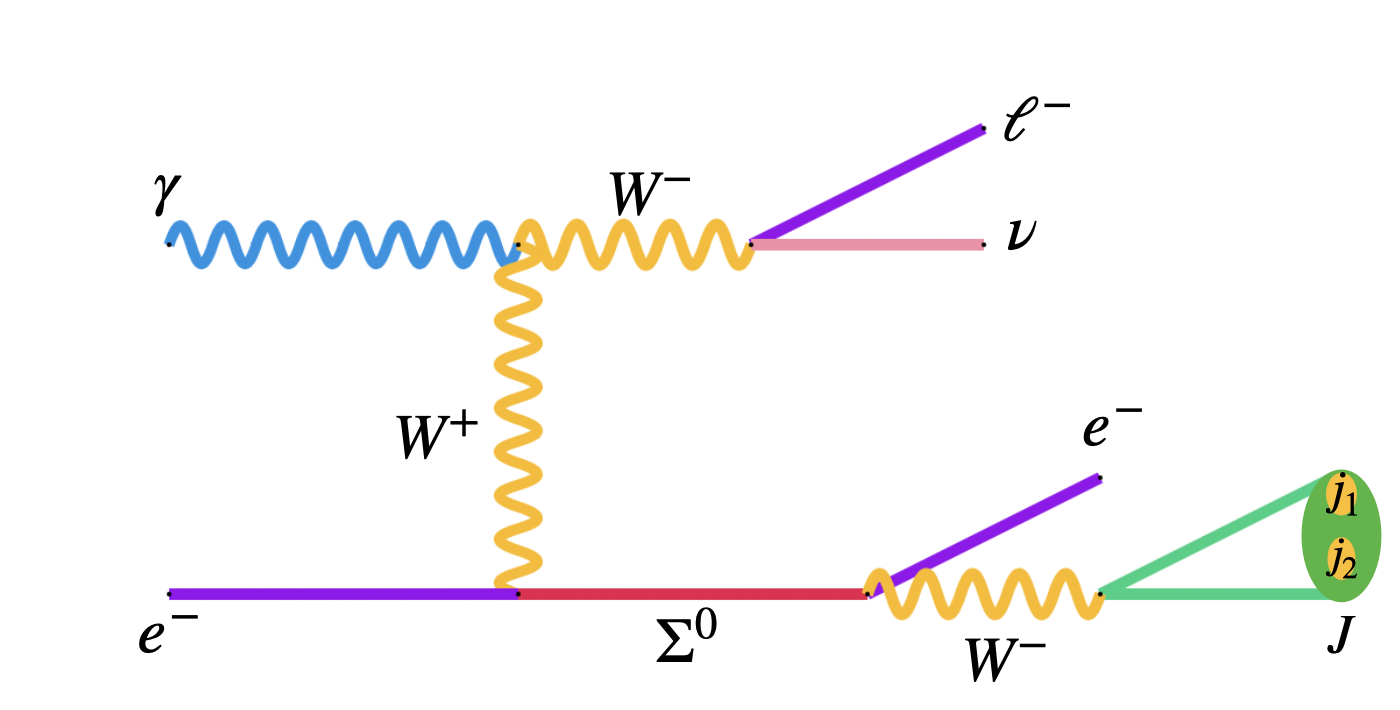}
\caption{SSDL+ J final state in association with missing momentum at $e^-\gamma$ collider where $\ell^-=e^-, \mu^-$. There is an additional $s$-channel process giving same final state which has been added in the analysis but not shown here.}
\label{lims-type-3-1-x1}
\end{figure}

We first consider the $W^- \Sigma^0$ case where $W^{-}$ decays leptonically and $\Sigma^{0}$ decays into the dominant mode $e^- W^+$ where $W^{+}$ decays hadronically. In this analysis we consider a heavy triplet $M_\Sigma \ge 1.2$ TeV. As a result $W^+$ from the $\Sigma^0$ decay can be boosted so that the hadronic jets could be collimated to produce a boosted object called fat-jet $(J)$. The fat jet topology is a very powerful tool to significantly reduce the SM backgrounds. In our analysis the jets are reconstructed by Cambridge-Achen algorithm \cite{Dokshitzer:1997in,Wobisch:1998wt} implemented in Fastjet~\cite{Cacciari:2005hq,Cacciari:2011ma} package with the radius parameter as $R = 0.8$. Hence we consider SSDL$+J$ signature in association with missing momentum. Again as in the case of Type-I seesaw we have two possibilities such as SSSF and SSDF depending on the flavour of $\ell^-$ coming from the associated $W^-$ decay. Hence we consider both the SSSF and SSDF signals with a fet-jet in association with missing momentum which can be useful to probe the Majorana nature of $\Sigma^0$. The corresponding Feynman diagram is given in Fig.~\ref{lims-type-3-1-x1}. 
In Fig.~\ref{pol-3}, we show the distributions of fat-jet transverse momentum $(p_T^J)$, transverse momenta of the leptons $(p_T^{\ell_{1,2}})$, fat-jet mass $(m_J)$, the polar angles of the leptons $(\cos\theta_{\ell_{1,2}})$ of the signal and corresponding backgrounds at $\sqrt{s}=3$ TeV. In view of these distributions we use the following cuts for the signal and generic background:
\begin{figure}[h]
\centering
\includegraphics[width=1\textwidth]{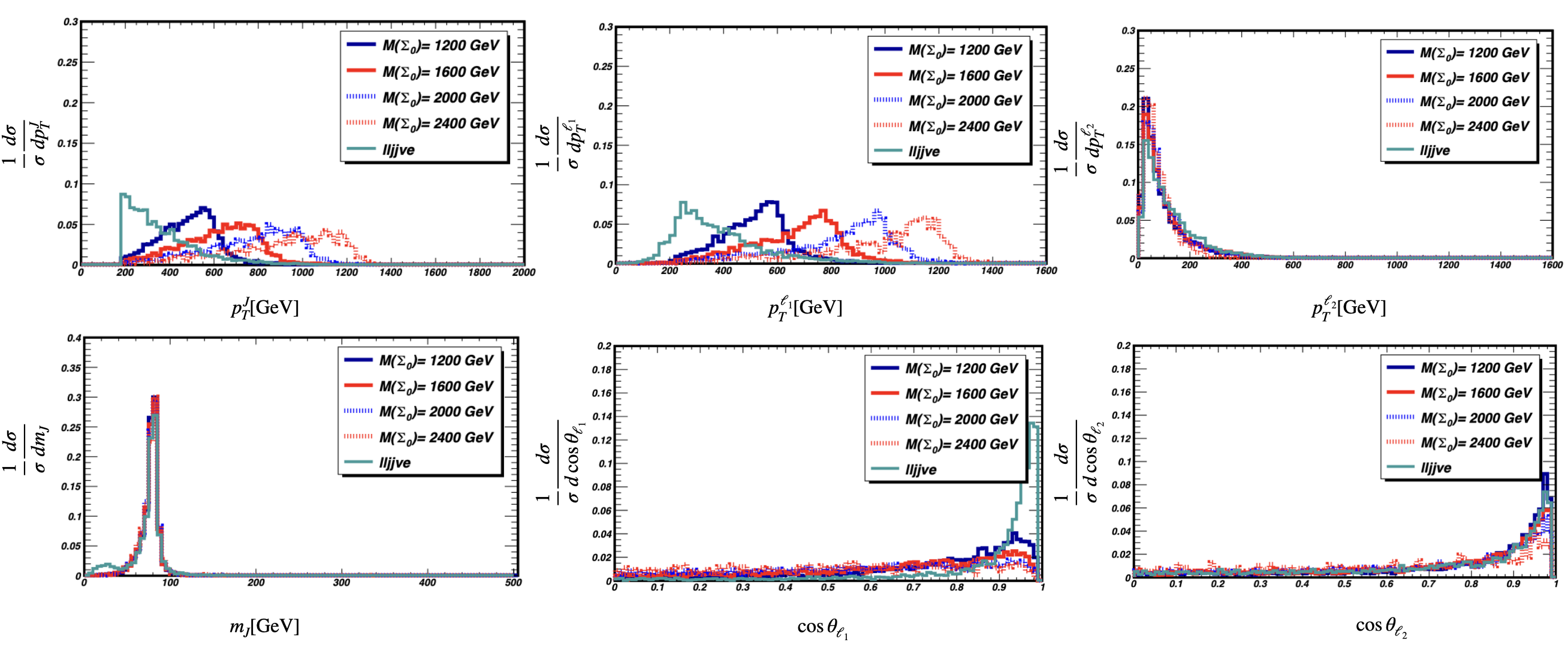}
\caption{The distributions of different kinematic variables from the SSDL signal from type-III seesaw and corresponding generic background from $e^-\gamma$ collider at $\sqrt{s}=$3 TeV.}
\label{pol-3}
\end{figure}
\begin{itemize}
\item[(i)] The events selected with at least one fat-jets $(J)$ and leptons $(\ell=e, \mu)$ having transverse momenta $p_T^{J} > 400$ GeV, $p_T^{\ell_1, \rm leading} > 400$  GeV and  $p_T^{\ell_2, \rm trailing} > 10$ GeV respectively for the neutral component of the triplet fermion with mass $M_{\Sigma^0}=1200$ GeV, 1400 GeV, 1600 GeV and 1800 GeV respectively. For $M_{\Sigma^0}=2000$ GeV, 2200 GeV and 2400 GeV,  the cuts are $p_T^{J} > 600$ GeV, $p_T^{\ell_1, \rm leading} > 600$  GeV and  $p_T^{\ell_2, \rm trailing} > 10$ GeV respectively.

\item[(ii)] The jet mass $(m_{J})$ of the fat-jet is considered to be within 60 GeV $< m_{J}<$ 100 GeV.

\item[(iii)] We reconstruct $\Sigma^0$ considering the electron and fat-jet system $(e^- J )$ and using the invariant mass cut $M_{\Sigma^0}-40$ GeV $< m_{\ell J} < M_{\Sigma^0}+40$ GeV where $M_{\Sigma^0}$ is the mass of the neutral component of the triplet fermion. 

\item[(iv)] The cosine of the polar angle of the leptons are considered to be $\cos\theta_{l_{1, \rm leading}} < 0.92$ and  $\cos\theta_{l_{2, \rm trailing}} < 0.9$ respectively.
\end{itemize}
The signal normalized by the square of the mixing and corresponding background cross sections for the SSDL final state at $\sqrt{s}=$3 TeV are given in Tab.~\ref{tab12}.
\begin{table}[h]
\begin{center}
\begin{tabular}{|c|c|c|c|c|c|}
\hline
\multirow{3}{*}{$\sqrt{s}$ (TeV)} &\multirow{3}{*}{$M_\Sigma$ (GeV)} & \multicolumn{2}{c|}{Signal} & %
    \multicolumn{2}{c|}{Background} \\
\cline{3-6}
&& before cuts (fb) & after cuts (fb) & before cuts (fb) & after cuts (fb)\\
\hline
&1200 &2716.82&312.4&95.8&0.204 \\
&1400&2527.68&377.15&95.8&0.243\\
3&1600&2317.24&394.13&95.8&0.252\\
&1800&2068.63&380.31&95.8&0.371\\
&2000&1763.19&711.1&95.8&0.128\\
&2200&1386.45&273.5&95.8&0.102\\
&2400&939.3&204.43&95.82&0.068\\
\hline
\end{tabular}
\caption{Cross sections of the signal (normalized by $|V_{e\Sigma}|^2$) and generic background before and after cuts for $e^- \gamma$ collider at $\sqrt{s}=$3 TeV for the SSDL final state.}
\label{tab12}
\end{center}
\end{table}
\subsubsection{Trilepton mode}
\begin{figure}[h]
\centering
\includegraphics[width=0.5\textwidth]{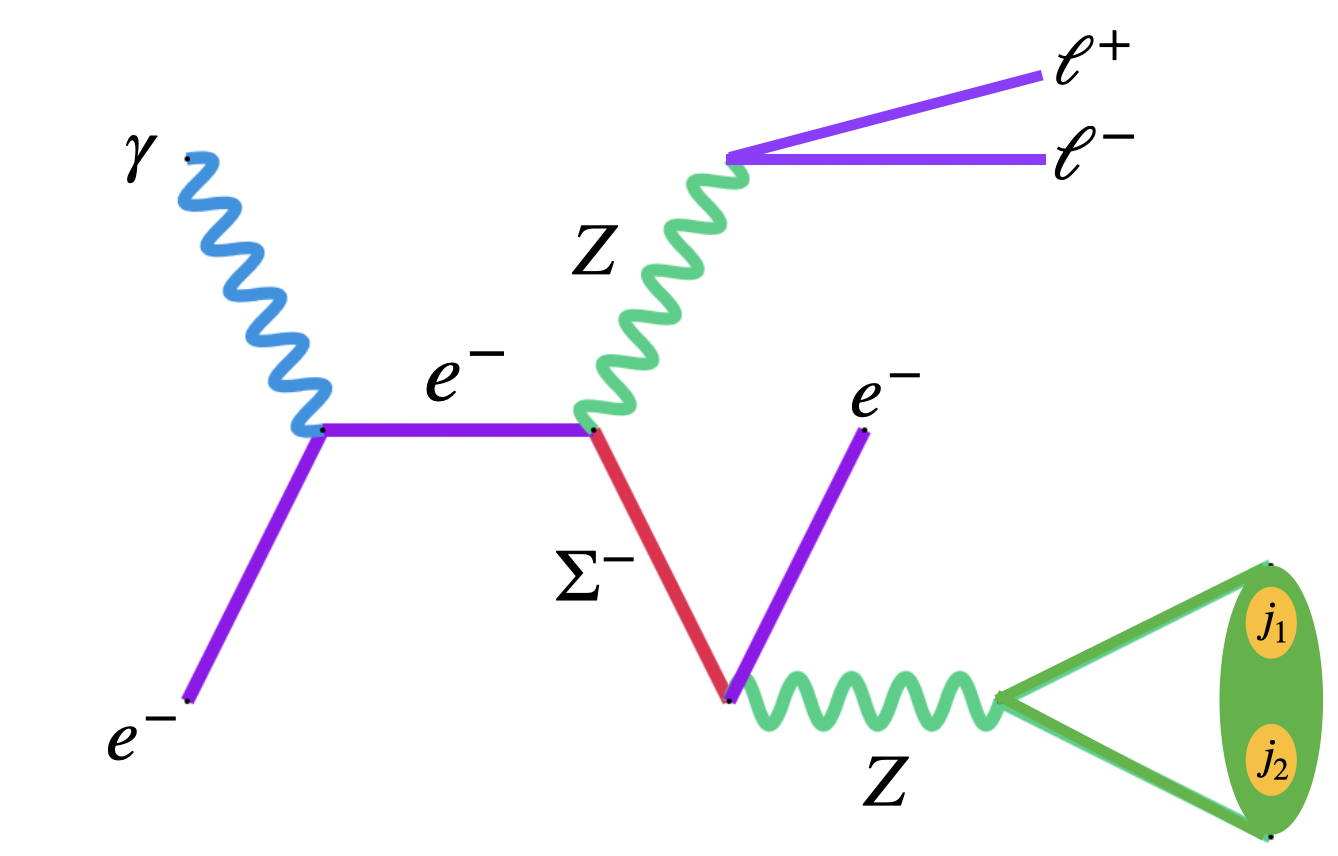}
\caption{Feynman diagram for the process $e^-\gamma\to e^-\ell^+\ell^-+J$. There is an additional $t$-channel process giving same final state which has been added in the analysis but not shown here.}
\label{lims-type-3-00}
\end{figure}

\begin{figure}[h]
\centering
\includegraphics[width=1\textwidth]{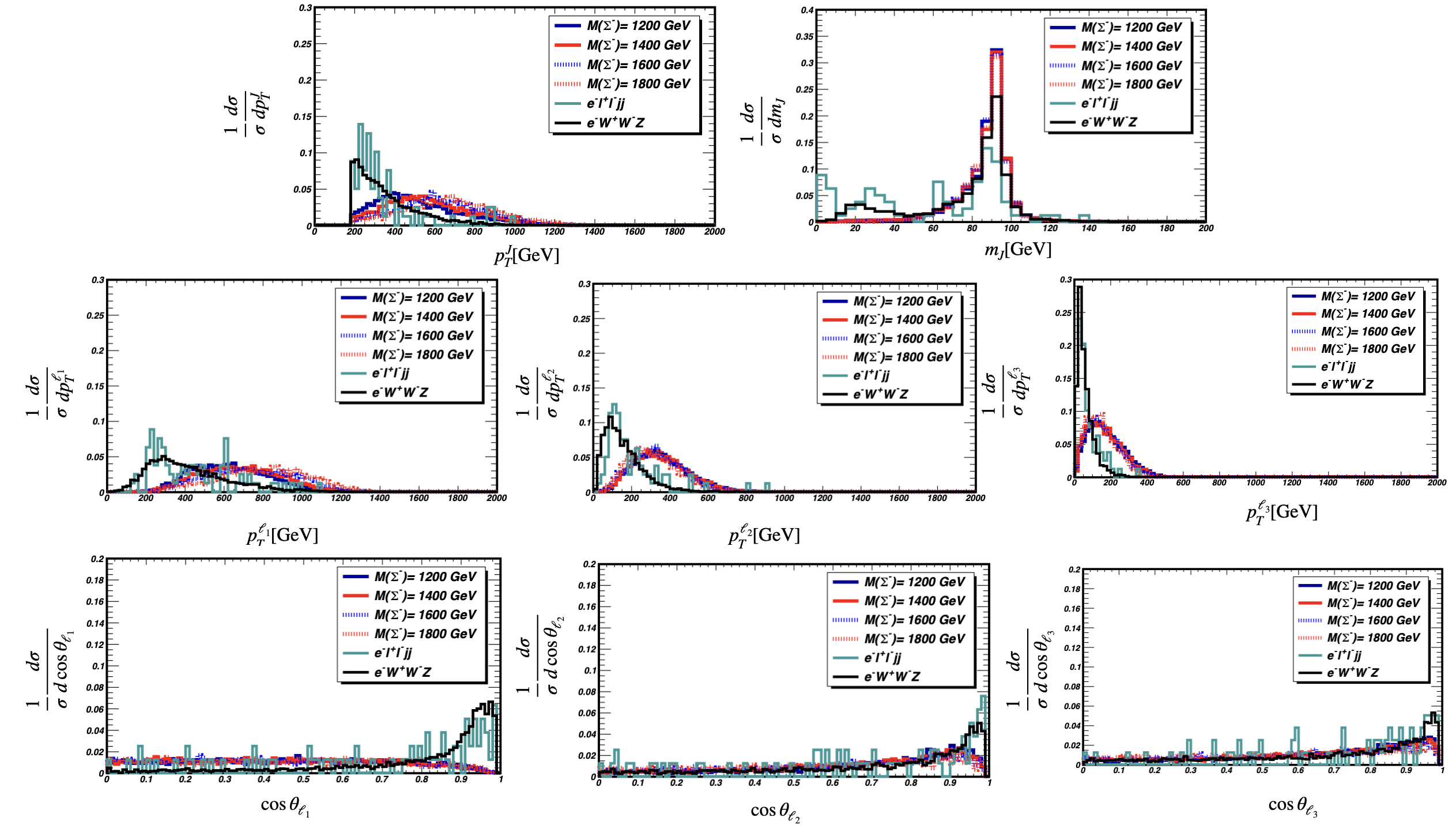}
\caption{The distributions of different kinematic variables from the trilepton plus fat-jet signal from type-III seesaw and corresponding backgrounds from $e^-\gamma$ collider at $\sqrt{s}=$3 TeV.}
\label{pol-3-1}
\end{figure}

The charged multiplets of the triplet fermion can also be produced at the $e^- \gamma$ collider in association with $Z$ boson. $\Sigma^{-}$ decays into $e^- Z$ followed by the hadronic decay mode of the $Z$ boson. Due to the heavy mass of the triplet fermion after the recent LHC limits we consider $M_{\Sigma} \ge 1 $ TeV hence it can produce a boosted $Z$ boson which can produce a fat-jet after the hadronic decay which could have a unique feature. The associated $Z$ boson decays into a pair of charged leptons. Hence we finally observe a trilepton plus fat-jet signal which could be an interesting signature at the $e^- \gamma$ colliders. In addition to this mode another interesting channel could be the $\Sigma^- h$ mode where $h$ can decay into b-jets dominantly followed by the $\Sigma^- \to e^- J$ mode from the charged multiplet of the triplet fermion. However, in this article we finally concentrate on the gauge boson associated triplet production processes not including the Higgs associated channel in further consideration.  The corresponding Feynman diagram is given in Fig.~\ref{lims-type-3-00}.
Generating the trilepton events with jets we show the distributions of fat-jet transverse momentum $(p_T^J)$, transverse momenta of the leptons $(p_T^{\ell_{1,2,3}})$, fat-jet mass $(m_J)$, the polar angles of the leptons $(\cos\theta_{\ell_{1,2,3}})$ of the signal and corresponding backgrounds in Fig.~\ref{pol-3-1}.
In view of these distributions we use the following cuts to generate trilepton plus fat-jet events for the signal and generic background:
\begin{itemize}
\item[(i)] The events selected with at least one fat-jet $(J)$ and three leptons having transverse momenta $p_T^{J} > 300$ GeV, $p_T^{\ell_1, \rm leading} > 400$  GeV, $p_T^{\ell_2, \rm trailing} > 150$ GeV and $p_T^{\ell_3, \rm trailing} > 75$ GeV respectively for $M_\Sigma^-=1200$ GeV, 1400 GeV, 1600 GeV and 1800 GeV respectively.

\item[(ii)] The jet mass $(m_{J})$ of the fat-jet is considered to be within 60 GeV $< m_{J}<$ 100 GeV.

\item[(iii)] We reconstruct $\Sigma^-$ considering the electron and fat-jet system $(e^- J )$ and using the invariant mass cut $M_{\Sigma^-}-40$ GeV $< m_{\ell J} < M_{\Sigma^-}+40$ GeV. 

\item[(iv)] The cosine of the polar angle of the leptons are considered to be $\cos\theta_{l_{1, \rm leading}} < 0.9$, $\cos\theta_{l_{2, \rm trailing}} < 0.92$ and $\cos\theta_{l_{3, \rm trailing}} < 0.94$ respectively.
\end{itemize}
The signal normalized by the square of the mixing and corresponding background cross sections for the trilepton plus fat-jet final state at $\sqrt{s}=$3 TeV are given in Tab.~\ref{tab13}.
\begin{table}[h]
\begin{center}
\begin{tabular}{|c|c|c|c|c|c|}
\hline
\multirow{3}{*}{$\sqrt{s}$ (TeV)} &\multirow{3}{*}{$M_\Sigma$ (GeV)} & \multicolumn{2}{c|}{Signal} & %
    \multicolumn{2}{c|}{Background} \\
\cline{3-6}
&& before cuts (fb) & after cuts (fb) & before cuts (fb) & after cuts (fb)\\
\hline
&1200 &24.67&3.63&&- \\
&&&&&0.0014\\
&1400&21.23&5.11&&-\\
&&&&&0.000741\\
3&1600&16.82&4.06&$e\ell \ell j j :0.491$&-\\
&&&&$eWWZ:$1.353&0.00115\\
&1800&12.1&2.73&&-\\
&&&&&0.00017\\
\hline
\end{tabular}
\caption{Cross sections of the signal $e^-\gamma\to e^-\ell^+\ell^-+J$ (normalized by $|V_{e\Sigma}|^2$) and generic background before and after cuts for $e^- \gamma$ collider at $\sqrt{s}=$3 TeV. In the background column `-' stands for extremely low cross section of the background $e\ell\ell jj$ compared to $eWWZ$.}
\label{tab13}
\end{center}
\end{table}
\subsubsection{Limits on the mixing angle}
\begin{figure}[h]
\centering
\includegraphics[width=0.8\textwidth]{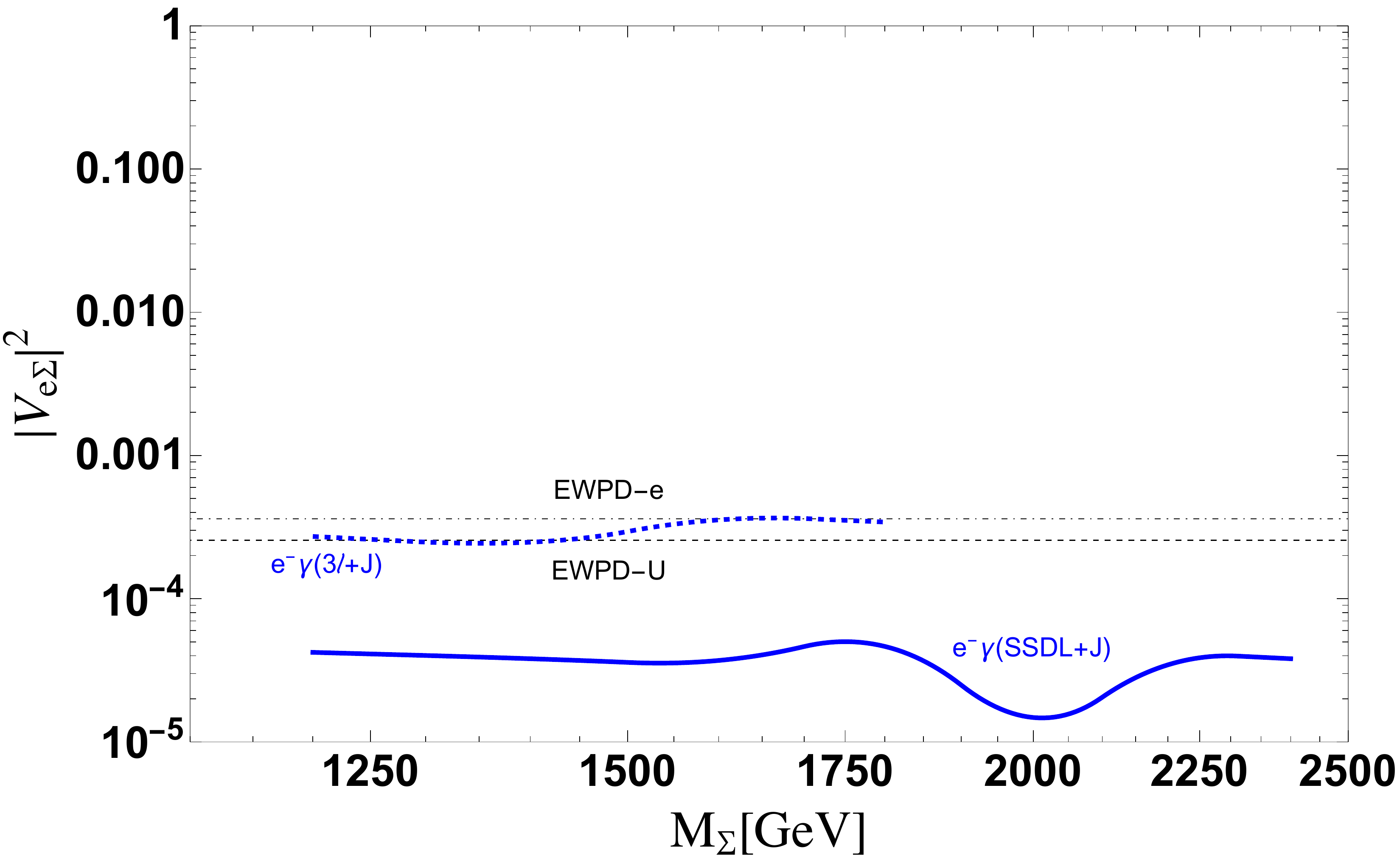}
\caption{$2\sigma$ limits on the mixing angle from SSDL+J in association with missing momentum (thick, blue) and $3\ell+J$ final state (dashed, blue) at $e^-\gamma$ collider with $\sqrt{s}=3$ TeV at 5 ab$^{-1}$ luminosity. We compare the results with the limits from EWPD for electron~(black dot-dashed) and universal EWPD~(black dashed).}
\label{lims-type-3-1}
\end{figure}

For both the case of SSDL$+J$ and $e^-\ell^+\ell^-+J$, studying the signal $(S)$ and the corresponding backgrounds $(B)$ we estimate the 2-$\sigma$ limits on the mixing using Eq.~\ref{signi} taking 5 ab$^{-1}$ of integrated luminosity as a benchmark. The limits coming from SSDL$+J$ and $e^-\ell^+\ell^-+J$ are represented by the thick blue and dashed blue line, respectively in Fig.~\ref{lims-type-3-1}. We compare the results with the limits from EWPD for electron $(V_{e\Sigma}=0.019)$ and universal EWPD $(V_{e\Sigma}=0.016)$ respectively from \cite{delAguila:2008pw}. We find that the 2-$\sigma$ bound on the mixing for the SSDL signal can reach $\mathcal{O}(10^{-5})$ for $M_{\Sigma}=2$ TeV which is well below the expected limits. The 2-$\sigma$ limit on the mixing from the trilepton plus fat-jet signal can reach up to $2.2\times 10^{-4}$ for $M_{\Sigma}=1.475$ TeV which is slightly stronger than the EWPD-U limit. We find that the limits almost remain flat through out the mass range we consider apart from some heavy masses where fat-jet signatures become stronger for the signal over the backgrounds. We do not probe the energy threshold because cross section sharply falls near that region which may provide comparatively weaker bounds on the mixing. 
\section{Discussions and conclusions}
\label{conc}
In this paper we study three types of electroweak scale seesaw scenarios at electron photon and photon photon colliders at different center of mass energies.
We have systemically calculated the cross sections of different production modes of the BSM particles at these colliders averaging over photon spectrum for fixed $\sqrt{s}$ as a function of the corresponding BSM particle mass. First we consider the production of a heavy Majorana neutrinos from the type-I seesaw scenario at the electron photon collider in association with $W$ boson.
From this channel we study $e^\pm+4j$ scenario where the heavy neutrino dominantly decays to $W^\mp e^\pm$, followed by the hadronic decay of 
W bosons produced in this channel. In addition to that we consider the leptonic decay of the associated $W$ boson so that SSDL and OSDL modes could be obtained. 
Considering these final states in association with two jets from the hadronic decay of the $W$ boson coming from the heavy neutrino we estimate the limits on the light-heavy mixing. Hence combining SSDL mode with $e^\pm+4j$ and OSDL mode with $e^\pm+4j$ we estimate limits on the mixing angle separately for comparison at $2\sigma$ significance. For completeness we compare the limits with the existing bounds and prospective limits at electron positron colliders at different center of mass energies. Depending on the center of mass energy, heavy neutrino mass and the choice of the final state signal, the limits on the light-heavy mixing obtained from electron photon colliders could be stronger compare to existing limits. The obtained limits are stronger than the limits obtained from the eletroweak precision study. Second we consider the type-II seesaw scenario where we produce doubly charged scalar at photon photon collider studying a variety of multilepton modes. Studying the signals and SM backgrounds we find that signals with 5$\sigma$ significance could be obtained at 3 TeV photon photon collider where the triplet scalar masses around 1 TeV. Like the singlet fermion we have studied triplet fermion production in electron photon collider from the type-III seesaw scenario as the third case. In this case we consider 3 TeV collider as the triplet mass is greater than 1.2 TeV due to the current LHC limit. We study SSDL mode from the neutral multiplet and trilepton mode from the charged multiplet of the triplet fermion in association with fat-jet. Studying the signal and SM backgrounds we estimate limits on the light-heavy mixings and find stronger bounds compared to those obtained from the eletroweak precision measurements. Finally we mention that a systematic study involving multivariate analysis may improve these results. When electron positron collider will be built in future, unique opportunities of constructing electron photon and photon photon colliders will be possible where one can test  electroweak scale seesaw scenarios. This will help to fix neutrino mass generation mechanism. In addition to that possible studies on Majorana or Dirac nature of neutrinos could also be possible.\\
Finally we would like to stress that in our study we have used the cut-based analyses where the cuts do depend on the mass of the heavy degrees of freedom. This search strategy is always less efficient than a ``bump-hunting" algorithm or a boosted decision tree~(BDT) based on general kinematic variables which are not based on search mass priors. Also we would like to again mention that if one consider polarized beam, the bound might be weaker compare to what we obtained.
\appendix
\section{Photon PDF}
\begin{lstlisting}[language=Fortran, caption=]
 ! bcompton(y,epol,apol) described below is the function which gives the energy spectrum of the photon coming from Compton back-scattering. ! y is the ratio between photon energy and electron energy. epol and apol are the electron and photon polarization.
 ! The bcompton(y,epol,apol) function described below is written based on the paper hep-ex/0108012 can be added at line number 101 of PhotonFlux.f file in Madgraph.

      real*8 function bcompton(y,epol,apol) 
      real*8 y,epol,apol    
      real*8 x,Dx,lee,laa,ymax,r,f
      data x/2.24034d0/

      lee=0.0d0*epol
      laa=apol

      Dx= (1d0-4d0/x-8d0/x**2)*dlog(1d0+x)+0.5d0+8d0/x-0.5d0/(1d0+x)**2
     &   +2d0*lee*laa*( (1d0+2d0/x)*dlog(1d0+x)-2.5d0+1d0/(1d0+x)
     &               -0.5/(1d0+x)**2 )

      ymax=x/(1d0+x)
      if (y.le.ymax) then
        r=y/x/(1d0-y)
        f=( 1d0/(1d0-y) +1d0-y -4d0*r*(1d0-r)
     &     -2d0*lee*laa*r*x*(2d0*r-1d0)*(2d0-y) ) /Dx
      else
        f=0d0
      endif
      bcompton = f
      !write(1,*)y,f
      end



\end{lstlisting}

\begin{acknowledgments}
The authors would like to thank Kentarou Mawatari for useful discussion and providing the codes for photon PDF implementation in Madgraph. The work of S.M. is supported by KIAS Individual Grants (PG086001) at Korea Institute for Advanced Study. The work of S.S. is supported by Funda\c{c}\~ao de Amparo \`a Pesquisa do Estadode S\~ao Paulo (FAPESP) grant 2021/09547-9.
\end{acknowledgments}

\bibliographystyle{utphys}
\bibliography{bibliography}
\end{document}